\documentclass[prd,twocolumn,showpacs,groupedaddress,amsfonts,floatfix]{revtex4}
\usepackage{graphicx}  % needed for figures
\usepackage{dcolumn}   % needed for some tables
\usepackage{bm}        % for math
\usepackage{amsmath}   % for math
\usepackage{multirow}
\newenvironment{description*}
{\vspace{-0.5\baselineskip}
\begin{description}
\setlength{\parskip}{0pt}
\setlength{\itemsep}{0pt plus 1pt minus 1pt}
\setlength{\parsep}{0pt plus 1pt minus 1pt}}
{\end{description}
\vspace{-0.5\baselineskip}
}

\begin{document}

% the following line is for submission, including submission to the arXiv!!
\hspace{5.2in} \mbox{FERMILAB-PUB-09-592-E}

\title{\boldmath Measurement of the $t\bar{t}$ cross section using high-multiplicity jet events}
% LIST_OF_AUTHORS_R2.TEX                 10/14/09           
%
\author{V.M.~Abazov$^{37}$}
\author{B.~Abbott$^{75}$}
\author{M.~Abolins$^{65}$}
\author{B.S.~Acharya$^{30}$}
\author{M.~Adams$^{51}$}
\author{T.~Adams$^{49}$}
\author{E.~Aguilo$^{6}$}
\author{M.~Ahsan$^{59}$}
\author{G.D.~Alexeev$^{37}$}
\author{G.~Alkhazov$^{41}$}
% ...

\affiliation{\vspace{0.1 in}(The D\O\ Collaboration)\vspace{0.1 in}}
% ...
\affiliation{$^{6}$University of Alberta, Edmonton, Alberta, Canada;
                Simon Fraser University, Burnaby, British Columbia, Canada;
                York University, Toronto, Ontario, Canada and
                McGill University, Montreal, Quebec, Canada}
% ...
\affiliation{$^{30}$Tata Institute of Fundamental Research, Mumbai, India}
% ...
\affiliation{$^{37}$Joint Institute for Nuclear Research, Dubna, Russia}
% ...
\affiliation{$^{41}$Petersburg Nuclear Physics Institute,
                St. Petersburg, Russia}
% ...
\affiliation{$^{49}$Florida State University, Tallahassee, Florida 32306, USA}
% ...
\affiliation{$^{51}$University of Illinois at Chicago,
                Chicago, Illinois 60607, USA}
% ...
\affiliation{$^{59}$Kansas State University, Manhattan, Kansas 66506, USA}
% ...
\affiliation{$^{65}$Michigan State University,
                East Lansing, Michigan 48824, USA}
% ...
\affiliation{$^{75}$University of Oklahoma, Norman, Oklahoma 73019, USA}
% ...
  % input Dzero author list
\date{November 22, 2009}

\begin{abstract}
  We present a measurement of the $t\bar{t}$ cross
  section using high-multiplicity jet events produced in $p\bar{p}$
  collisions at $\sqrt{s}=1.96$~TeV.  These data were recorded at
  the Fermilab Tevatron collider with the D0 detector.  Events with at
  least six jets, two of them identified as $b$ jets, were selected
  from a 1~fb${}^{-1}$ data set.  The measured cross section, assuming
  a top quark mass of $175$~GeV$/c^2$, is $6.9\pm 2.0$~pb, in
  agreement with theoretical expectations.
\end{abstract}

\maketitle 

\section{Introduction}

\begin{figure}[b!]
\centering
\includegraphics[width=0.47\textwidth]{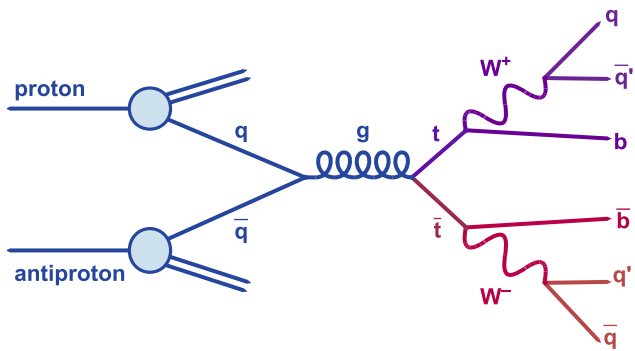}
\caption{\label{fig:alljets_event_signature} Dominant Feynman diagram
  for $t\bar{t}$ production in the all-hadronic decay channel.  The
  $t$ decays into a $W^+b$ and the $W^+$ decays into either $u\bar{d}$
  or $c\bar{s}$ (represented by the $q$ and $\bar{q}^\prime$ in the
  figure); the $\bar{t}$ and $W^-$ decay into the charge conjugates.
  The event signature consists of two $b$ jets and at least four other
  jets.}
\end{figure}

The top quark is the most massive fundamental particle ever observed.
Its mass, $m_t=173.1\pm1.3$~GeV/$c^2$~\cite{world-average-mass}, is
approximately twice that of the next heaviest elementary particle, the
$Z$ boson, and is approximately $35$~times that of its weak-isospin
partner, the bottom quark.
Top quarks are primarily produced in pairs at the Fermilab Tevatron
$p\bar{p}$ collider via the $q\bar{q}\rightarrow t\bar{t}$
($\approx85\%$) and $gg\rightarrow t\bar{t}$ ($\approx 15\%$) quantum
chromodynamic (QCD) processes.
They decay to a $W$~boson and a $b$~quark with a branching fraction
near one according to the standard model (SM).  The $W$~boson
subsequently decays into a lepton and a neutrino or into a
quark-antiquark pair.  The decay products of the $W$ bosons are used
to classify the top quark decay channel.  The all-hadronic decay
channel, with a branching fraction of $46\%$~\cite{PDG}, has a final
state containing two $b$~quarks and four lighter quarks and is shown
schematically in Fig.~\ref{fig:alljets_event_signature}.
The top quark might also decay into non-SM particles (e.g., a charged
Higgs boson) and the decay products of these new particles can change
the branching fractions of the leptonic and all-hadronic $t\bar{t}$
decay channels~\cite{theory}.  Comparing the $t\bar{t}$ production
cross section between different decay channels directly constrains the
existence of beyond the standard model particles lighter than the top
quark.

In this paper, we present a new measurement of the
$p\bar{p}\rightarrow t\bar{t}+X$ cross section using events containing
at least six jets, two of them identified as $b$ jets.  The data
sample corresponds to $\approx 1$~fb${}^{-1}$ acquired by the D0
experiment at a center-of-mass energy $\sqrt{s}=1.96$~TeV.
D0 previously published a measurement of the $t\bar{t}$ cross section
in multijet events with $0.4$~fb${}^{-1}$ of integrated luminosity and
obtained $4.5^{+2.0}_{-1.9}\;({\rm stat.})^{+1.4}_{-1.1}\;({\rm
  sys.})\pm0.3\;({\rm lum.})$~pb~\cite{oldAlljet}.  CDF published a
similar measurement with $1$~fb${}^{-1}$ and obtained
$8.3\pm1.0\;({\rm stat.})^{+2.0}_{-1.5}\;({\rm sys.})\pm0.5\;({\rm
  lum.})$~pb~\cite{CDFAlljet}.
Both measurements assumed $m_t=175$~GeV$/c^2$ and agree with the cross
section measurement presented in this paper and with the SM
expectation of $6.90^{+0.44}_{-0.62}$~pb~\cite{moch,kidonakis}.
The dominant source of background in the all-hadronic channel is QCD
multijet production.  Rather than relying on event generators such as
{\sc pythia}~\cite{pythia}, {\sc herwig}~\cite{herwig}, or {\sc
  alpgen}~\cite{alpgen} to reproduce all characteristics of events
with six or more jets, we instead derived a background sample from the
triggered data (Sec.~\ref{sec:background}).
The background was suppressed compared to signal by requiring at least
two of the jets be identified as $b$ jets (Sec.~\ref{sec:b-jets}).
The $t\bar{t}$ signal was simulated by the {\sc alpgen} event
generator that used {\sc pythia} with the tune A~\cite{tuneA}
parameter settings for the parton shower, hadronization, and
underlying event aspects.
Kinematic selection criteria were applied to
further improve the signal-to-background ratio to approximately $1:7$
(Sec.~\ref{sec:selection}).
The $t\bar{t}$ production cross section was extracted using signal and
background templates for a likelihood discriminant constructed from
topological and kinematic observables.  (Sec.~\ref{sec:results}).

\section{Detector and Reconstruction}

\subsection{Detector}

The D0 detector~\cite{run2det} has a central-tracking system
consisting of a silicon microstrip tracker (SMT) and a central fiber
tracker (CFT), both located within a 2~T superconducting solenoidal
magnet, with designs optimized for tracking and
vertexing at pseudorapidities $|\eta|<3$ and $|\eta|<2.5$,
respectively~\cite{pseudorapidity}.  Central and forward preshower
detectors are positioned just outside of the superconducting coil.
The liquid-argon and uranium calorimeter has a central section (CC)
covering pseudorapidities $|\eta|\lesssim1.1$ and two end
calorimeters (EC) that extend coverage to $|\eta|\approx 4.2$, with
all three housed in separate cryostats~\cite{run1det}.  
Each calorimeter contains a four-layer electromagnetic (EM) section 
closest to the interaction region, followed by finely- and
coarsely-segmented hadronic sections.
Scintillators between the CC and EC cryostats provide sampling of
developing showers at $1.1\lesssim|\eta|\lesssim1.4$.  The luminosity is
measured using scintillators placed in front of the EC
cryostats~\cite{luminosity}.
An outer muon system, covering $|\eta|<2$, consists of a layer of
tracking detectors and scintillation trigger counters in front of
1.8~T iron toroids, followed by two similar layers beyond the toroids.
The trigger and data acquisition systems were designed to accommodate
the high luminosities of Tevatron Run~II.

\subsection{\label{sec:trigger}Trigger}

The events used in this analysis were collected using a multijet
trigger.  The first level of the trigger used dedicated hardware and
preliminary information from the calorimeter to identify multijet
events.  This selection was refined in a second level with more
complex algorithms.  The third trigger level employed a fast
reconstruction of the event with a simple cone jet
algorithm~\cite{d0jets}.  This selection was further refined using the
final reconstruction algorithms which included the midpoint cone jet
algorithm~\cite{d0jets}.  Kinematic and jet multiplicity requirements
were applied at each stage to reduce the overall data rate.

\begin{figure}[t]
\includegraphics[width=0.49\textwidth]{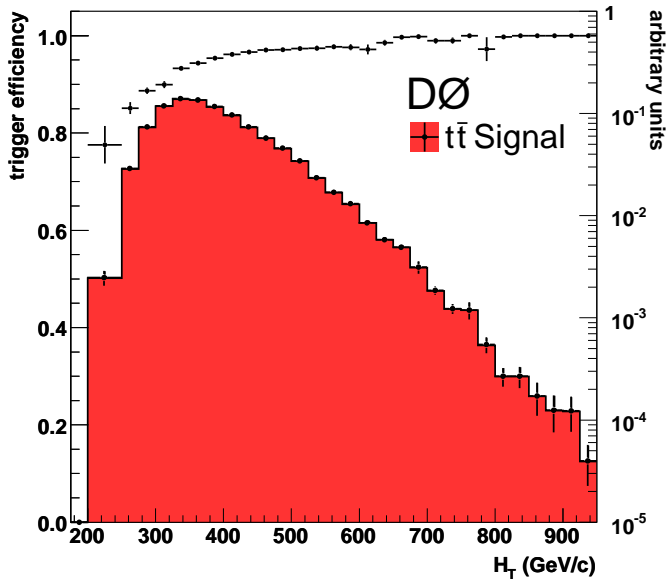}
\caption{\label{fig:trigeff} Average trigger efficiency for simulated
  all-hadronic $t\bar{t}$ events as a function of $H_T$.  The
  untriggered $t\bar{t}$ $H_T$ distribution, normalized to unit area,
  is also shown (scale shown on the right.)  Displayed error bars
  represent statistical uncertainties only.  }
\end{figure}

The trigger required at least four reconstructed jets.  The specific
requirements on the jets, particularly the energy thresholds, were
changed several times during data collection to cope with the
increasing instantaneous luminosity delivered by the Tevatron.
Efficiencies were measured independently for each trigger epoch and
combined together weighted by the integrated luminosity of each epoch.
Rather than correcting the data for inefficiencies in the trigger, the
simulated $t\bar{t}$ signal was weighted by the trigger efficiency.
The average trigger efficiency for $t\bar{t}$ signal events that
passed all selection criteria used in this analysis is shown in
Fig.~\ref{fig:trigeff} as a function of $H_T$ where $H_T=\sum p_T$
over all jets and $p_T$ is the transverse momentum of a jet.  The
background sample was created from the triggered data (see
Sec.~\ref{sec:background}) and therefore need no additional
corrections.

\subsection{\label{sec:vertex}Tracks and Vertices}

\begin{figure}[t!]
\includegraphics*[width=0.49\textwidth,trim=0 0 0 50]{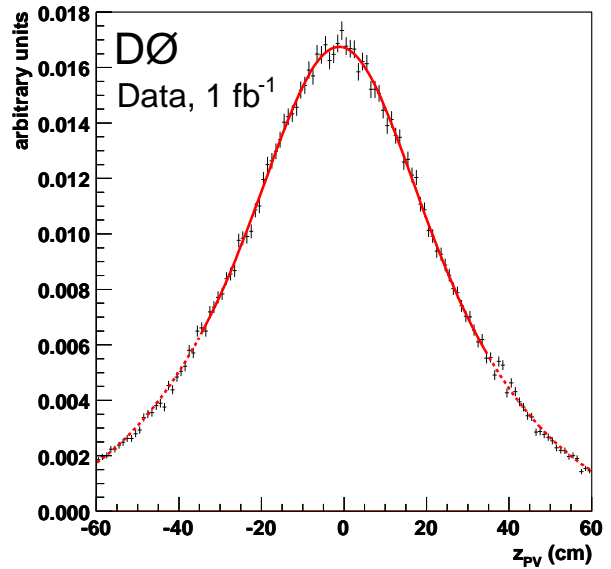}
\caption{\label{fig:pvz_extrapolation} 
The distribution of the primary vertex $z$ position 
with respect to the center of the detector in the triggered data.
The solid line is a fit to the
  region with $|z_{\rm PV}|<35$~cm, while the dotted line is an
  extrapolation of the fit outside that region. 
Displayed error bars represent statistical uncertainties only.
The distribution is normalized to unit area.
}
\end{figure}

Tracks were reconstructed from hit information in the SMT and CFT.
The location of the hard-scatter interaction point was reconstructed
by means of an adaptive primary vertex algorithm~\cite{adaptive-pv,lepjetPRD}.
Only vertices constructed from at least three tracks were considered
in this analysis; ${\cal O}(40)$ tracks are associated, on
average, with primary vertices in simulated all-hadronic $t\bar{t}$
events.
A distribution of the location of primary vertices along the $z$~axis
in triggered events is displayed in Fig.~\ref{fig:pvz_extrapolation}.
The primary interaction vertex was required to be within $35$~cm of
the center of the detector along the $z$ axis to keep it within the
fiducial volume of the SMT~\cite{lepjetPRD}. 
The distribution in Fig.~\ref{fig:pvz_extrapolation} was fitted within
the $|z_{\rm PV}| < 35$~cm range with the sum of two Gaussians.  The
fit extrapolation outside this range is also shown.  The total primary
vertex acceptance was $79.5\pm2.0\%$.
%Integrating the
%fit result yields a primary vertex acceptance of $79.5\pm2.0\%$.

\subsection{Jets}

Jets were reconstructed from energy deposits in calorimeter cells
using the Run~II midpoint cone algorithm~\cite{d0jets} with a cone
radius $\mathcal{R}=\sqrt{(\Delta\phi)^2+(\Delta
  y)^2}=0.5$~\cite{yphi}.
Noisy calorimeter cells were suppressed by only including cells that
had energies $\geq4\sigma$ above the average electronic noise and that
also had adjacent cells with energies $\geq2\sigma$ above noise.
Jets were required to have $<40\%$ of their energy in the coarse
hadronic calorimeter, have at least half the remaining transverse
energy matched to energy depositions identified by the hardware
trigger, and have between~$5$\% and~$95$\% of their energy in the EM
calorimeter.  These requirements were for jets reconstructed in the
CC; they were looser at forward rapidities.

Jet energies were corrected for the energy response of the
calorimeter, for the effect of particles showering outside the jet
cone, for overlaps due to multiple interactions and event pile-up, and
for calorimeter noise~\cite{JES}.
The calorimeter response was measured using the $p_T$ imbalance in
$\gamma+$jet and dijet events; the response of the calorimeter to
electromagnetic showers was calibrated using the $Z\rightarrow e^+e^-$
mass peak and a detailed accounting of the material between the
calorimeter and the interaction point.
The jet energy calibration also used $Z+$jet events and events
acquired using low bias triggers.
Jets that contained muons, assumed to originate from $c$- or
$b$-hadron decays, were corrected to account for the energy of the
muon and the accompanying neutrino.  Muons with $p_T>60$ GeV$/c$ were
treated as having $p_T=60$~GeV$/c$ to avoid the impact from poorly
reconstructed muon momenta.
Jet energies were calibrated independently in the data and in the
simulation using the same methodology.  Jets in the simulation
required additional corrections to reproduce the reconstruction
efficiency and energy resolution in the data. The uncertainty on the
jet energy calibration is $\approx1.5\%$.

\begin{figure}[t!]
  \includegraphics[width=0.5\textwidth,trim=20 20 20 40,clip=true]{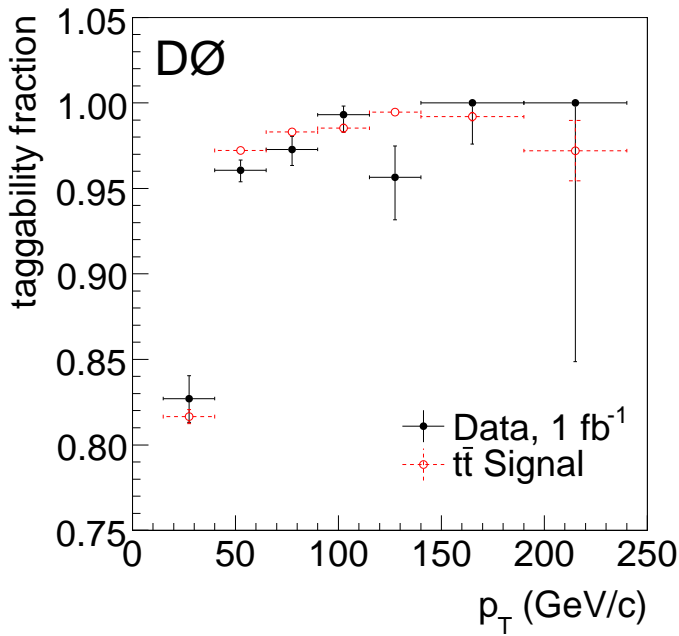}
  \caption{Comparison of taggability fraction in selected multijet data after
    selection with that in the $t\bar{t}$ simulation as a
    function of jet $p_T$.  Displayed error bars represent statistical
    uncertainties only.}
  \label{fig:taggability}
\end{figure}

Jets were further required to be matched with at least two good
quality tracks having $p_T>1$~GeV$/c$ and $p_T>0.5$~GeV$/c$,
respectively, that included SMT hits and pointed to the primary
vertex.  These requirements are termed ``taggability'' and are
important for identifying heavy-flavor jets (Sec.~\ref{sec:b-jets})
and to reject jets produced by overlapping $p\bar{p}$ collisions.
The taggability fraction depends nominally on the jet $p_T$, jet
rapidity, $z_{\rm PV}$, ${\rm sign}(z_{\rm PV}\times \eta_{\rm jet})
\times |z_{\rm PV}|$, and the flavor of the jet~\cite{lepjetPRD}.
The fraction of jets that were taggable was measured using the
selected sample of multijet events (Sec.~\ref{sec:selection}) and is
shown in Fig.~\ref{fig:taggability} binned in jet $p_T$.
Differences between the taggability determined with multijet data and
with the $t\bar{t}$ signal simulation could bias the cross section
measurement. 
The $t\bar{t}$ simulation yielded the same taggability fraction as a
function of jet $p_T$ and $\eta$ as the multijet data within the
statistical uncertainties (Fig.~\ref{fig:taggability}).
The uncertainty on the relative difference between data and simulation
is $2\%$ and is dominated by the limited statistics in the comparison.

\subsection{\label{sec:b-jets}$b$ Jets}

Jets that contain a $b$~hadron are called ``$b$ jets'' as they
typically originate from a $b$ quark.  $b$~hadrons have
relatively long lifetimes and so usually travel several millimeters
before they decay.  Secondary vertices, displaced from the primary
vertex, are usually formed by the tracks associated with the decay
products of the $b$ hadron.

An artificial neural network (NN) was used to identify $b$
jets~\cite{thesis-scanlon}.  Selected characteristics of secondary
vertices and tracks associated with $b$ hadron decays were used as
inputs to the NN.  These included aspects of the secondary vertex such
as its decay length significance, goodness of fit, number of tracks,
mass of the system of particles associated with the vertex, and the
number of secondary vertices found in the jet.  Additionally, the
weighted combination of track impact parameter significances and the
probability that the jet originated from the primary vertex were also
input into the NN.

The probability to identify a $b$ jet, the tag rate function, was
measured in data and parametrized as a function of the jet $p_T$ and
$\eta$.  Similar functions were determined for charm jets.  The fake
rate, the probability to assign a $b$ tag to a non-$b$ jet, was
dominated by light jets and long-lived particles (e.g., $K_s^0$,
$\Lambda^0$).
The $b$-tagging efficiency is $(57\pm2)\%$, the tagging efficiency for
charm is $(15\pm1)\%$, and the fake rate is $(0.57\pm0.07)\%$ for
the NN output threshold used in this analysis at
$p_T=40$~GeV$/c$~\cite{thesis-scanlon}.

\section{Analysis Techniques}

\subsection{\label{sec:datasample} Data Sample}

The data used for this analysis were collected between August 2002 and
February 2006 with the four-jet trigger described in
Sec.~\ref{sec:trigger}.
Quality requirements were imposed on the selected data; runs or parts
of runs in which detector systems essential to this analysis had
problems or significant noise were discarded.
The integrated luminosity of the data sample, including these trigger
and quality requirements, is $0.97\pm0.06$~fb${}^{-1}$.

\begin{figure*}[!t]
%0.49
%\includegraphics*[width=0.49\textwidth,trim=30 5 20 40]{\PATHsmallfonts/background_extra/bpt_plots/15/plot_bpt_explanation_bpt15_drb0b1_lin.eps}
\includegraphics*[width=0.49\textwidth,trim=30 5 20 40]{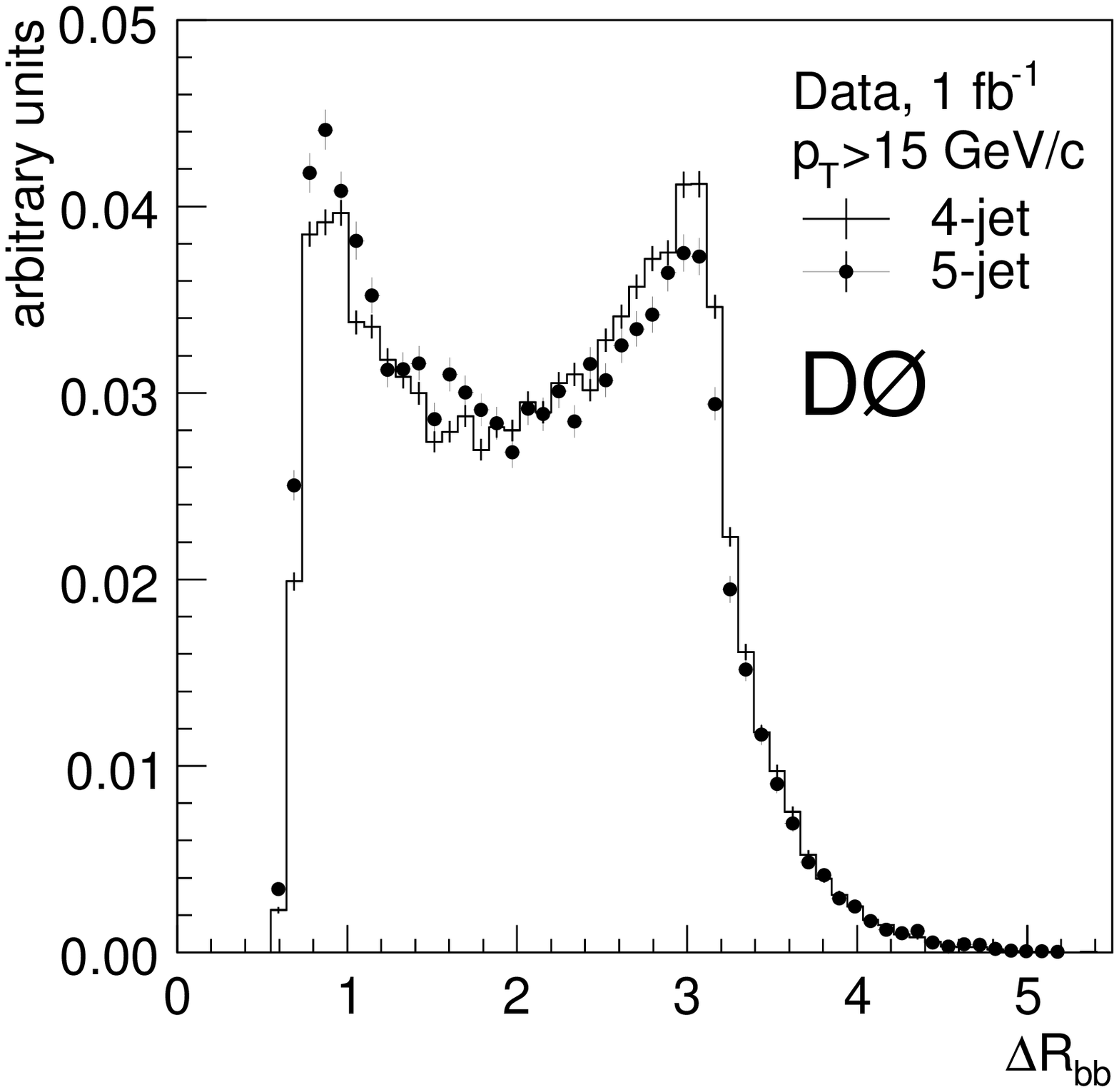}
\put(-205,220){\textsf{\textbf{(a)}}}%
\includegraphics*[width=0.49\textwidth,trim=30 5 20 40]{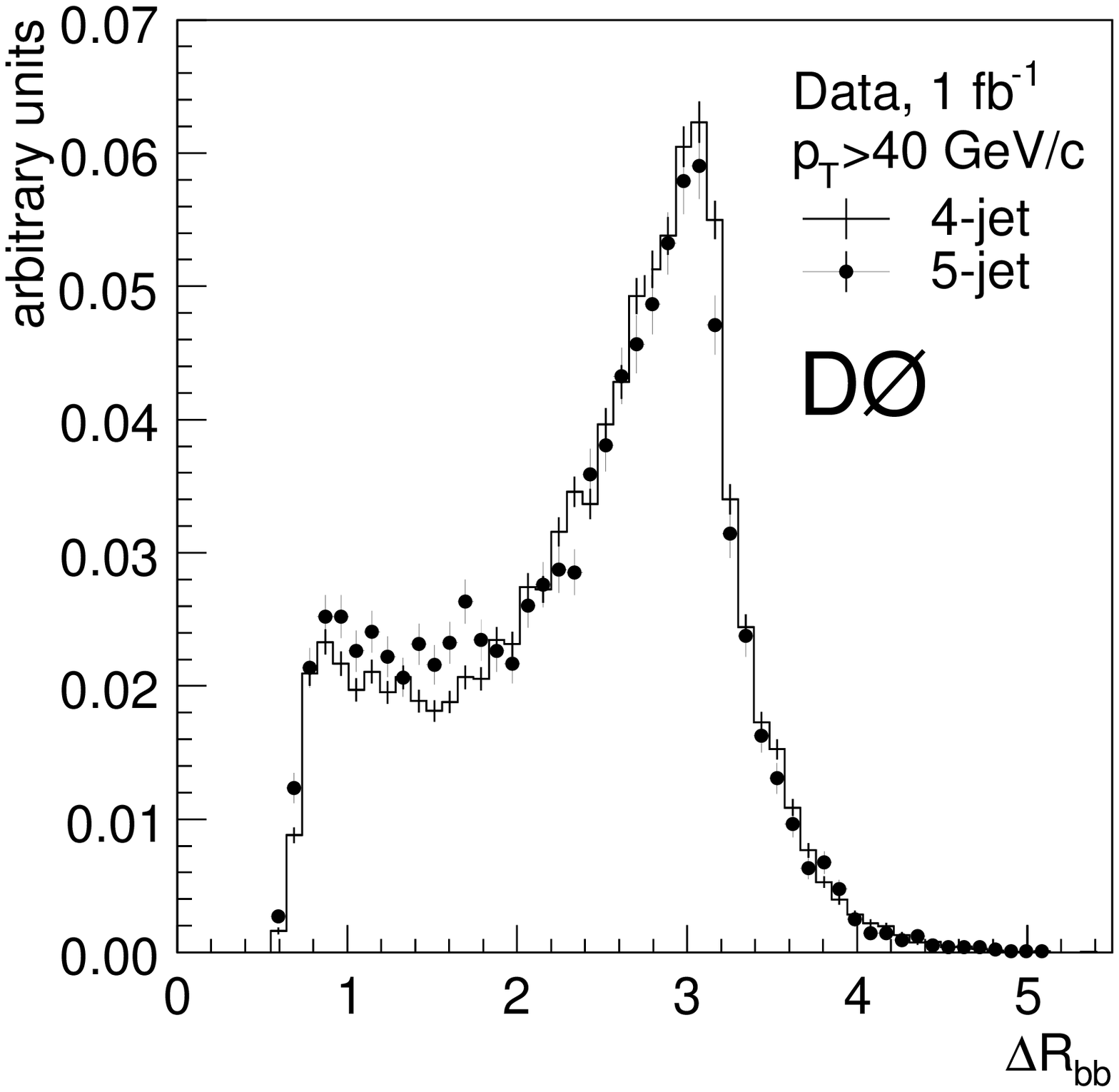}
\put(-205,220){\textsf{\textbf{(b)}}}%
\caption{\label{fig:deltarbb} $\Delta R$ between the two leading
  $b$-tagged jets in 4-jet and 5-jet events with (a) $p_T>15$~GeV$/c$;
  (b) $p_T>40$~GeV$/c$.  The peak near~$\Delta R\approx\pi$ is
  dominated by direct $b\bar{b}$ production while the peak near
  $\Delta R\approx1$ (twice the jet radius) is mainly $g\rightarrow
  b\bar{b}$. 
Displayed error bars represent statistical uncertainties only.
Distributions are normalized to unit area. }
\end{figure*}

\subsection{\label{sec:background}Background Model}

QCD multijet events that have at least two heavy-flavor jets are the
dominant source of background to $t\bar{t}$ production in the
all-hadronic decay channel.
This large background is distinguished from the $t\bar{t}$ signal by
exploiting differences between the kinematic and topological
distributions of jets in $t\bar{t}$ and multijet events.  Correlations
between jets, particularly for $b$ jets,
must be reproduced for the observables used in this analysis.

The background sample was created using triggered data events.  Signal
contamination in the background sample was minimized by selecting
events with two $b$-tagged jets and low jet multiplicities.  Samples of
events with at least four taggable jets having $p_T>15$~GeV$/c$ were
selected from the triggered data.  The $b$-jet identification criteria
described in Sec.~\ref{sec:b-jets} were applied to these samples;
events were kept if there were at least two tagged jets.  
The background sample was then created by attaching low-$p_T$ jets
selected from events with six or more jets to events with four or five
jets.  A reasonable distribution of the jets in the available phase
space was ensured using a set of matching criteria.

One concern with basing the background distributions on a lower
jet-multiplicity sample was that the relative contributions of
different production diagrams might depend strongly on jet
multiplicity.  This was tested by examining distributions of the
$\Delta R$ between the $b$ jets.  We expect a peak near~$\pi$ for
$b\bar{b}$ produced in $2\rightarrow 2$ hard scatters, whereas we
expect a peak near one (twice the jet radius) for $b\bar{b}$ produced
via gluon splitting, $g\rightarrow b\bar{b}$.  This is illustrated for
four and five jet events in Fig.~\ref{fig:deltarbb}.
Figure~\ref{fig:deltarbb}(a) shows $\Delta R_{bb}$ for $b$~jets with
$p_T>15$~GeV$/c$ while Fig.~\ref{fig:deltarbb}(b) is the $\Delta
R_{bb}$ for $b$~jets with $p_T>40$~GeV$/c$. The relative height of the
two peaks depends strongly on the $p_T$ requirement, but there is
little difference between four- and five-jet events.  The
gluon-splitting contribution is significantly suppressed by increasing
the $b$-jet $p_T$ requirement from~$15$ to~$40$~GeV$/c$.

The scheme for creating a background sample was developed in a
relatively pure QCD multijet context.  A ``background'' sample was
constructed by adding the lowest $p_T$ jet from five-jet events to
four-jet events.  The two sources of jets were matched together to
ensure compatible phase-space configurations.  The leading jets in
each sample were required to have a difference in $p_T$ ($\Delta p_T$)
within 1~GeV$/c$.
Matches resulting in unphysical configurations (e.g., spatially
overlapping jets) were rejected.
The background event statistics were enhanced by running twenty times
over the four- and five-jet samples.  In each step the $\Delta p_T$
requirement was relaxed by $1$~GeV$/c$.
\begin{figure*}[!t]
  \includegraphics*[width=0.32\textwidth,trim=30 10 30 40,clip=true]{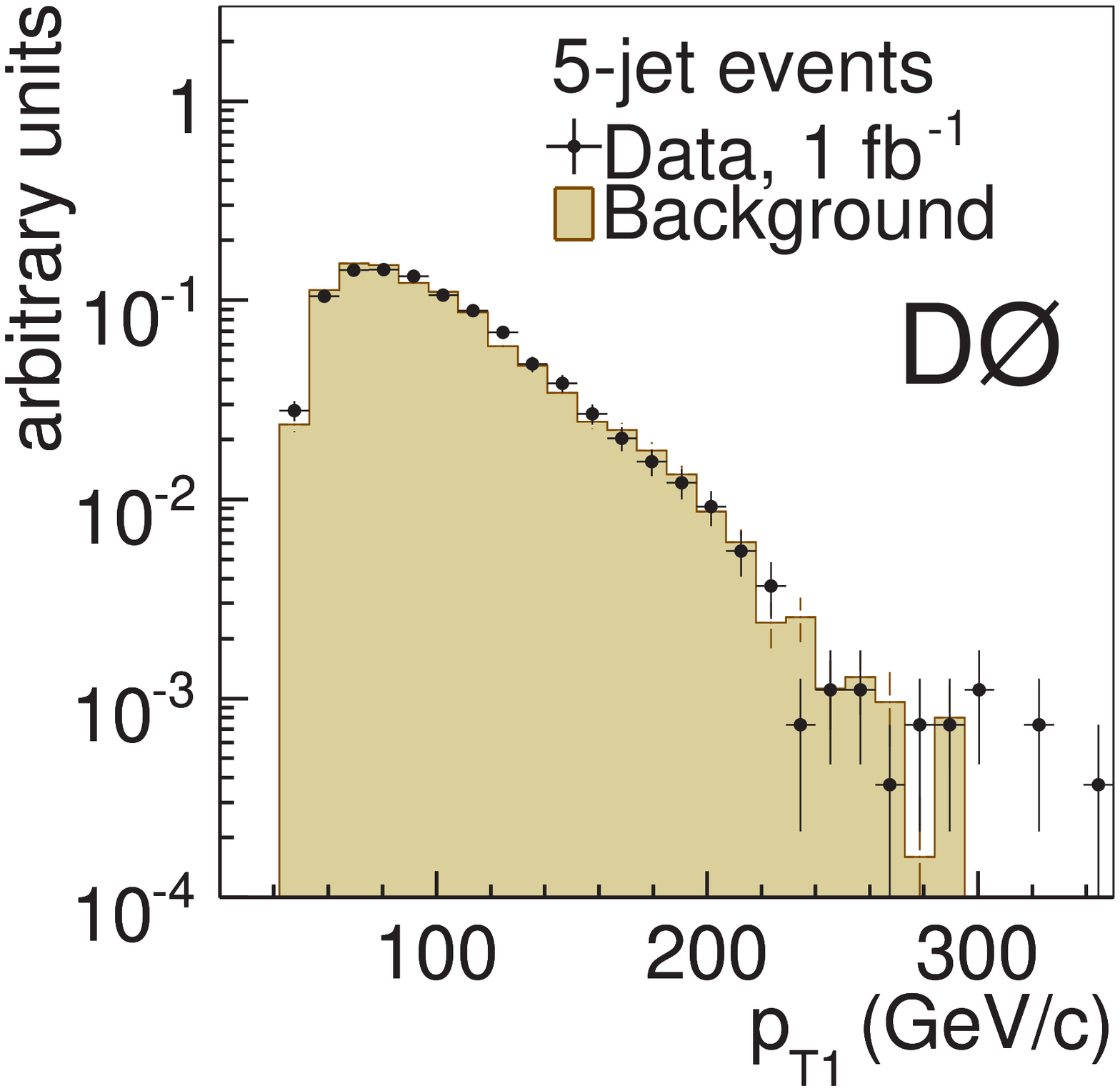}
\put(-130,145){\textsf{\textbf{(a)}}}%
\includegraphics*[width=0.32\textwidth,trim=30 10 30 40,clip=true]{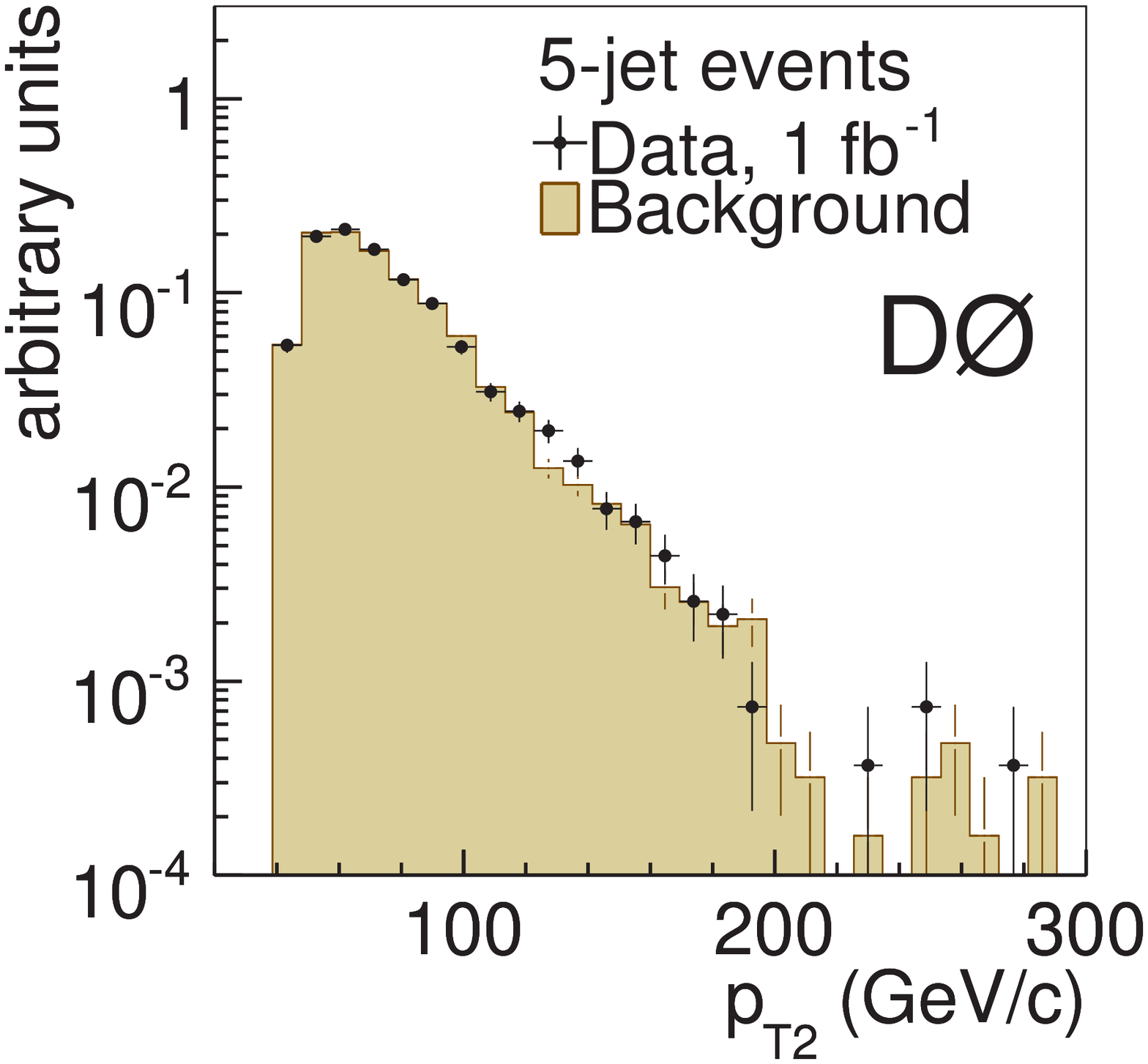}
\put(-130,145){\textsf{\textbf{(b)}}}%
\includegraphics*[width=0.32\textwidth,trim=30 10 30 40,clip=true]{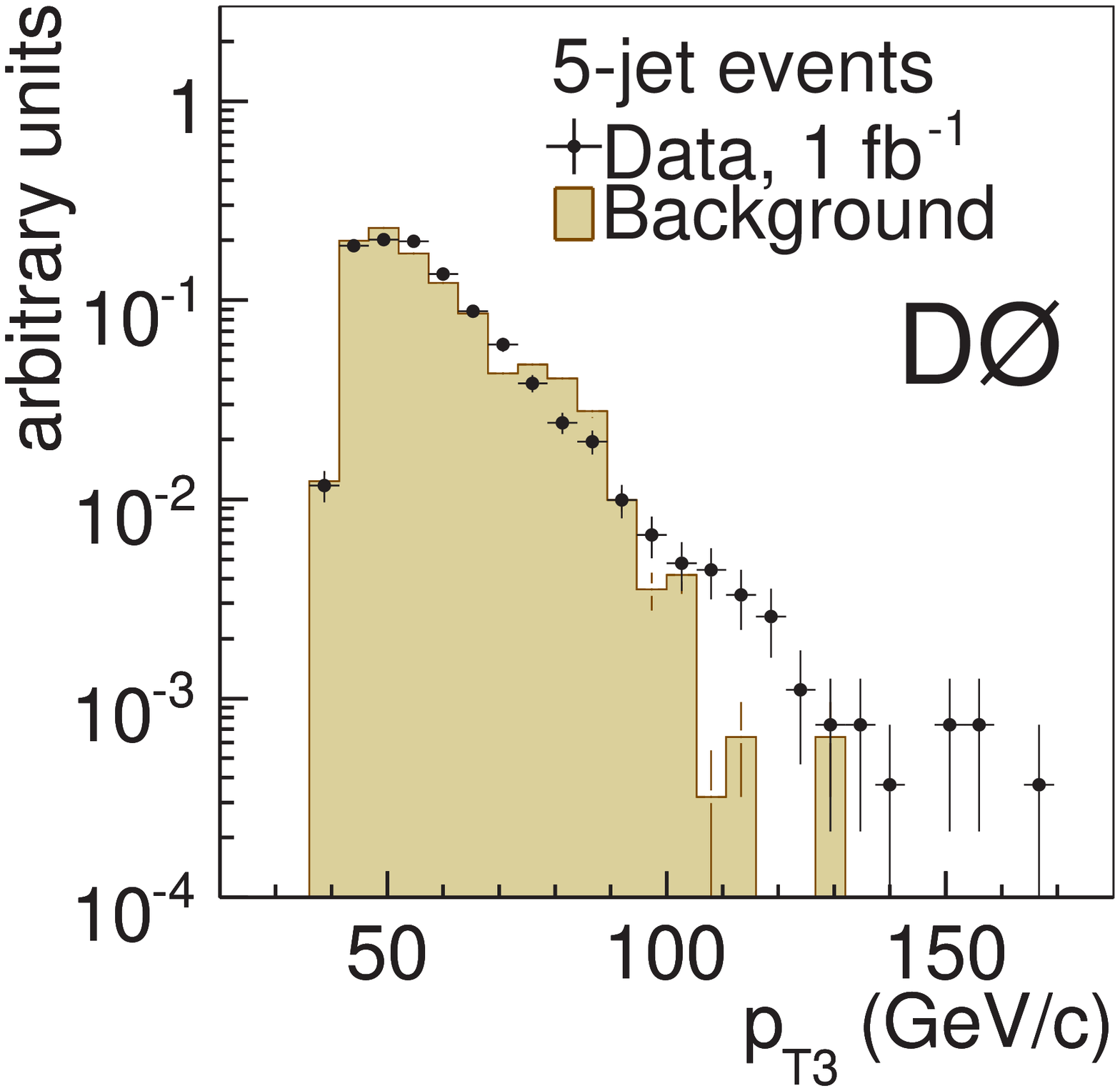}
\put(-130,145){\textsf{\textbf{(c)}}}%
\\
\includegraphics*[width=0.32\textwidth,trim=30 10 30 40,clip=true]{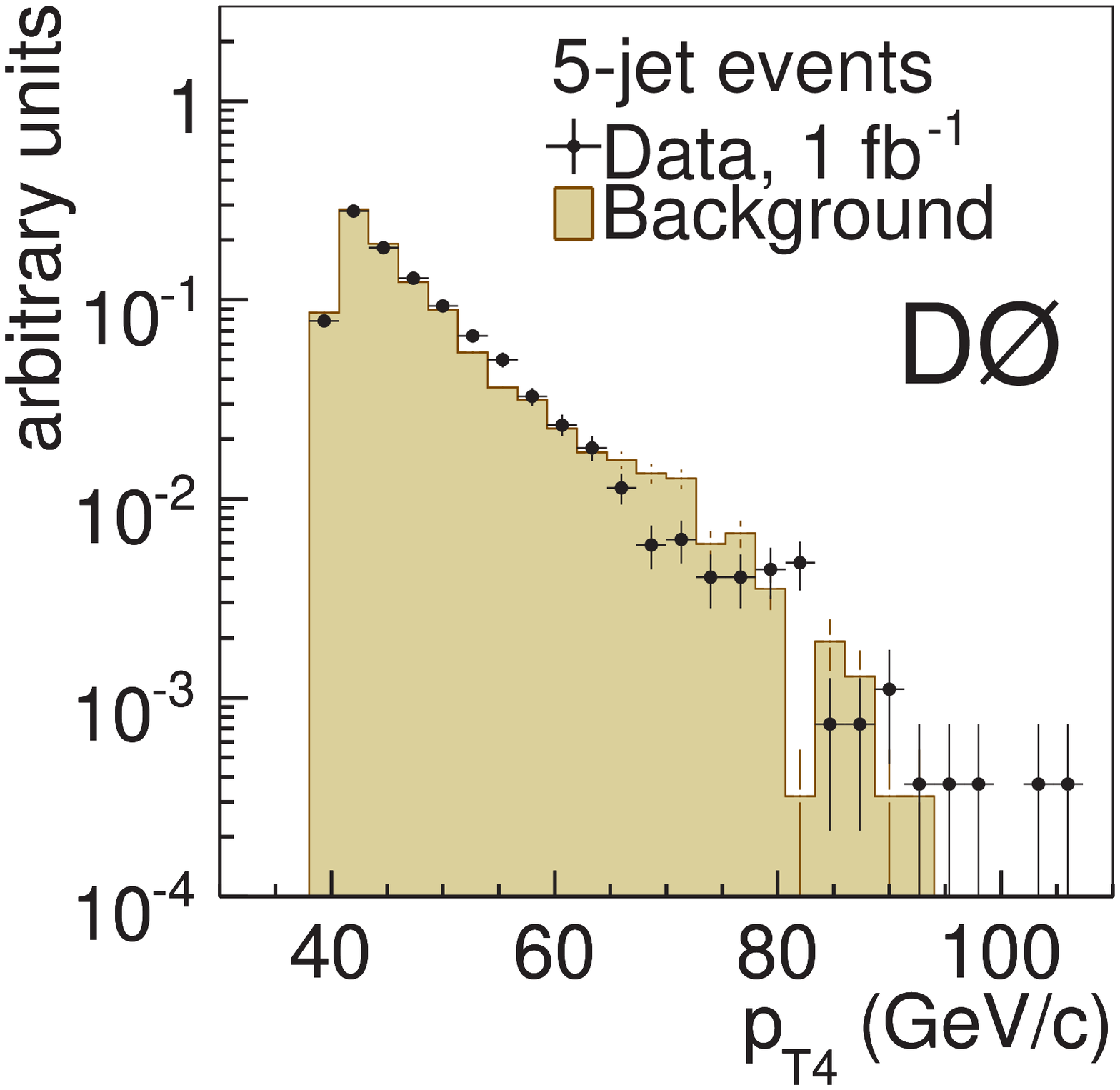}
\put(-130,145){\textsf{\textbf{(d)}}}%
\includegraphics*[width=0.32\textwidth,trim=30 10 30 40,clip=true]{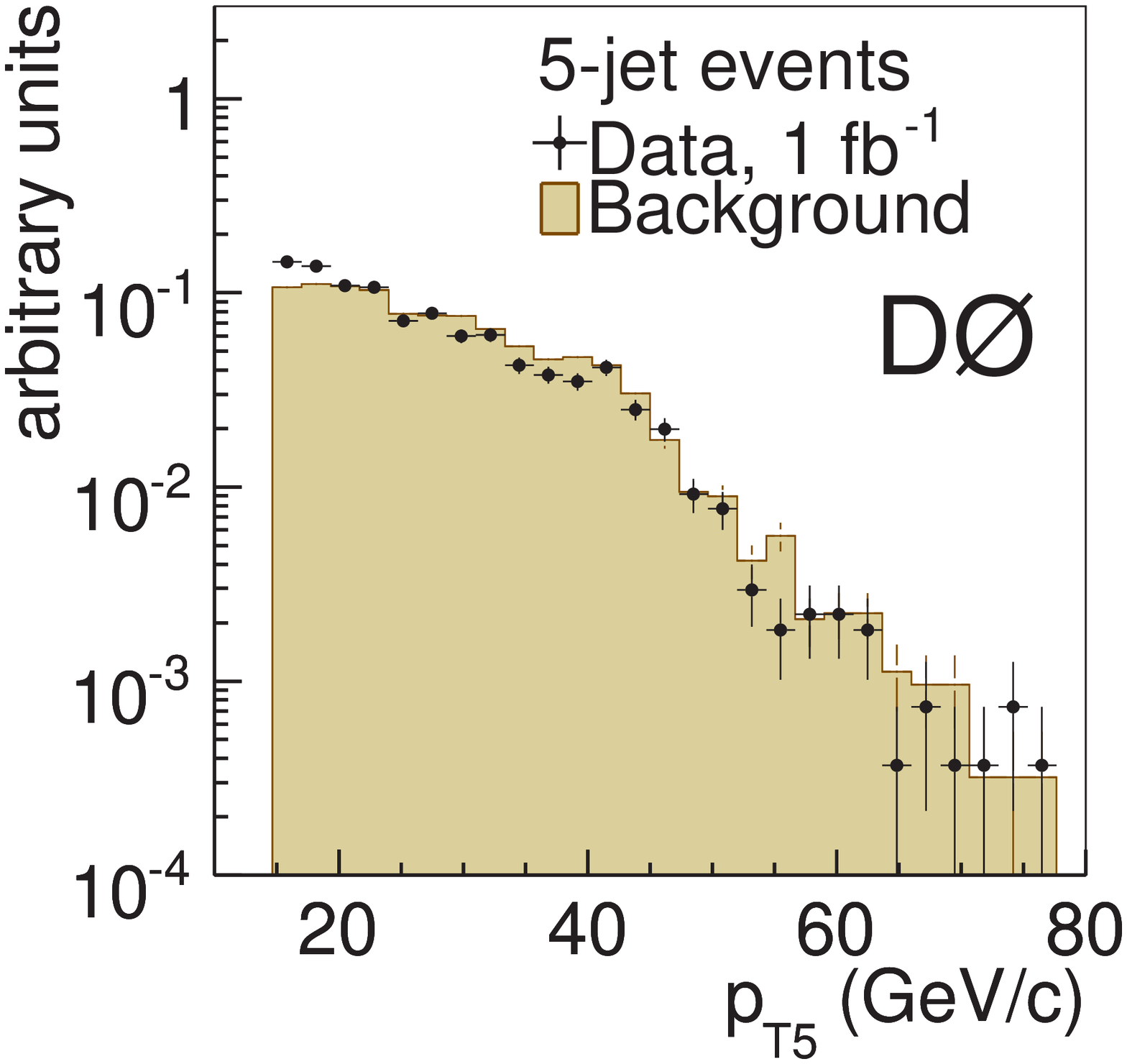}
\put(-130,145){\textsf{\textbf{(e)}}}%
\includegraphics*[width=0.32\textwidth,trim=30 10 30 40,clip=true]{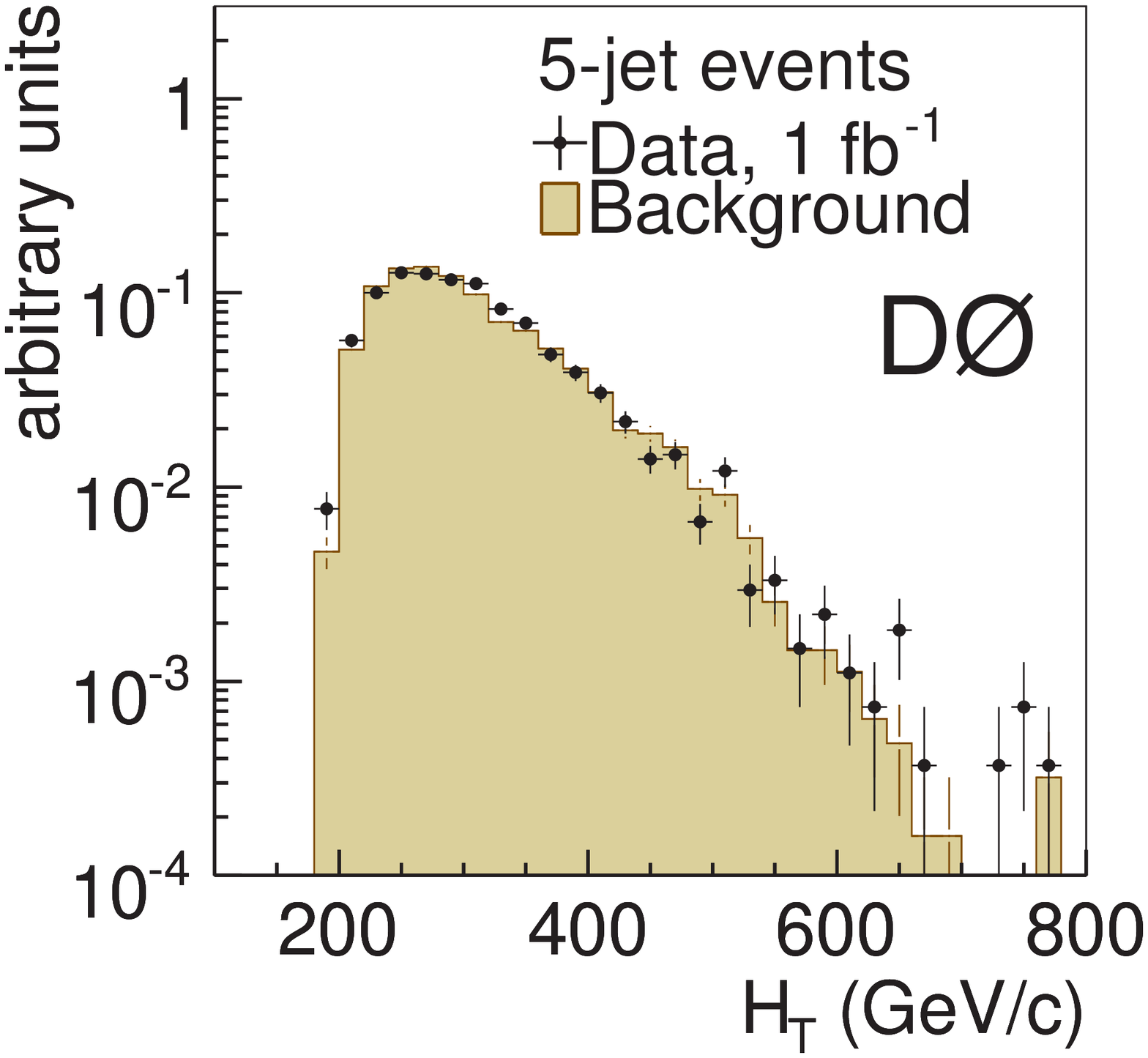}
\put(-130,145){\textsf{\textbf{(f)}}}%
\caption{\label{fig:4vs5pt} Comparisons between the five-jet data and
  the background created from four-jet data for the $p_T$
  distributions of each jet ($p_{T1}$ through $p_{T5}$) and for $H_T$.
  The leading four jets were required to have $p_T>40$~GeV$/c$.
Displayed error bars represent statistical uncertainties only.
Distributions are normalized to unit area.}
\end{figure*}

One issue with this matching scheme is that an initial four-jet event
might not have sufficient phase space for an additional jet.  Since
QCD multijet events are not expected to contain significant missing
transverse energy ($\not\!\!E_T$), the presence of $\not\!\!E_T$
implies the presence of unreconstructed or mismeasured jets which
makes these events more suitable for use in the background sample.
However, badly mis-reconstructed events or events containing hard
neutrinos can skew the phase space.  Requiring the ratio of
$\not\!\!E_T$ to $H_{T4}\equiv\sum_{i=1}^4 p_{Ti}$ to be small reduced
these contributions.  Agreement between the ``signal'' (unadulterated)
and ``background'' five-jet samples was best with
$\not\!\!\!E_T>5$~GeV$/c$ and $\not\!\!\!E_T/H_{T4}<0.1$.  Variations
in this additional phase space selection were included in the
systematic uncertainty evaluation~\cite{variations}.

\begin{figure*}[!t]
\includegraphics*[width=2.25in,trim=20 5 25 35,clip=true]{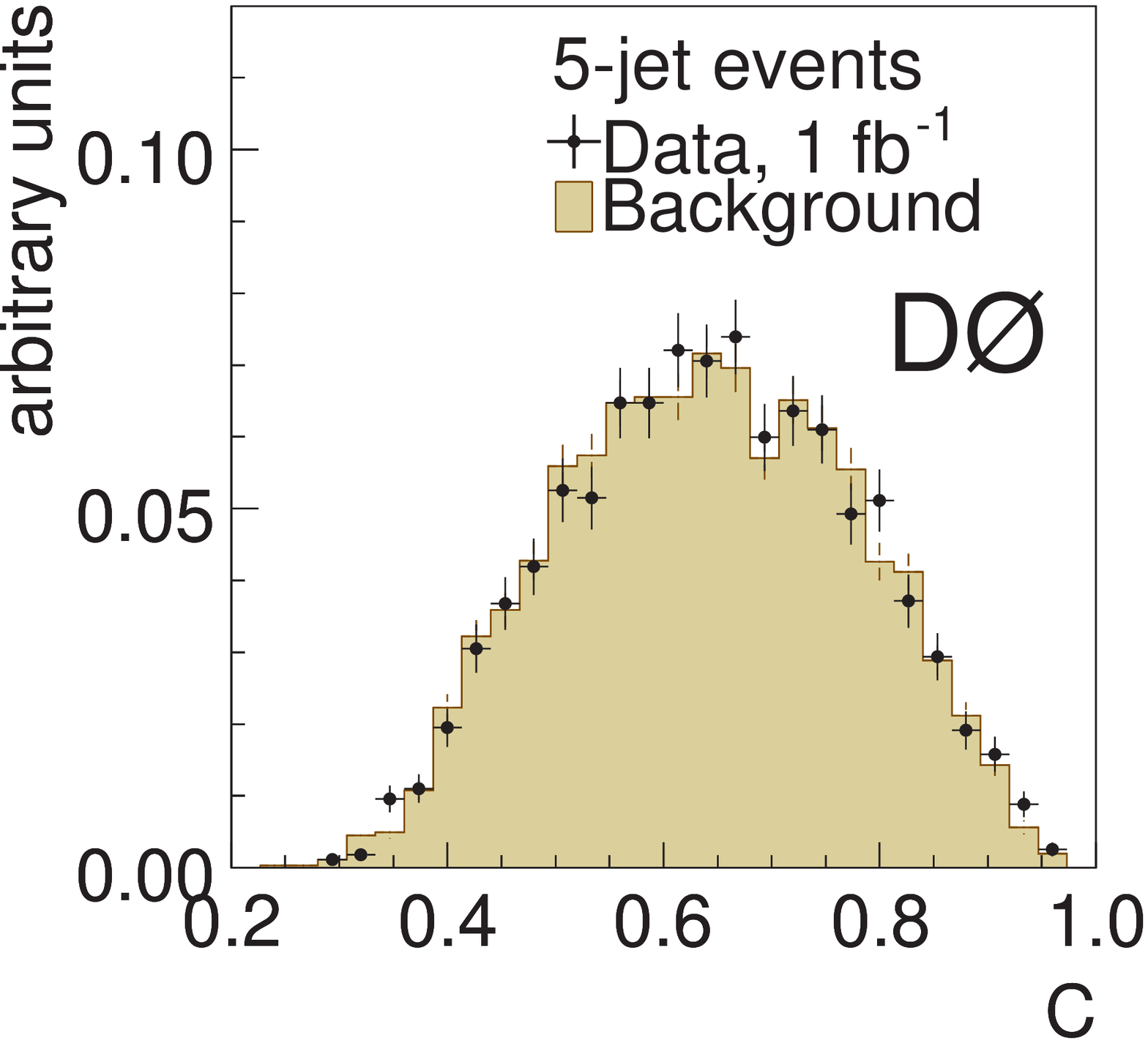}
\put(-130,140){\textsf{\textbf{(a)}}}%
\includegraphics*[width=2.25in,trim=20 5 25 35,clip=true]{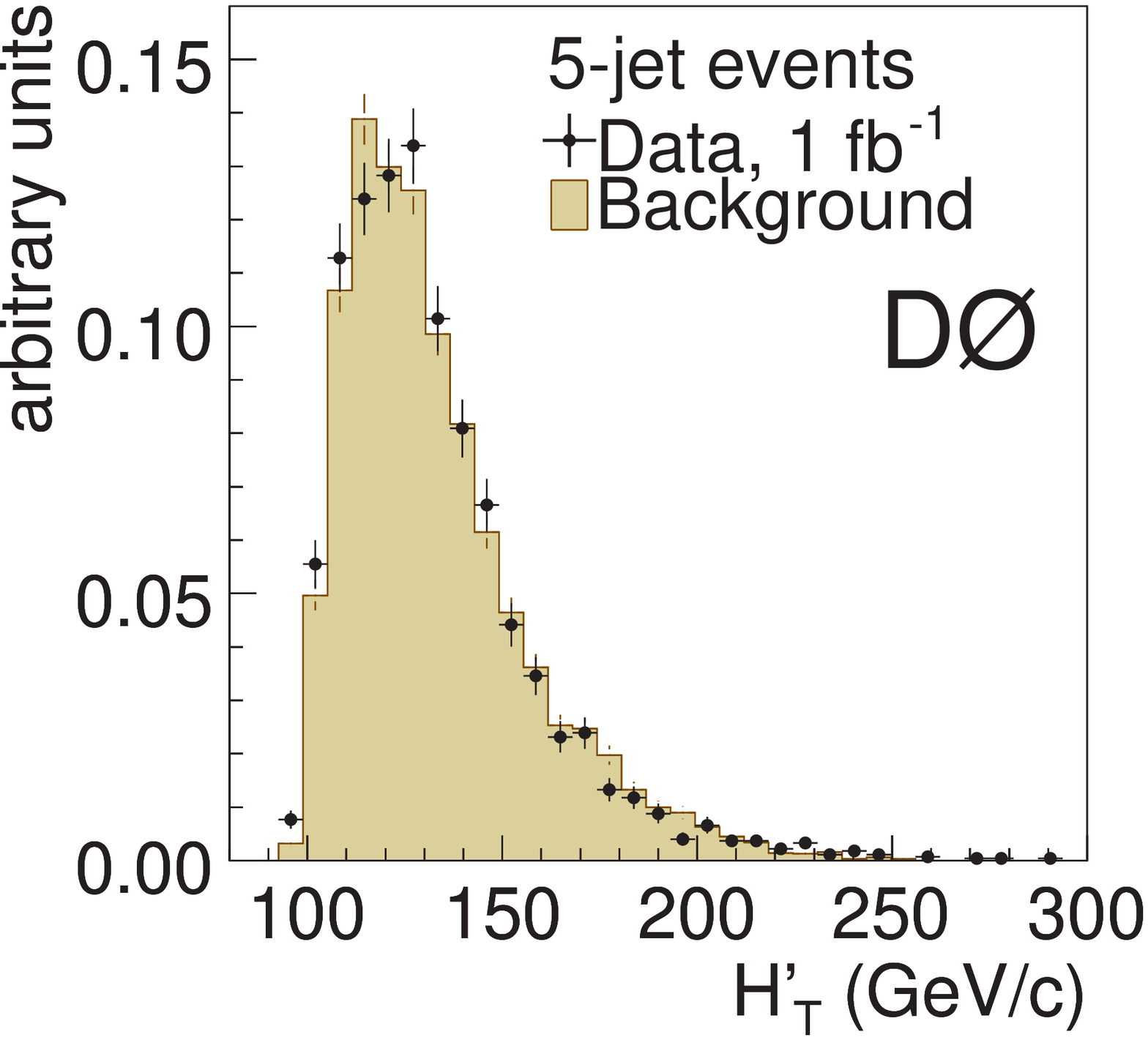}
\put(-130,140){\textsf{\textbf{(b)}}}%
\includegraphics*[width=2.25in,trim=20 5 25 35,clip=true]{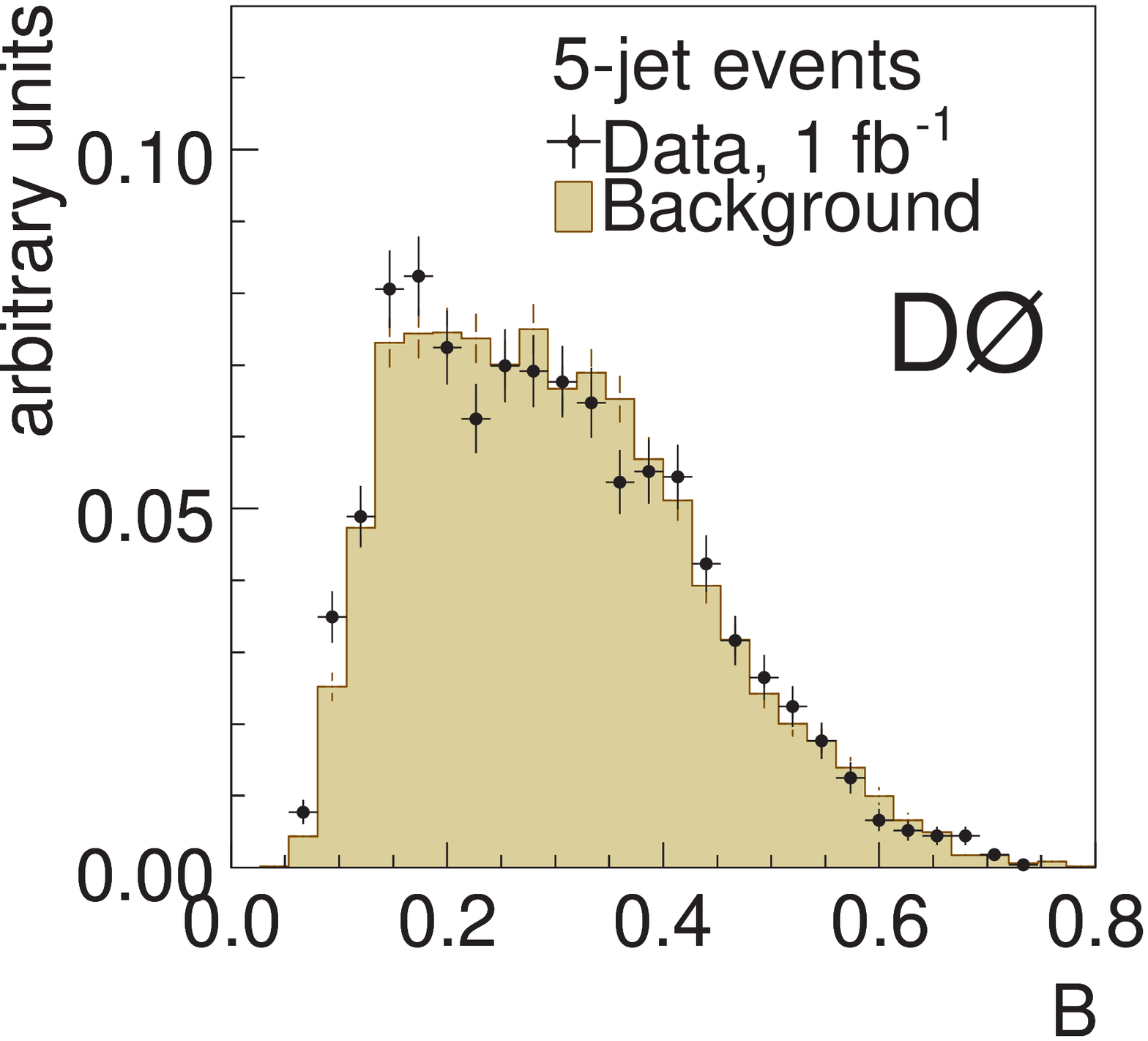}
\put(-130,140){\textsf{\textbf{(c)}}}%
\\
\includegraphics*[width=2.25in,trim=20 5 25 35,clip=true]{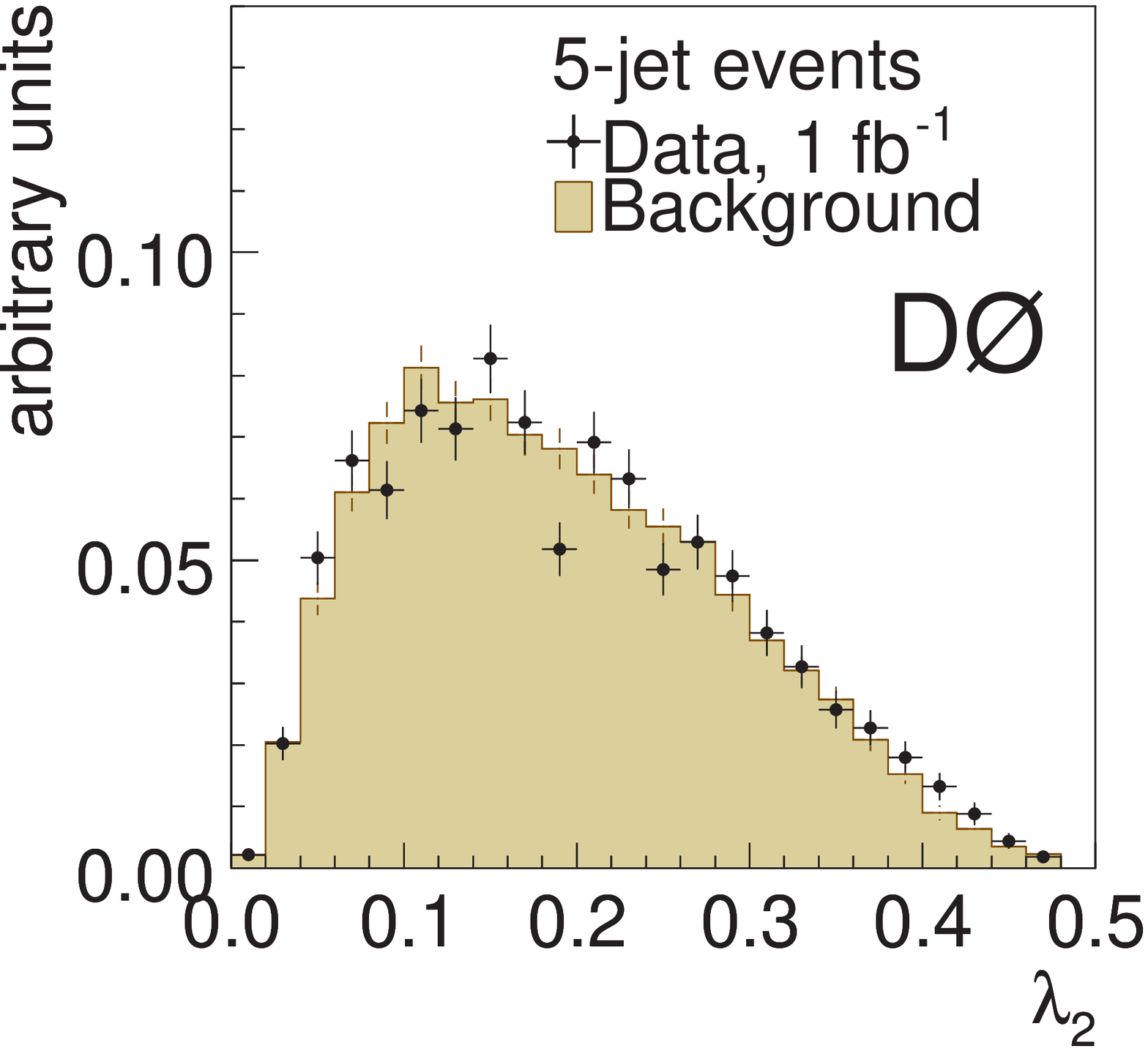}
\put(-130,140){\textsf{\textbf{(d)}}}%
\includegraphics*[width=2.25in,trim=20 5 25 35,clip=true]{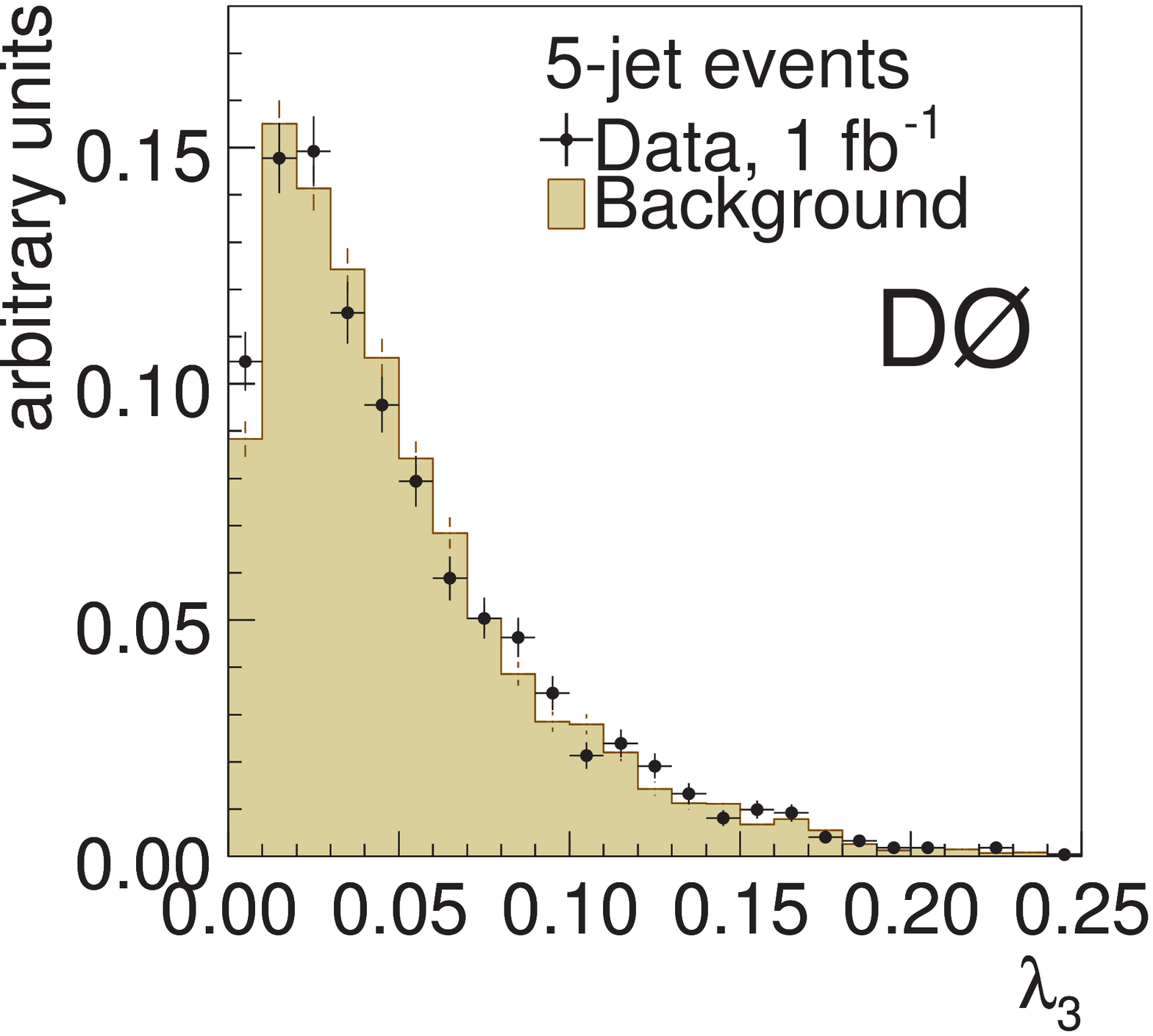}
\put(-130,140){\textsf{\textbf{(e)}}}%
\includegraphics*[width=2.25in,trim=20 5 25 35,clip=true]{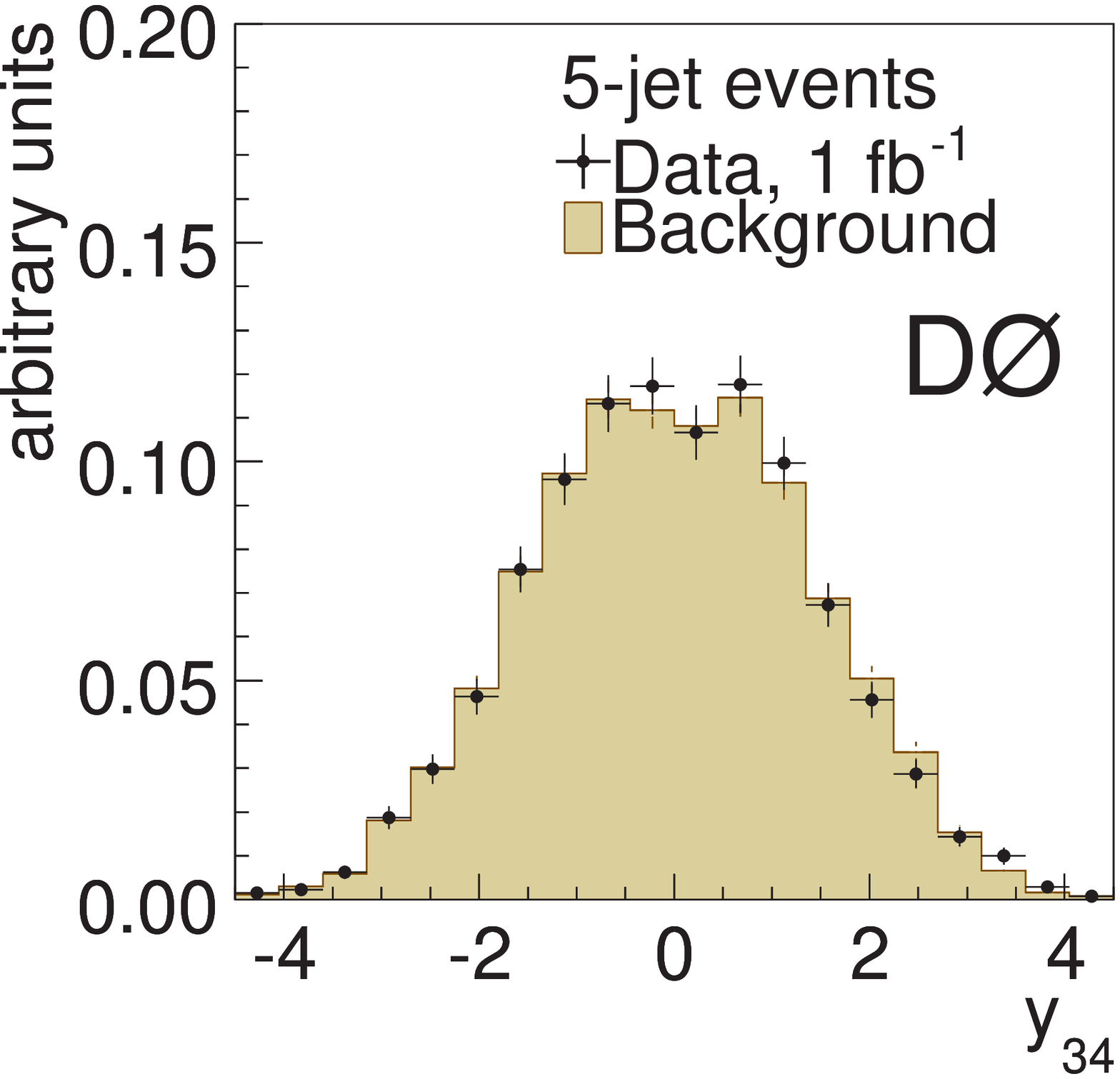}
\put(-130,140){\textsf{\textbf{(f)}}}%
\\
\includegraphics*[width=2.25in,trim=20 5 25 35,clip=true]{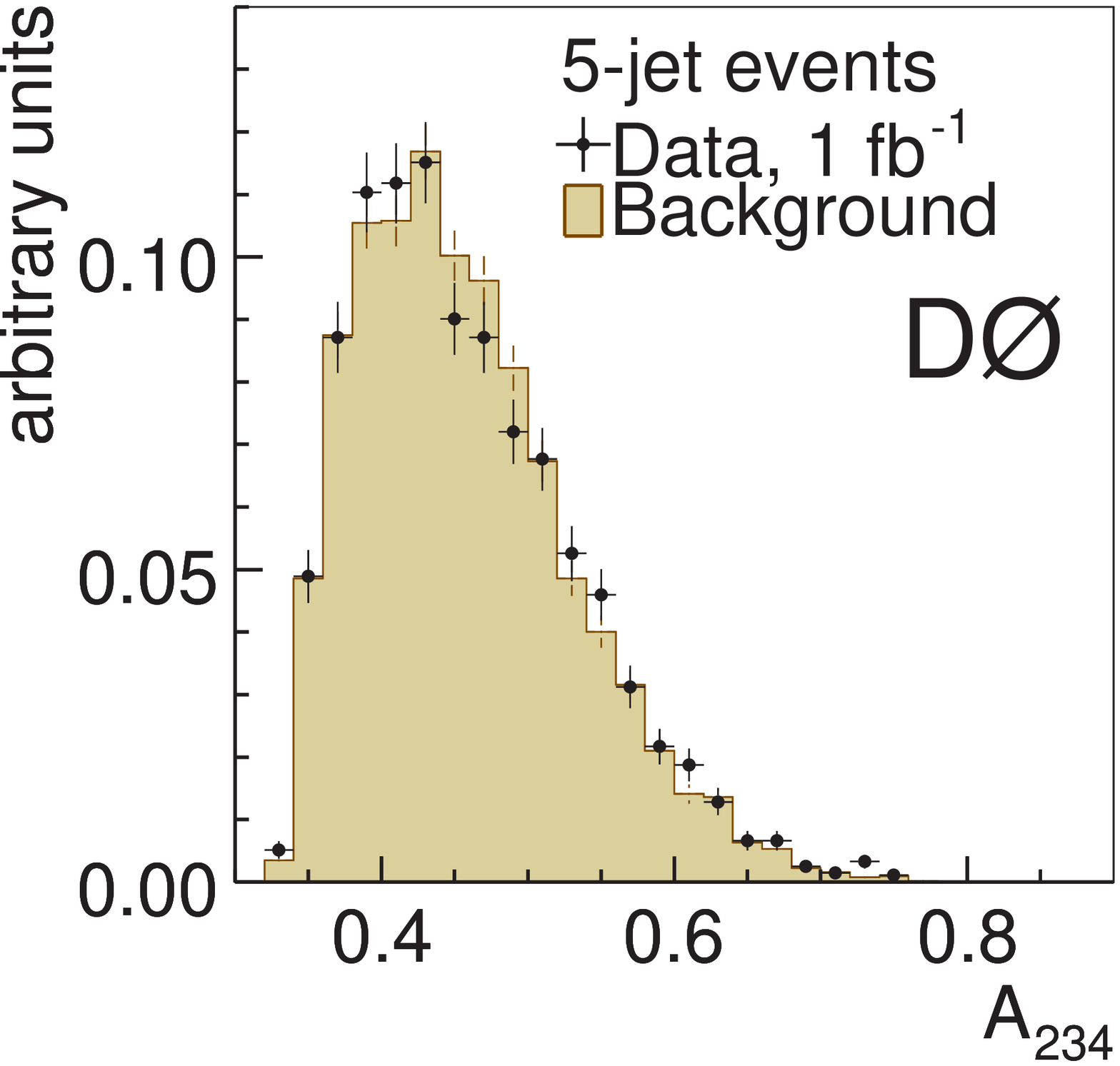}
\put(-130,140){\textsf{\textbf{(g)}}}%
\includegraphics*[width=2.25in,trim=20 5 25 35,clip=true]{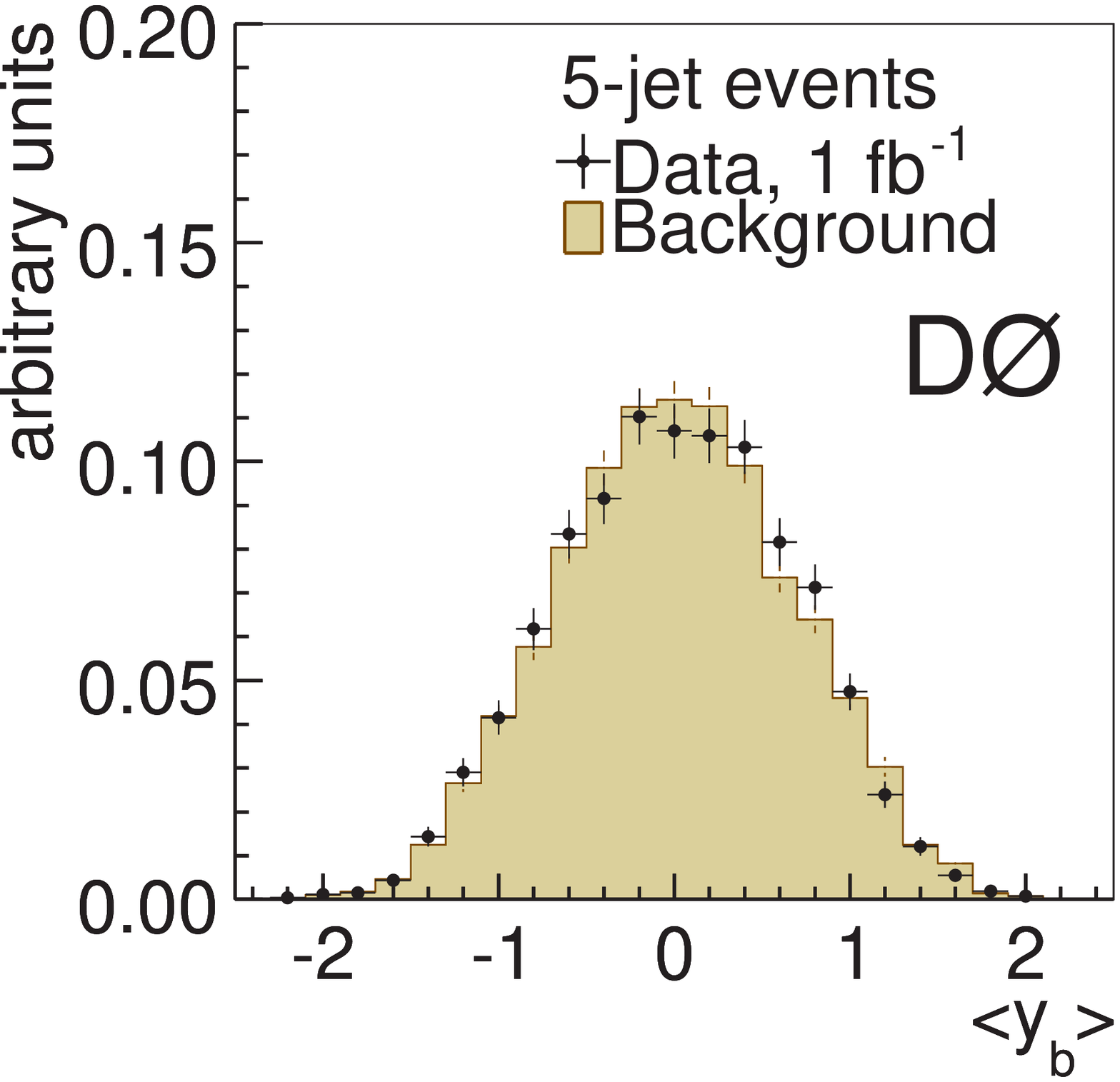}
\put(-130,140){\textsf{\textbf{(h)}}}%
\includegraphics*[width=2.25in,trim=20 5 25 35,clip=true]{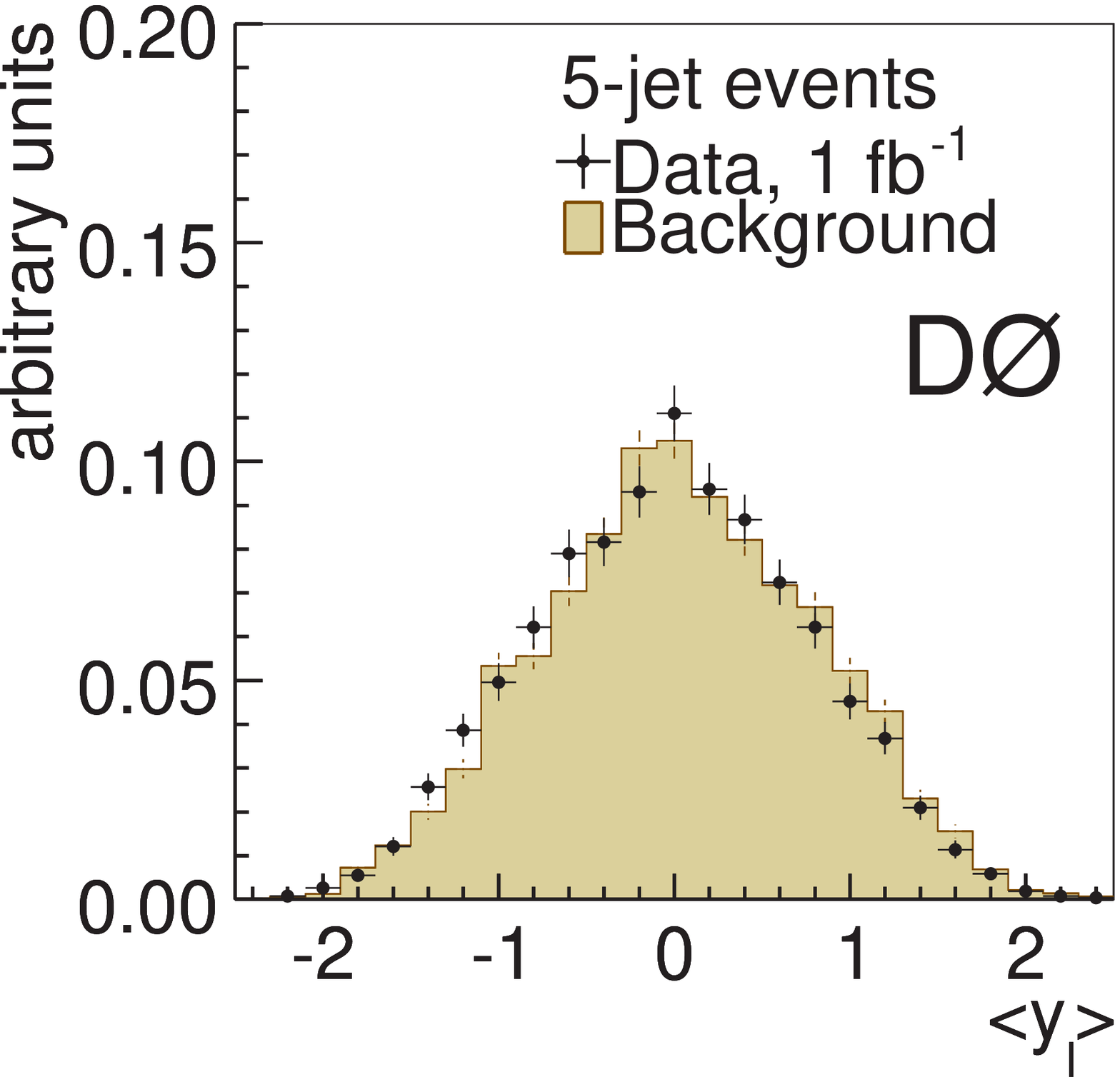}
\put(-130,140){\textsf{\textbf{(i)}}}%
\caption{\label{fig:4vs5likelihood} Comparisons between the five-jet
  data and the background created from four-jet data for variables
  used in the likelihood discriminant.  The leading four jets were required
  to have $p_T>40$~GeV$/c$.
Displayed error bars represent statistical uncertainties only.
Distributions are normalized to unit area.}
\end{figure*}

The resulting events were compared with the five-jet sample as
illustrated in Fig.~\ref{fig:4vs5pt}.  Reasonable agreement was
achieved with the individual jet $p_T$ distributions and with their
sum.  These manufactured background events are also compared against
the five-jet events for several topological variables (defined in
Sec.~\ref{sec:likelihood}) in Fig.~\ref{fig:4vs5likelihood}.

\begin{figure*}[!t]
%0.49
%\includegraphics*[width=0.49\textwidth,trim=30 10 30 40,clip=true]{\PATHsmallfonts/background_syst/4+40/4+2_vs_5+1/check_background_4+40_background_42_vs_51_ht_log.eps}
\includegraphics*[width=0.49\textwidth,trim=30 10 30 40,clip=true]{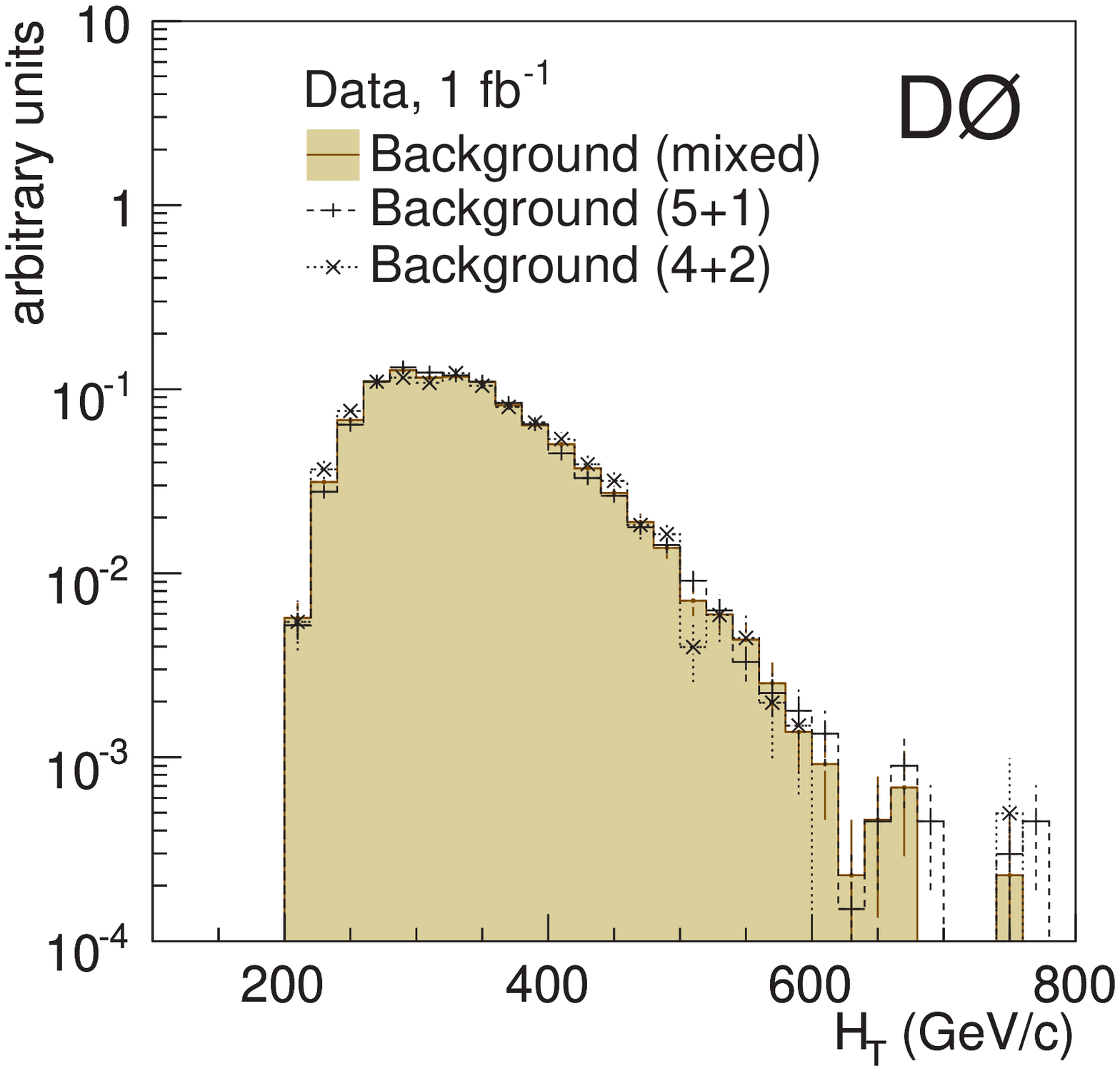}
\put(-200,225){\textsf{\textbf{(a)}}}%
\includegraphics*[width=0.49\textwidth,trim=30 10 30 40,clip=true]{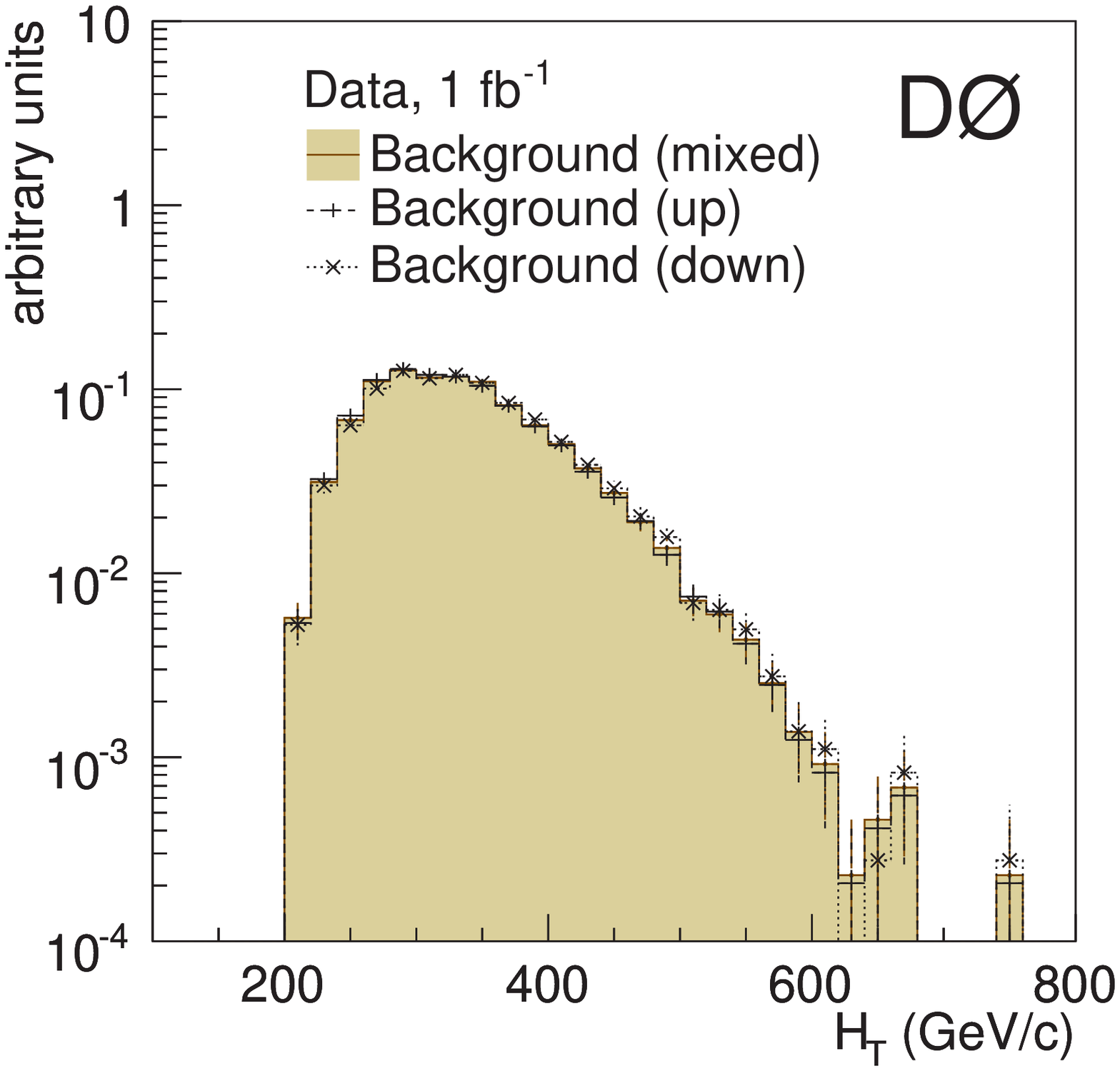}
\put(-200,225){\textsf{\textbf{(b)}}}%
\caption{\label{fig:42vs51ht} Systematic variations in the
  background sample with six or more jets as a function of $H_T$.  (a) Comparisons with the
  background samples created using only four-jet (4+2) or five-jet
  (5+1) events.  (b) Comparisons including one-sigma systematic
  variations in the phase-space matching criteria.  The leading four
  jets were required to have $p_T>40$~GeV$/c$.
 Distributions are normalized to unit area.}
\end{figure*}

Both the original four-jet sample used to create these five-jet
``background'' events and the ``signal'' five-jet sample to which it
was compared had little contamination from $t\bar{t}$ ($0.2\%$ and
$0.7\%$, respectively), so this tests our ability to use one multijet
sample to create a representation of a higher-multiplicity sample.
This scheme was extended to produce the background sample for events
with six or more jets.  In this case, the lowest $p_T$ jets were added
to either four-jet (fifth and lower $p_T$ jets) or five-jet (sixth and
lower $p_T$ jets) samples.  There was no reason to prefer the
four-jet-initiated background over the one built from a five-jet
sample.  Instead, an equal mix of the two was used for the final
background sample and the difference between the two separate
background samples and the mixed sample was used when evaluating
systematic uncertainties.  Variations between the two samples as a
function of $H_T$ are shown in Fig.~\ref{fig:42vs51ht}.  Also shown is
the change in the background due to systematic variations in the
phase-space matching criteria described above.
\begin{table*}[t]
  \caption{The number of events after each selection requirement.  
    Each selection is inclusive of the ones above it.
    Shown are the criteria, the number of events that pass the selection, the efficiency
    of the selection ($\varepsilon$), and the cumulative selection efficiency
    ($\varepsilon_{\rm{cum}}$) for all-hadronic $t\bar{t}$, all other $t\bar{t}$ decay channels, 
    and the data-based 
    background.  The $m_t=175$~GeV$/c^2$ sample was used for the signal expectation.  
    Signal and background numbers have been adjusted, using the 12.5\% signal fraction 
    measured in this analysis, to sum to the number of candidate events selected in the data.  
    Statistical uncertainties are included for the overall signal efficiency.
  }
  \label{tab:cutflow}
\begin{ruledtabular}
\begin{tabular}{l|rrr|rrr|rrd|l} 
\multirow{3}{*}{Selection} &
\multicolumn{3}{c|}{All Hadronic $t\bar{t}$} & 
\multicolumn{3}{c|}{Other $t\bar{t}$} & 
\multicolumn{3}{c|}{Background} &
\multicolumn{1}{c}{\multirow{2}{*}{Approx.}}\\
& 
\multicolumn{1}{c}{Num.} & \multicolumn{1}{c}{$\varepsilon$} & \multicolumn{1}{c|}{$\varepsilon_{\rm cum}$} & 
\multicolumn{1}{c}{Num.} & \multicolumn{1}{c}{$\varepsilon$} & \multicolumn{1}{c|}{$\varepsilon_{\rm cum}$} & 
\multicolumn{1}{c}{Num.} & \multicolumn{1}{c}{$\varepsilon$} & \multicolumn{1}{c|}{$\varepsilon_{\rm cum}$} &
\multicolumn{1}{c}{\multirow{2}{*}{S:B}}\\
&&\multicolumn{1}{c}{(\%)}&\multicolumn{1}{c|}{(\%)}
&&\multicolumn{1}{c}{(\%)}&\multicolumn{1}{c|}{(\%)}
&&\multicolumn{1}{c}{(\%)}&\multicolumn{1}{c|}{(\%)}\\
\hline
Total                                   & $3024$ & $100.0$ & $100.0$& $3712$ & $100.0$ & $100.0$&&&& \\
Trigger, vertex, $\geq4$ jets with $p_T>15$ GeV$/c$& $1663$ & $55.0$ & $55.0$& $ 773$ & $20.8$ & $20.8$& $ 18856263$ & $100.0$ & 100.0 & 1:7700\\
Lepton veto                             & $1662$ & $100.0$ & $55.0$& $ 558$ & $72.2$ & $15.0$& $ 12679185$ & $67.2$ & 67.3 & 1:5700 \\
$\geq6$ jets with $p_T>15$ GeV$/c$          & $ 913$ & $55.0$ & $30.2$& $ 165$ & $29.6$ & $4.5$& $  1734595$ & $13.7$ & 9.2 & 1:1600\\
$\geq6$ taggable jets with $p_T>15$ GeV$/c$ & $ 628$ & $68.8$ & $20.8$& $  60$ & $36.3$ & $1.6$& $   506277$ & $29.2$ & 2.7 & 1:740\\
$\geq2$ b-tagged jets with $p_T>40$ GeV$/c$ & $ 150$ & $23.8$ & $4.9$& $  13$ & $21.8$ & $0.4$& $     2562$ & $0.5$ & 0.014 & 1:16 \\
$\geq3$ jets with $p_T>40$ GeV$/c$          & $ 147$ & $98.1$ & $4.9$& $  12$ & $95.2$ & $0.3$& $     2059$ & $80.4$ & 0.011 & 1:13\\
$\geq4$ jets with $p_T>40$ GeV$/c$          & $ 122$ & $83.2$ & $4.0$& $   9$ & $70.3$ & $0.2$& $      920$ & $44.7$ & 0.0049 & 1:7 \\
\hline
Efficiency & \multicolumn{3}{c|}{$(4.04\pm0.02)\%$}& \multicolumn{3}{c|}{$(0.24\pm0.01)\%$} & \\
Inclusive $t\bar{t}$ Efficiency & \multicolumn{6}{c|}{$(1.94\pm0.01)\%$}\\
\end{tabular}
\end{ruledtabular}
\end{table*}

\subsection{\label{sec:signal} Signal Model}

The $t\bar{t}$ signal was simulated with the {\sc alpgen} event
generator. Two inclusive $t\bar{t}$ samples were used in this
analysis: one with $m_t=170$~GeV$/c^2$ and one with
$m_t=175$~GeV$/c^2$~\cite{topmassrange}.  {\sc pythia}, with the tune
A parameter settings, was used for the parton shower, hadronization,
and underlying event aspects.  The resulting events were processed
through a {\sc geant}~\cite{geant} simulation of the D0 detector and
underwent the full reconstruction and analysis procedure.  Information
from data events selected by a random beam crossing trigger were
overlayed on the simulated events to reproduce experimental conditions
including detector noise and overlapping $p\bar{p}$ interactions.  The
instantaneous luminosity distribution of the simulated events was
weighted to match that of the triggered data.

Several additional corrections were applied to the simulated events.
First, the event generator used the leading order (LO) parton
distribution functions (PDF) from CTEQ6L1~\cite{CTEQ6a,CTEQ6b}.
Events were reweighted to correspond to the CTEQ6.5M~\cite{CTEQ6.5}
PDF.  Second, the default heavy-flavor fragmentation function in {\sc
  pythia} was reweighted to one that described the LEP $e^+e^-$
data~\cite{LEPtune}.
Additionally, the resolutions of reconstructed objects in the
simulation were slightly better than those in the data.
so the energies of jets, muons, and electrons were smeared to reproduce
the resolutions observed in data~\cite{singletopb}.
The jet identification efficiency is slightly higher in the simulation
than in data.  Therefore, jets in the simulation were randomly removed
to make the efficiencies agree.

\subsection{\label{sec:selection}Event Selection}

Selection criteria were applied to triggered events to minimize
background while retaining a relatively high signal efficiency.
The selection criteria, together with the number of events after each
cut, the cut efficiency $\varepsilon$, and the cumulative selection
efficiency $\varepsilon_{\rm{cum}}$, are presented in
Table~\ref{tab:cutflow}.  Values are given for the all-hadronic
$t\bar{t}$ signal, for signal in all other $t\bar{t}$ decay channels,
and for the data-based background.
The signal fraction in the final selected sample corresponded to a
purity of $12.5\%$ (as found in Sec.~\ref{sec:signalfraction}).
As the background was derived from triggered data, the minimum set of
requirements on that sample, which also included a reconstructed
primary vertex with $|z_{\rm PV}|<35$~cm and $\geq4$~jets having
$p_T>15$~GeV$/c$, are listed as the second line in
Table~\ref{tab:cutflow}.  This corresponded to a starting
signal-to-background ratio of approximately $1:7700$.

Events with isolated high-$p_T$ electrons and muons were removed to
avoid overlap with other D0 $t\bar{t}$ cross section
measurements~\cite{lepjet1fb,dilepton1fb}.  This requirement had
little effect on the all-hadronic $t\bar{t}$ signal, but did remove a
considerable number of events from the background.

Events considered in this analysis were required to have at least six
jets.  Each jet was required to be taggable, have $p_T>15$~GeV$/c$,
and $|\eta|<2.5$.  Furthermore, at least four of the jets were
required to have $p_T>40$~GeV$/c$.  At least two of these high-$p_T$
jets were required to be $b$ tagged.  These additional jet
requirements improve the signal-to-background ratio by a factor
of~$100$.

In total, 1051 data events satisfy the selection criteria.  The
efficiency for all-hadronic $t\bar{t}$ events with $m_t=175$~GeV$/c^2$
is $(4.04\pm0.02)\%$ while the overall efficiency for inclusive
$t\bar{t}$ events is $(1.94\pm0.01)\%$ (statistical uncertainties
only).  The equivalent efficiencies with $m_t=170$~GeV$/c^2$ are
$(3.65\pm0.04)\%$ and $(1.76\pm0.02)\%$, respectively.
Given these efficiencies and the standard model branching fractions,
$\approx 93\%$ of the selected $t\bar{t}$ events are from the
all-hadronic decay channel.  The surviving leptonic $t\bar{t}$ events
were primarily from the $\ell+$jets ($\approx 60\%$) and $\tau+$jets
($\approx 40\%$) decay channels.  Few dileptonic events survived the
full selection criteria ($\approx 0.05\%$ of $t\bar{t}$).

The expected signal-to-background ratio, given the $12.5\%$ signal purity
extracted during the cross section measurement, is $1:7$.
\begin{figure*}
\includegraphics*[width=2.25in,trim=20 5 20 50,clip=true]{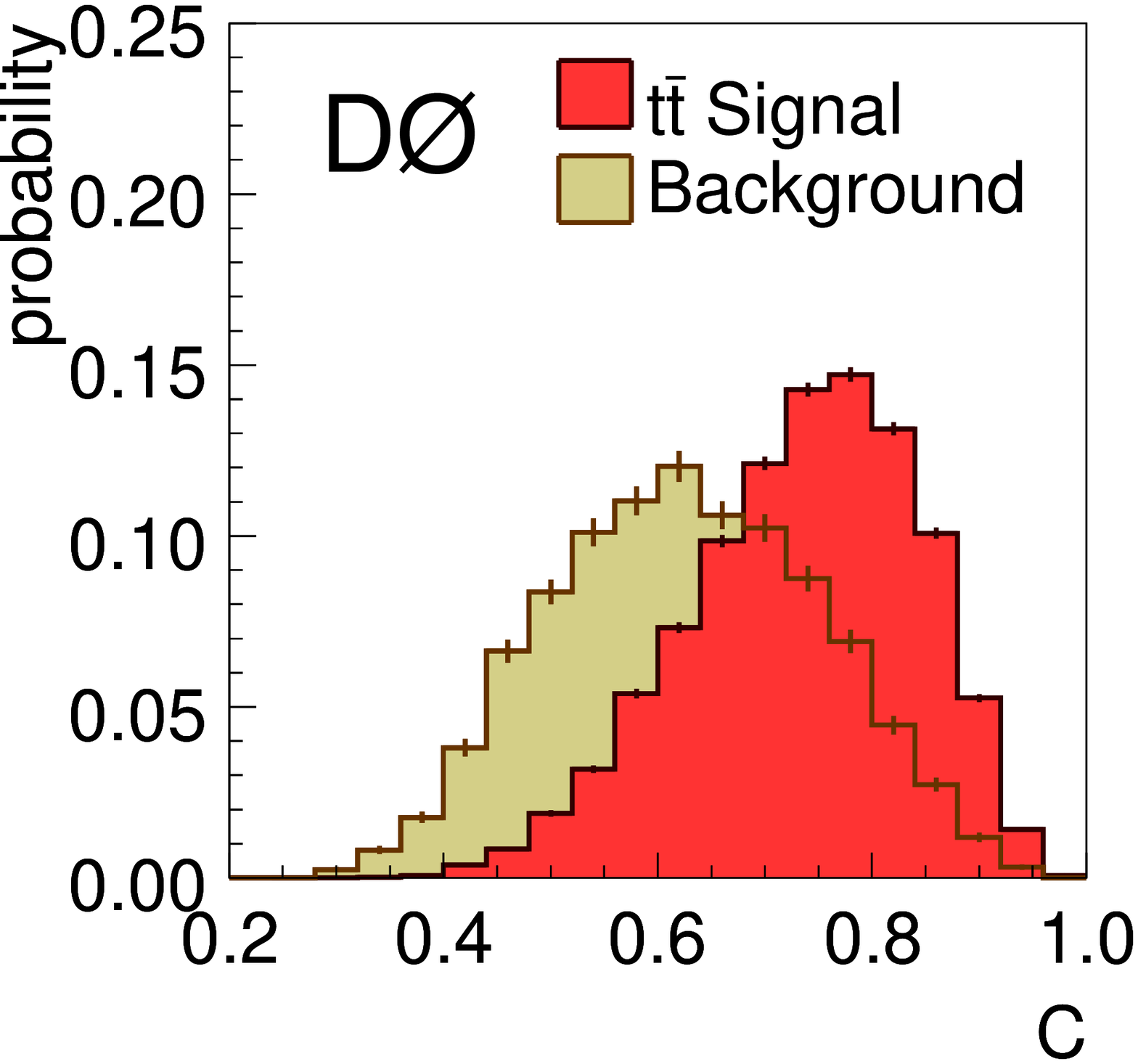}
\put(-30,140){\textsf{\textbf{(a)}}}%
\includegraphics*[width=2.25in,trim=20 5 20 50,clip=true]{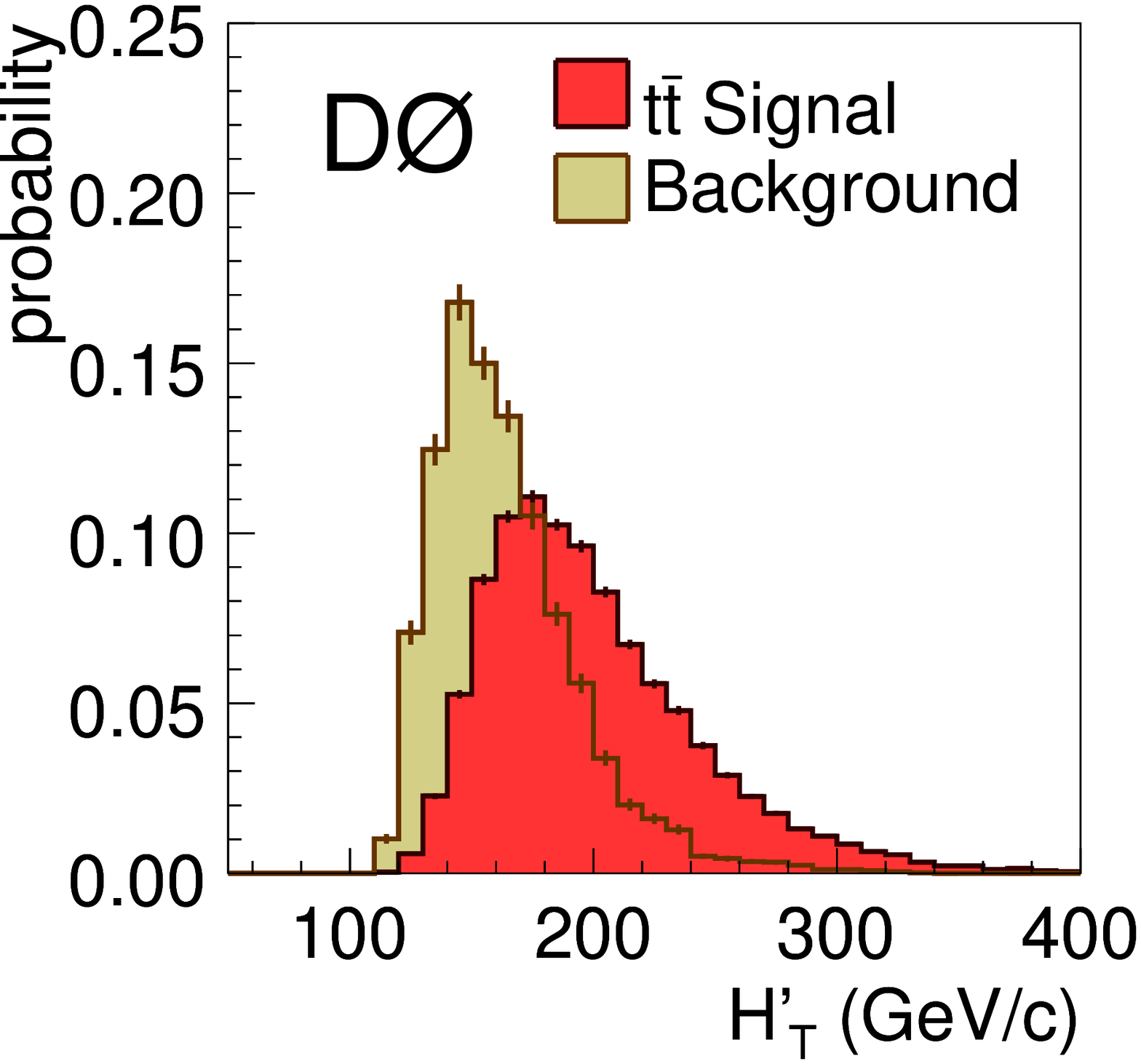}
\put(-30,140){\textsf{\textbf{(b)}}}%
\includegraphics*[width=2.25in,trim=20 5 20 50,clip=true]{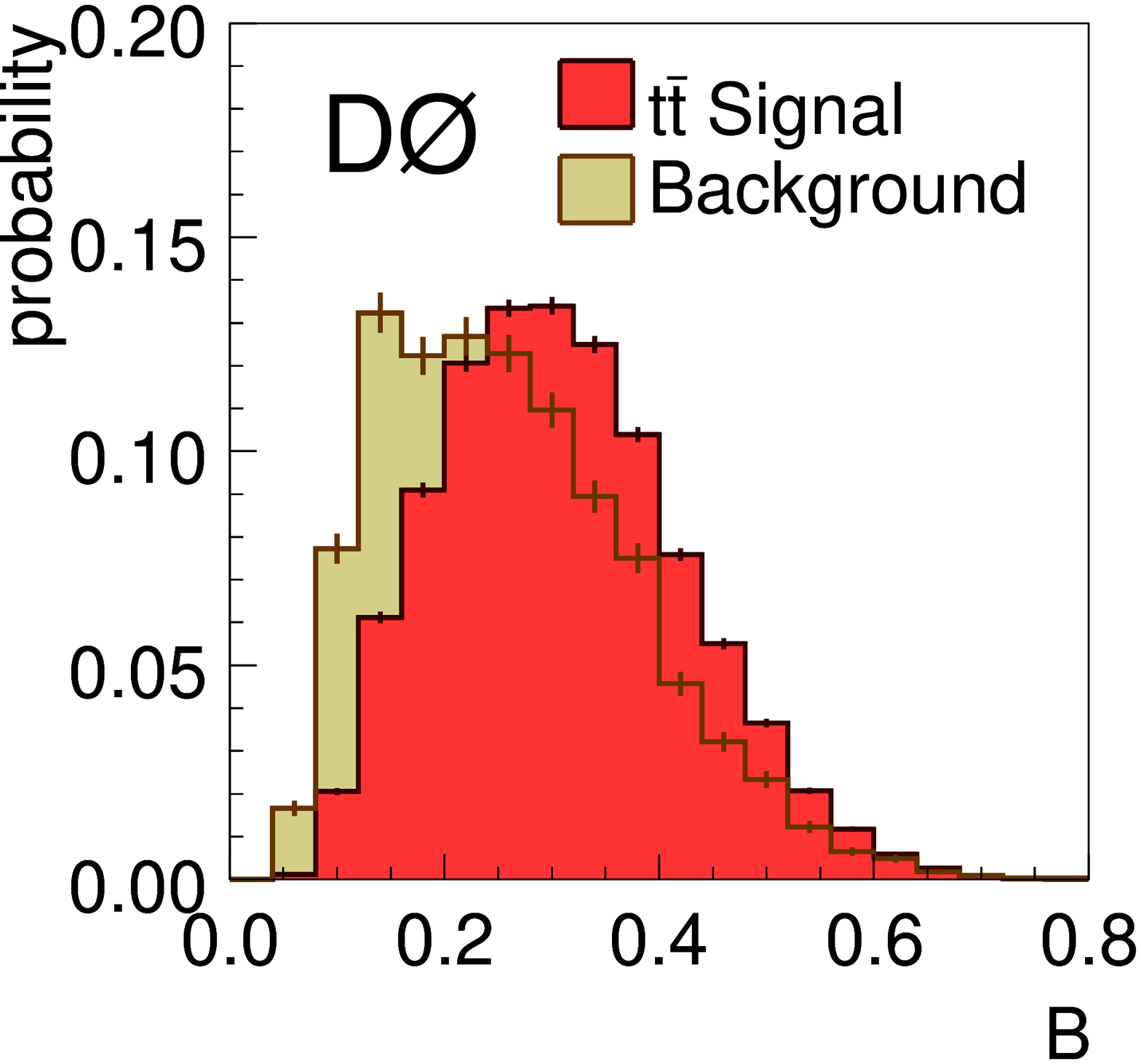}
\put(-30,140){\textsf{\textbf{(c)}}}%
\\
\includegraphics*[width=2.25in,trim=20 5 20 50,clip=true]{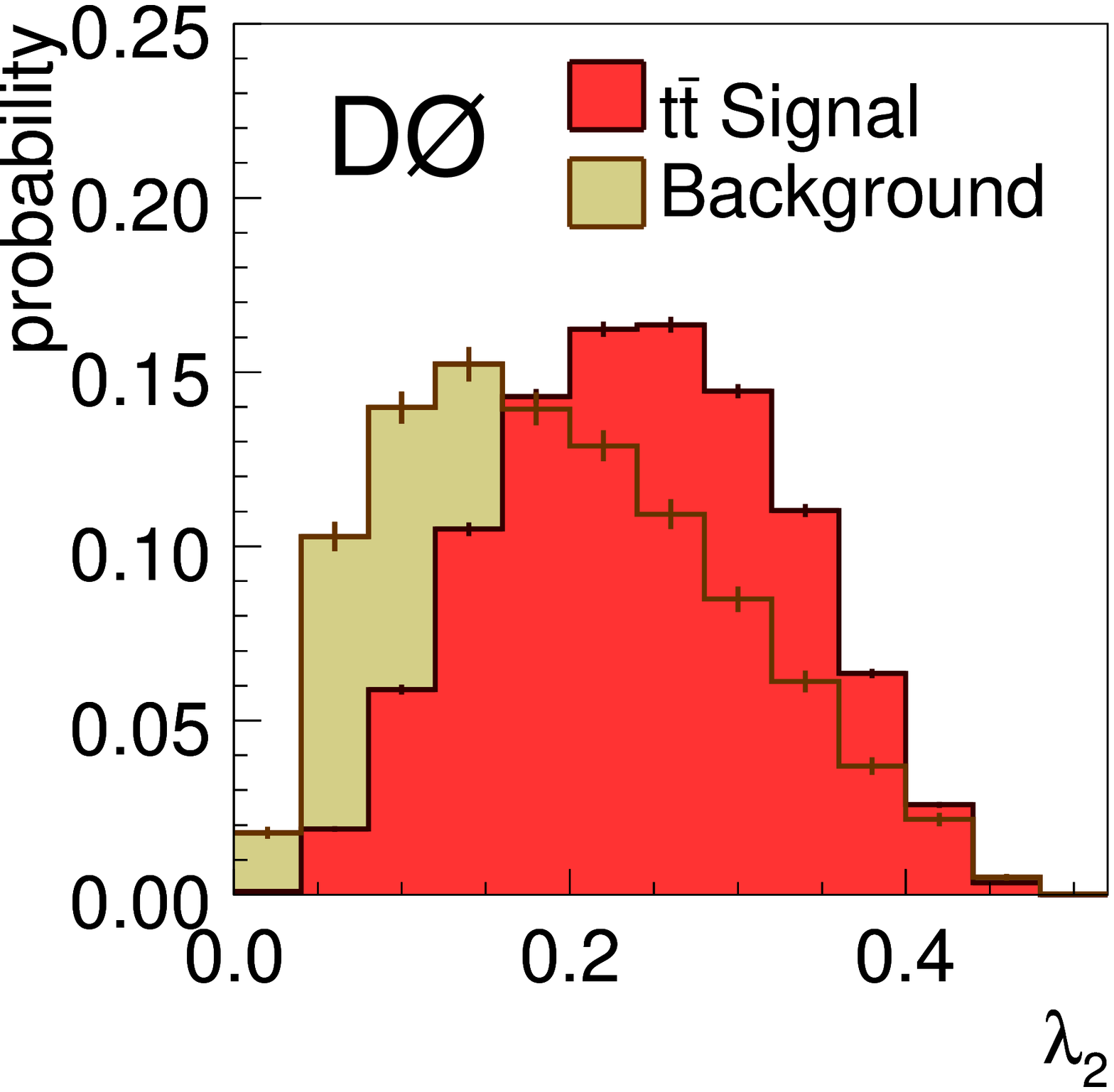}
\put(-30,140){\textsf{\textbf{(d)}}}%
\includegraphics*[width=2.25in,trim=20 5 20 50,clip=true]{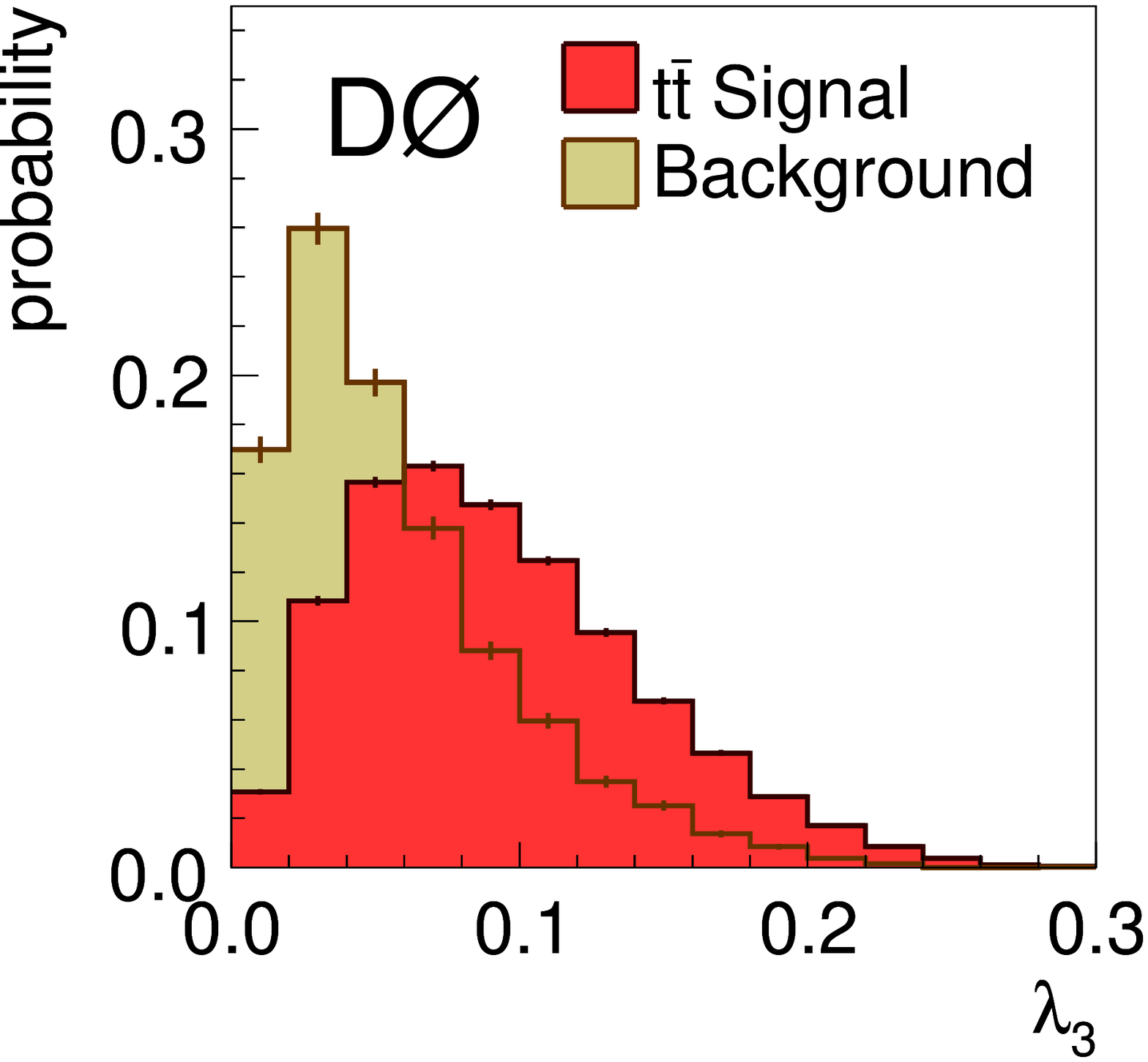}
\put(-30,140){\textsf{\textbf{(e)}}}%
\includegraphics*[width=2.25in,trim=20 5 20 50,clip=true]{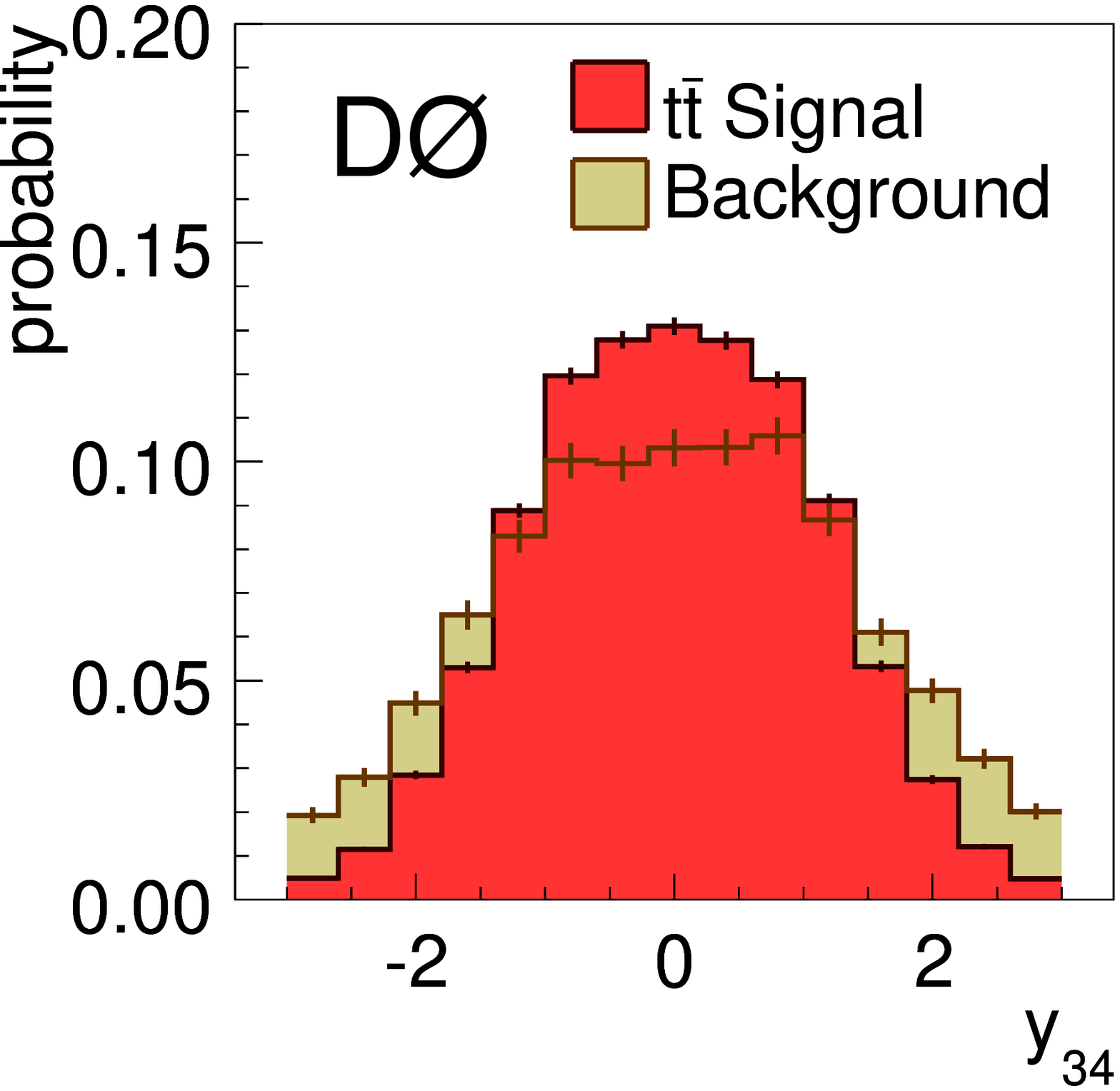}
\put(-30,140){\textsf{\textbf{(f)}}}%
\\
\includegraphics*[width=2.25in,trim=20 5 20 50,clip=true]{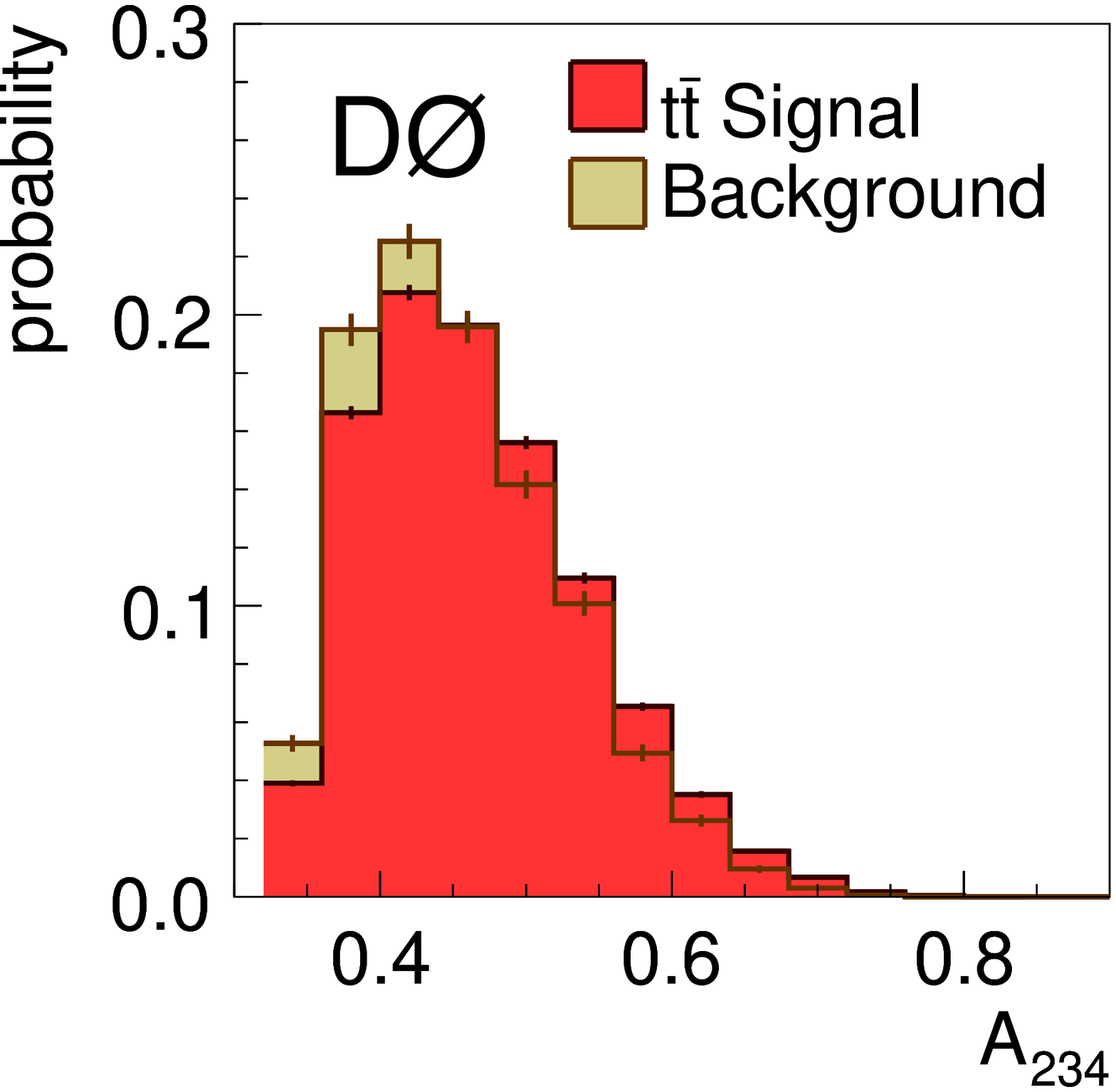}
\put(-30,140){\textsf{\textbf{(g)}}}%
\includegraphics*[width=2.25in,trim=20 5 20 50,clip=true]{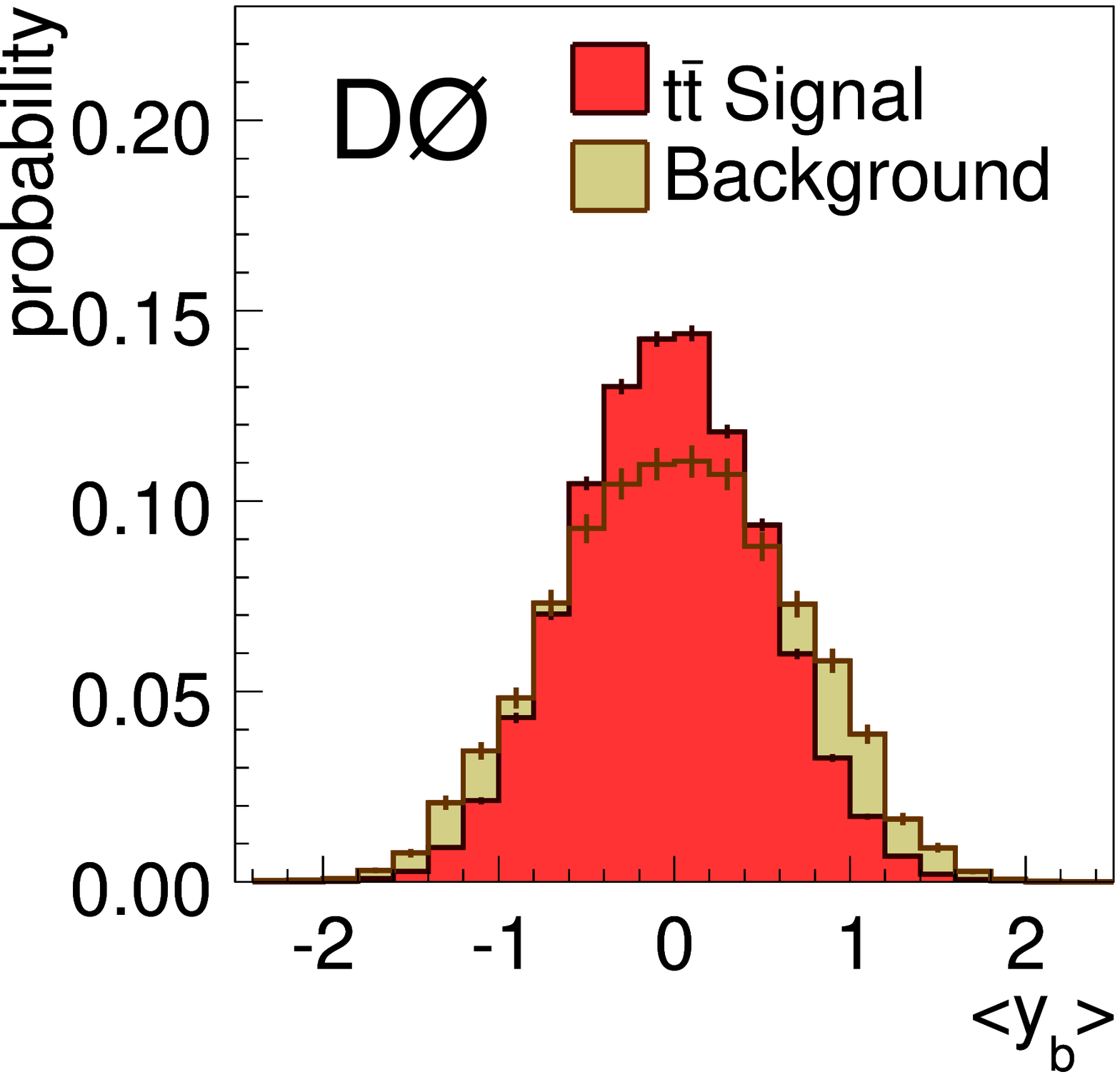}
\put(-30,140){\textsf{\textbf{(h)}}}%
\includegraphics*[width=2.25in,trim=20 5 20 50,clip=true]{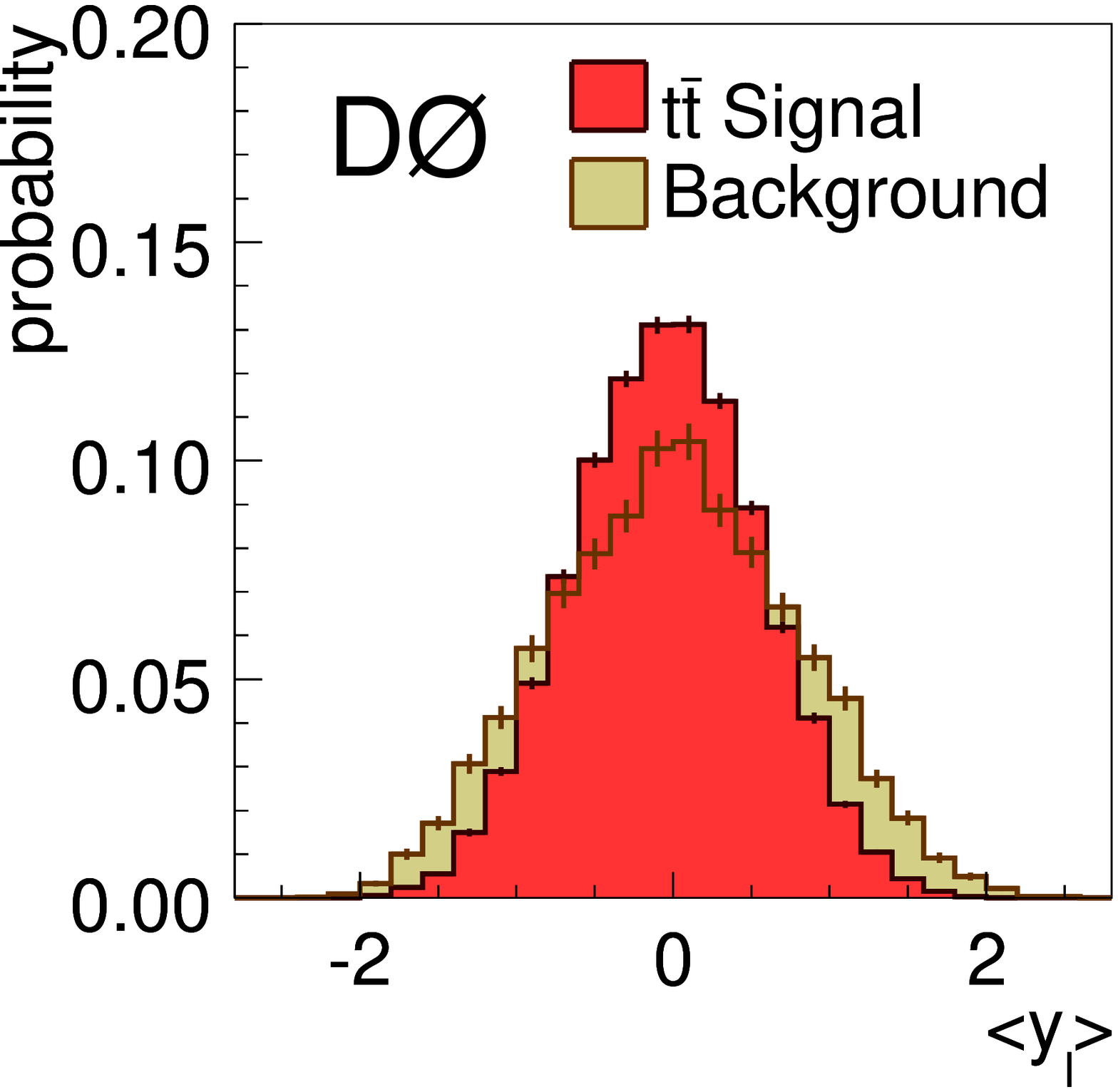}
\put(-30,140){\textsf{\textbf{(i)}}}%
\caption{\label{fig:likelihood_input} Probability distributions for
  the variables input into the likelihood ratio.  The signal
  distributions were extracted from the sample with
  $m_t=175$~GeV$/c^2$.  Displayed error bars represent statistical
  uncertainties only.  }
\end{figure*}
\begin{figure*}[t!]
%0.49
%\includegraphics[width=0.49\textwidth,trim=15 20 20 30,clip=true]{\PATHsmallfonts/mva_plots/combination41/output/tmva_dlh_combination41_tt_170_alpgen_alljet_40b-6+j-4+40-2+15_all2b_actasjesmucapsmeared_bidnn0_65_skim4jt_tgbl_deta25_llh_out_lin.eps}
\includegraphics[width=0.49\textwidth,trim=15 20 20 30,clip=true]{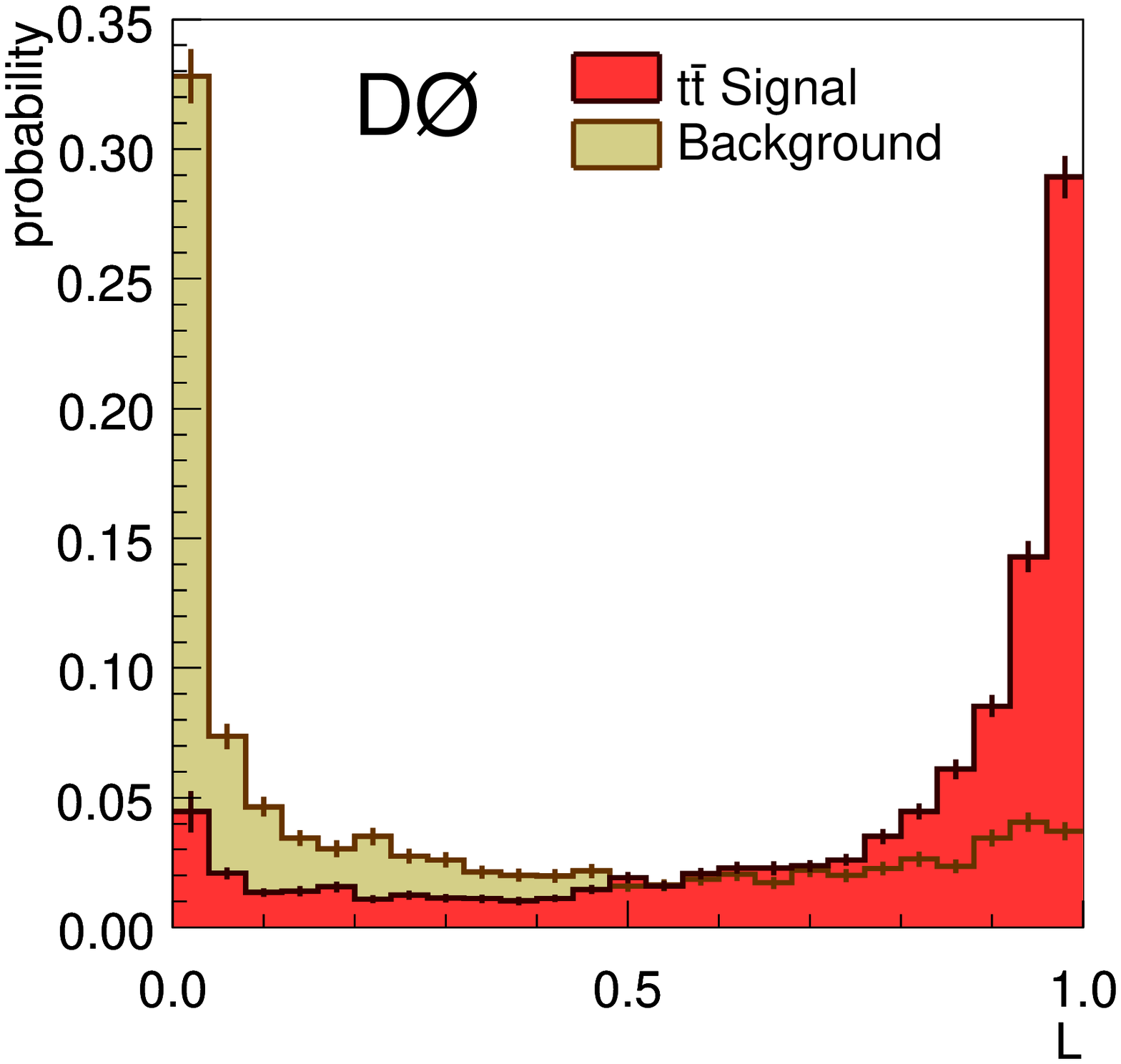}
\put(-40,208){\textsf{\textbf{(a)}}}
\includegraphics[width=0.49\textwidth,trim=15 20 20 30,clip=true]{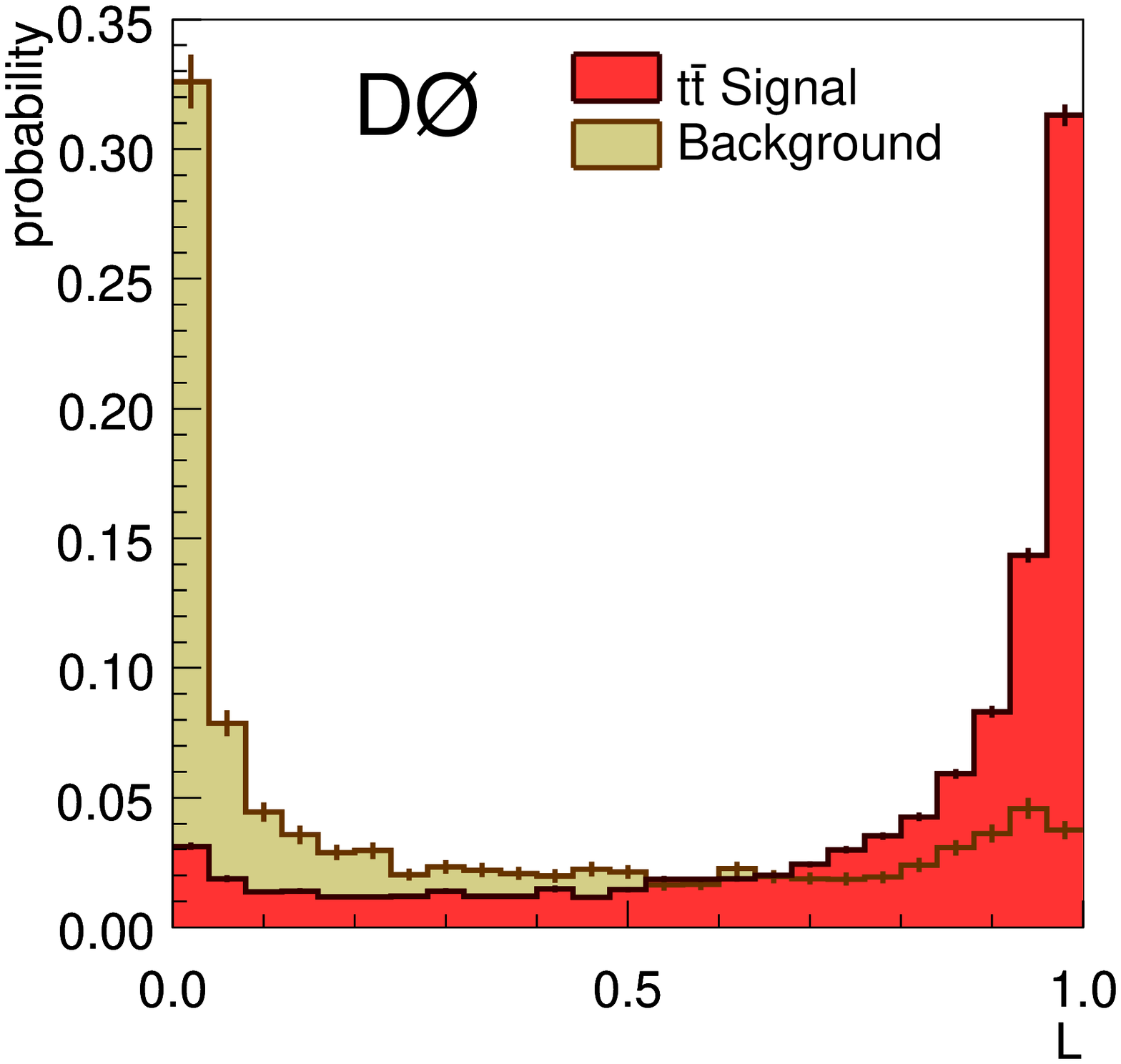}
\put(-40,208){\textsf{\textbf{(b)}}}
\caption{Probability distributions from the likelihood function, $L$, for
  $t\bar{t}$ signal and the data-derived
  background.  Displayed error bars represent statistical uncertainties only.
(a) Signal sample with $m_t=170$~GeV$/c^2$; (b) $m_t=175$~GeV$/c^2$.}
\label{fig:tmva_output}
\end{figure*}

\subsection{\label{sec:likelihood}Maximum Likelihood}

A likelihood discriminant based on topological observables was
constructed to separate the all-hadronic $t\bar{t}$ signal from the
multijet background.  The likelihood ratio, $L$, for an event $i$ is
defined as
\begin{equation*}
L = \frac{{L}_S(i)}{{L}_S(i)+{L}_B(i)},
\end{equation*}
where
\begin{equation*}
{L}_S(i)=\prod_{k=1}^{n_{\rm var}}{\cal P}_{S,k} [x_k(i)]
\end{equation*}
for signal and similarly for background.  Here, ${\cal P}_{S,k}$ is
the signal probability density function, normalized to unit area, for
the $k$th input variable $x_k$, and $n_{\rm var}$ is the number of
variables.
The {\sc tmva}~\cite{TMVA} package was used to build the probability
distributions and the resulting likelihood ratio.

The criteria for selection of observables to be input into the
likelihood were: separation between signal and background, reasonable
agreement in the five-jet background validation, little correlation
with other chosen variables, and little dependence on jet energies (to
minimize systematic uncertainty due to jet energy calibration).  The
following nine variables were used in the likelihood determination and
are shown for simulated signal and data-based background events in
Fig.~\ref{fig:likelihood_input}:
\begin{description*}
\setlength\labelwidth{3.0em}
\setlength\itemindent{0em}
\setlength\leftskip{3.0em}
\item[$C$] is the centrality defined as the scalar sum of jet $p_T$ 
divided by the sum of jet energies;
\item[$H_T^\prime$] is the scalar sum of jet $p_T$ excluding
  the two highest $p_T$ jets;
\item[$B$] is the ratio of the dijet mass of the two leading
  $b$-tagged jets to the total mass of all the jets;
\item[$\lambda_2$, $\lambda_3$] are the smallest two eigenvalues of
  the momentum tensor $M^{\alpha, \beta} = {\sum_i
    p^{\alpha}_i p^{\beta}_i}/{\sum_i |\vec{p}_i|^2}$ where $i$ runs
  over the number of jets and $\alpha, \beta = 1, 2, 3$ denote the
  three spatial components of the jet momenta~\cite{lambda};
\item[$y_{34}$] is the rapidity difference between the third and 
fourth leading jets;
\item[$A_{234}$] is the $p_T$ asymmetry between the 
second and third jet and the fourth jet defined as $({p_T}_2 + {p_T}_3 -
{p_T}_4) / ({p_T}_2 + {p_T}_3 + {p_T}_4)$;
\item[$\langle y_b\rangle$] is the $p_T$-weighted average of
the rapidities of the leading two $b$-tagged jets;
\item[$\langle y_l\rangle$] is the $p_T$-weighted average of
the rapidities of the leading two light (not $b$ tagged) jets.
\end{description*}
Comparisons are shown in Fig.~\ref{fig:4vs5likelihood} for these
variables in the five-jet background validation sample.  The combined
probability distributions for signal and background are shown in
Fig.~\ref{fig:tmva_output}.  The probability distributions and
likelihoods were extracted independently for the $m_t=170$~GeV$/c^2$
and $175$~GeV$/c^2$ samples.
\begin{figure*}[t!]
%0.49
%\includegraphics*[width=0.49\textwidth,trim=15 0 20 30,clip=true]{\PATHsmallfonts/mva_plots/combination41/control_plots/tmva_dlh_combination41_tt_170_alpgen_alljet_40b-6+j-4+40-2+15_all2b_actasjesmucapsmeared_bidnn0_65_skim4jt_tgbl_deta25_llh_res_llh_mtop_170_lin.eps}
\includegraphics*[width=0.49\textwidth,trim=15 0 20 30,clip=true]{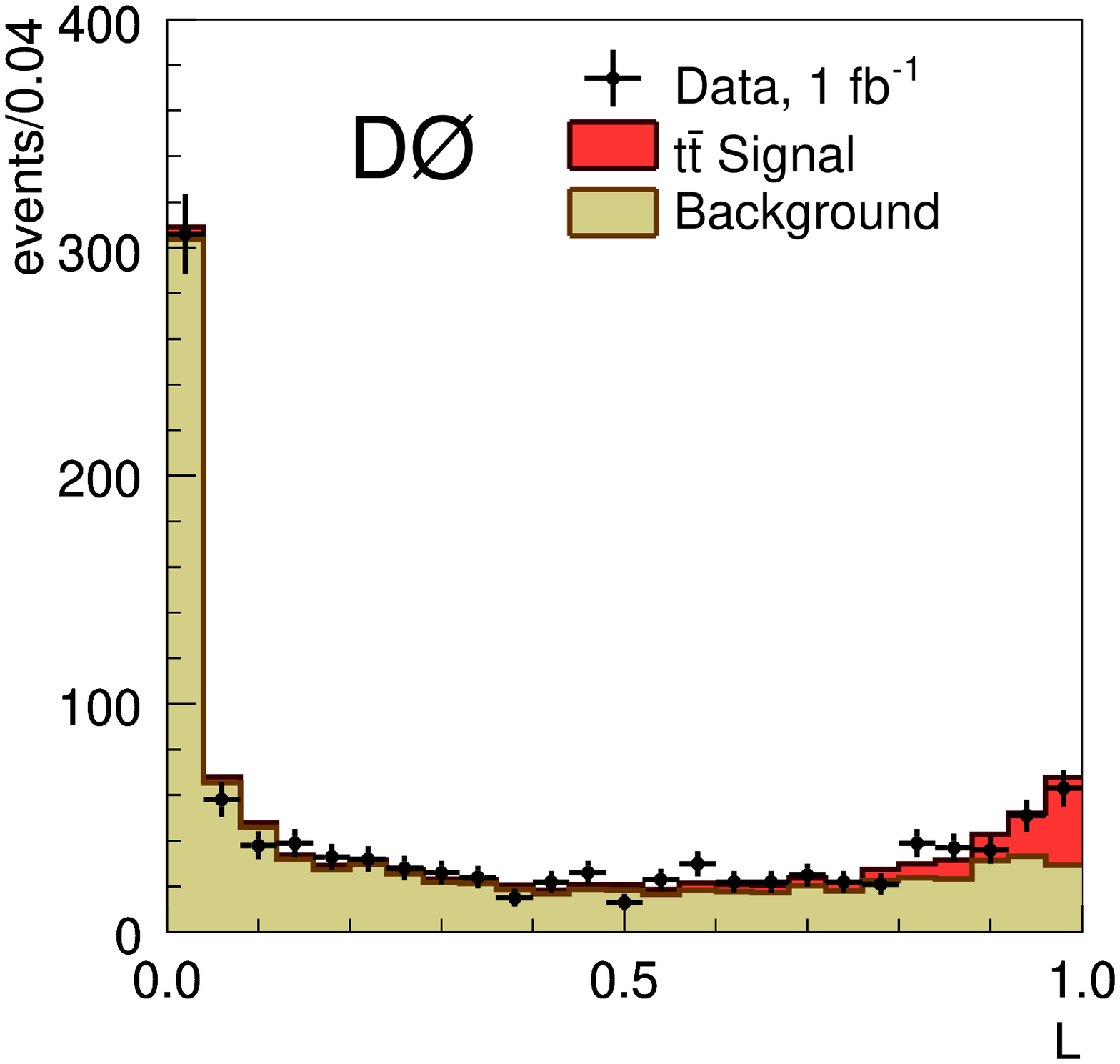}
\put(-40,215){\textsf{\textbf{(a)}}}%
\includegraphics*[width=0.49\textwidth,trim=15 5 20 30,clip=true]{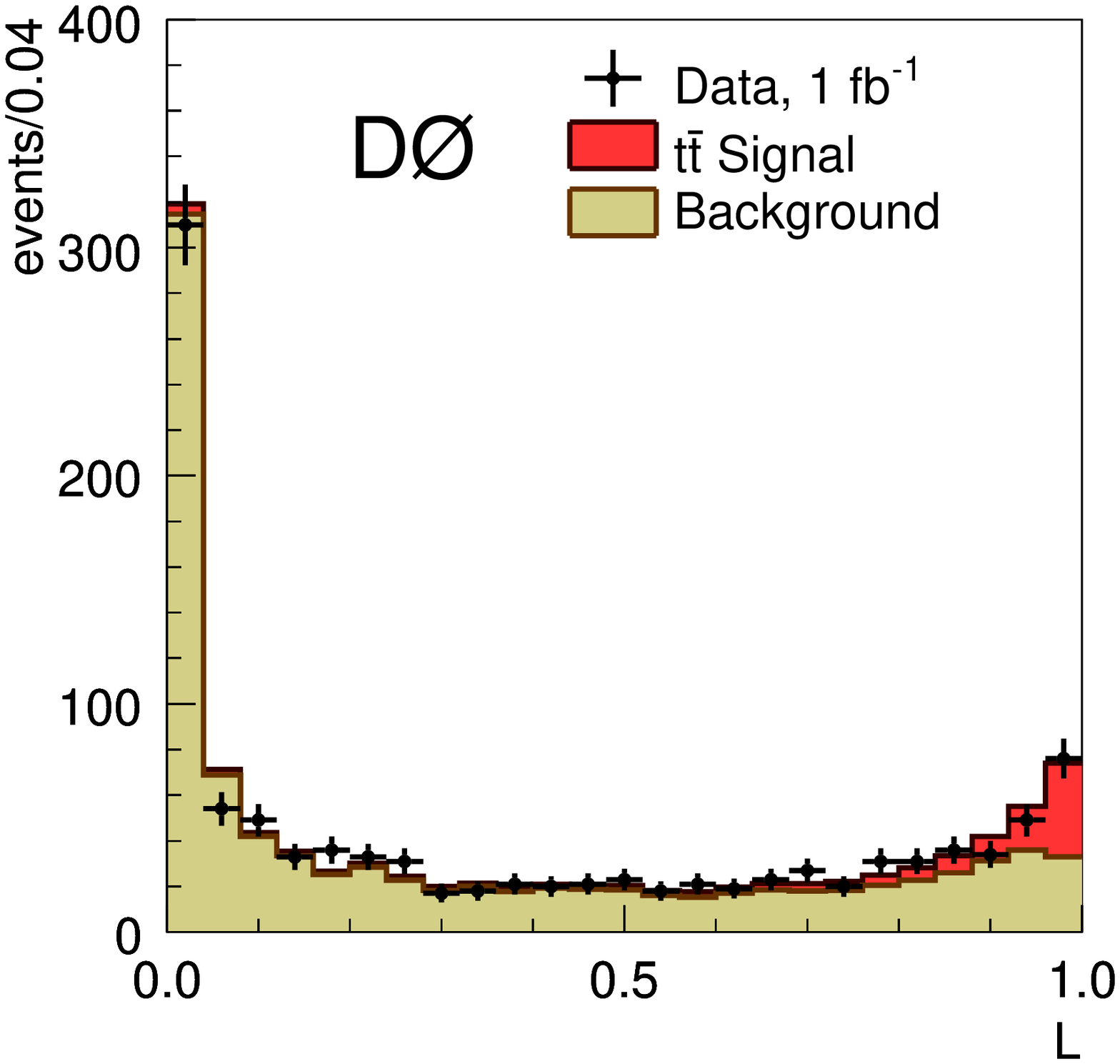}
\put(-40,215){\textsf{\textbf{(b)}}}%
\caption{Comparison of the distributions of likelihood output values, $L$, for the selected 
data candidates (points) with those from the $t\bar{t}$ signal and data-based background samples.
Signal and background were fit to the data candidates and are presented with a normalization equal
to the fit purity times the number of candidate events for the signal.
Displayed error bars represent statistical uncertainties only.
(a) Signal sample with $m_t=170$~GeV$/c^2$; (b) $m_t=175$~GeV$/c^2$.
\label{fig:llh_result}}
\end{figure*}

\section{Results\label{sec:results}}

\subsection{Signal Fraction\label{sec:signalfraction}}

\begin{figure*}
\includegraphics*[width=2.25in,trim=20 5 15 45,clip=true]{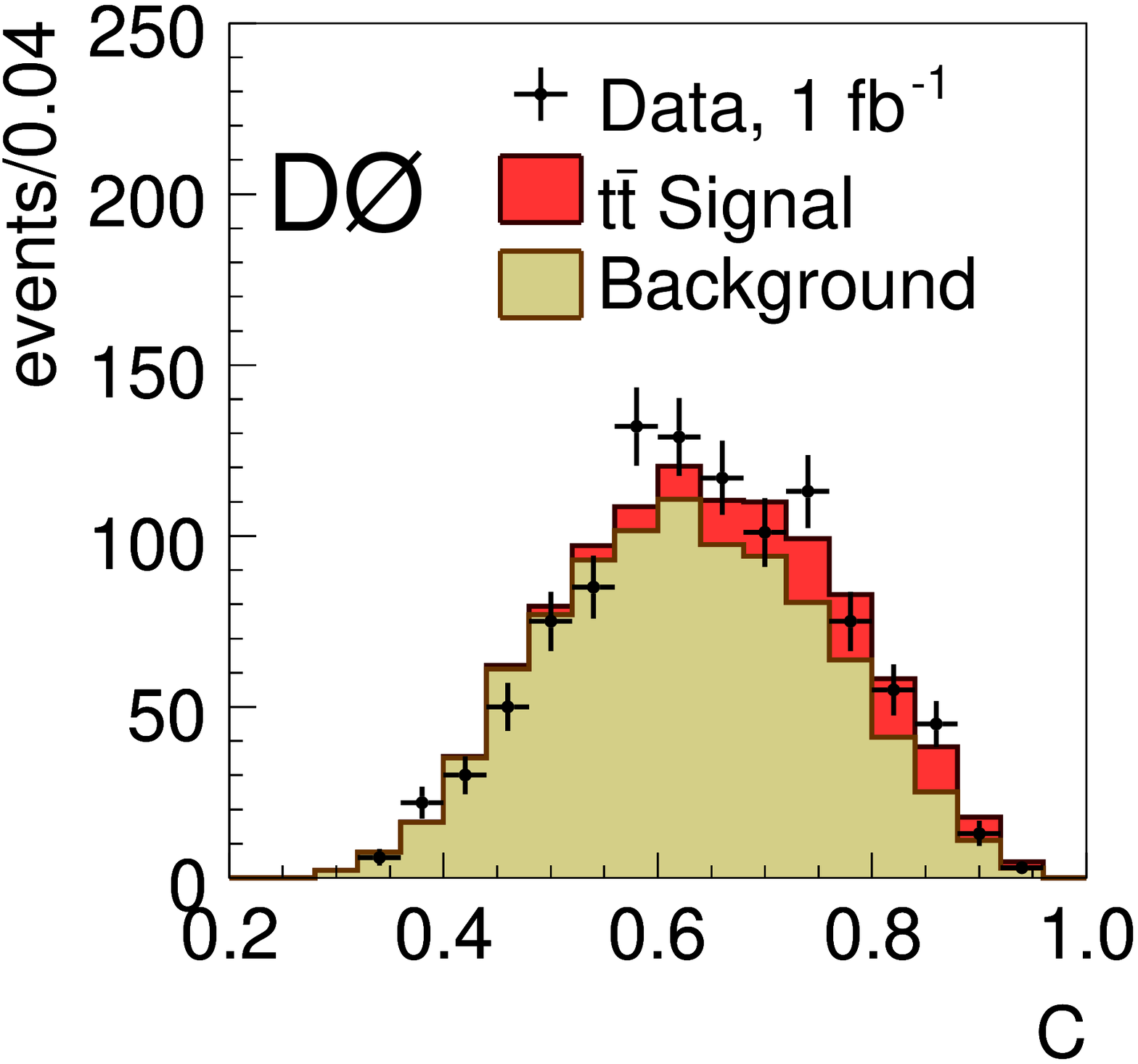}
\put(-130,138){\textsf{\textbf{(a)}}}%
\includegraphics*[width=2.25in,trim=20 5 15 45,clip=true]{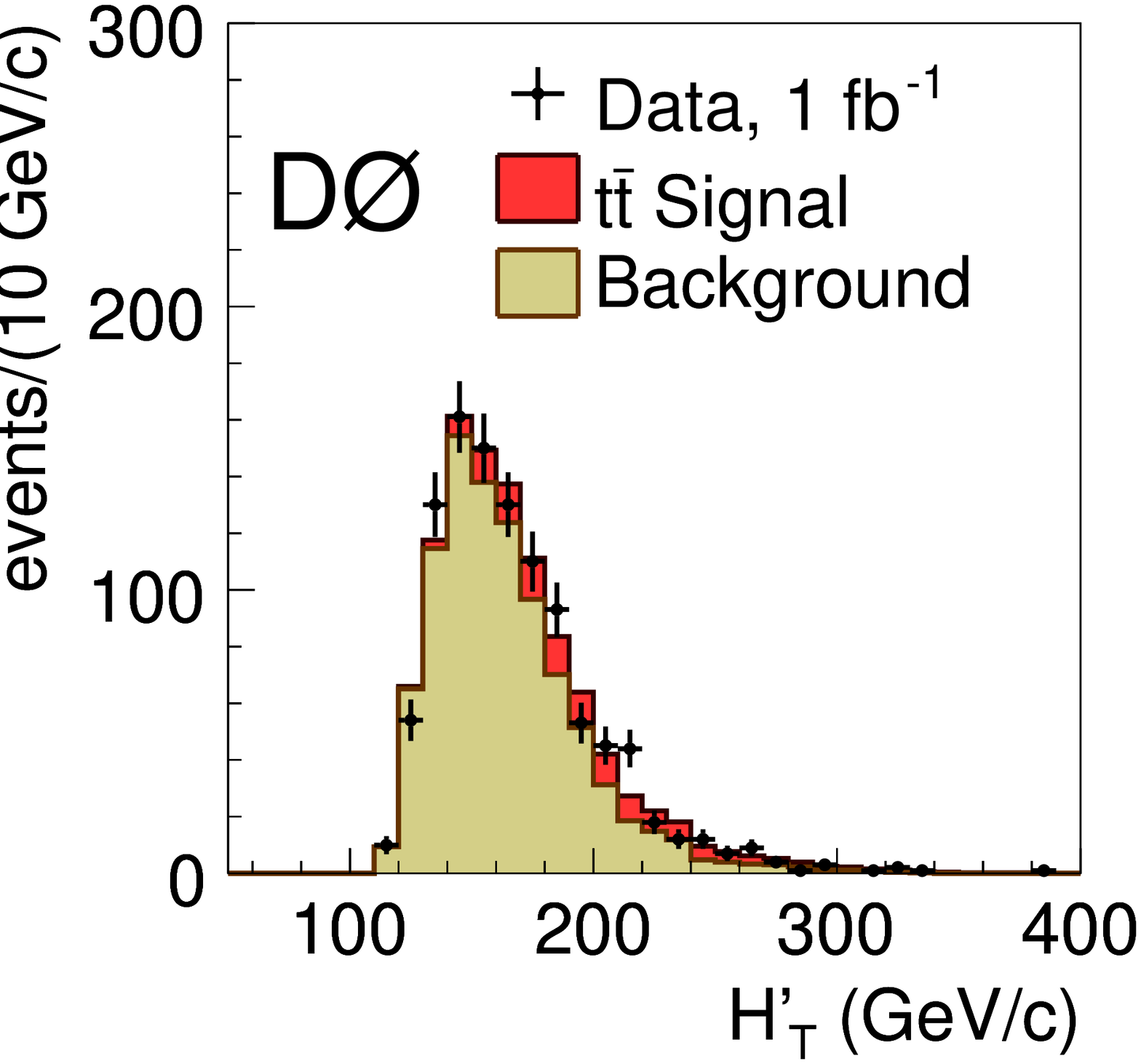}
\put(-130,138){\textsf{\textbf{(b)}}}%
\includegraphics*[width=2.25in,trim=20 5 15 45,clip=true]{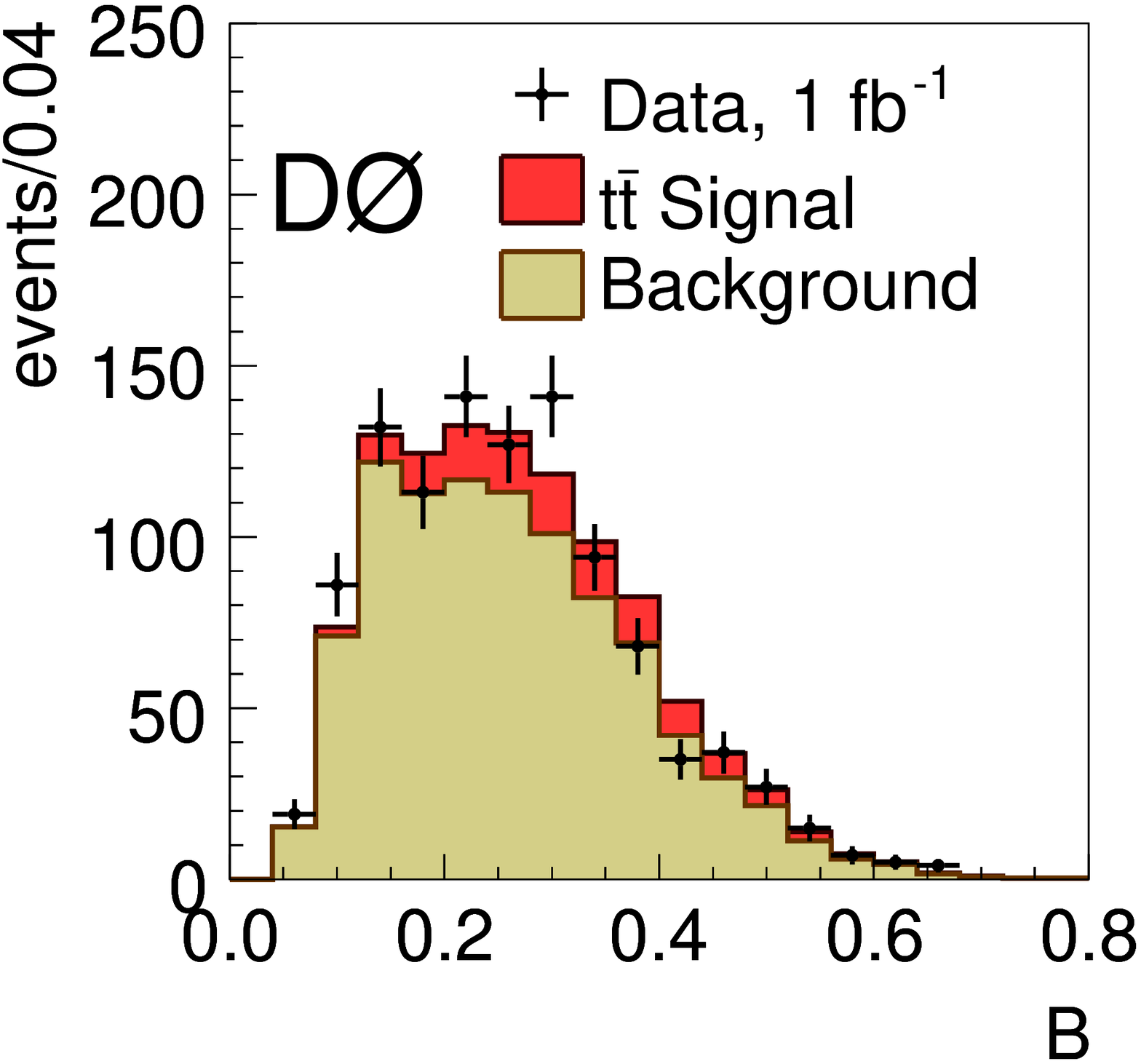}
\put(-130,138){\textsf{\textbf{(c)}}}%
\\
\includegraphics*[width=2.25in,trim=20 5 15 45,clip=true]{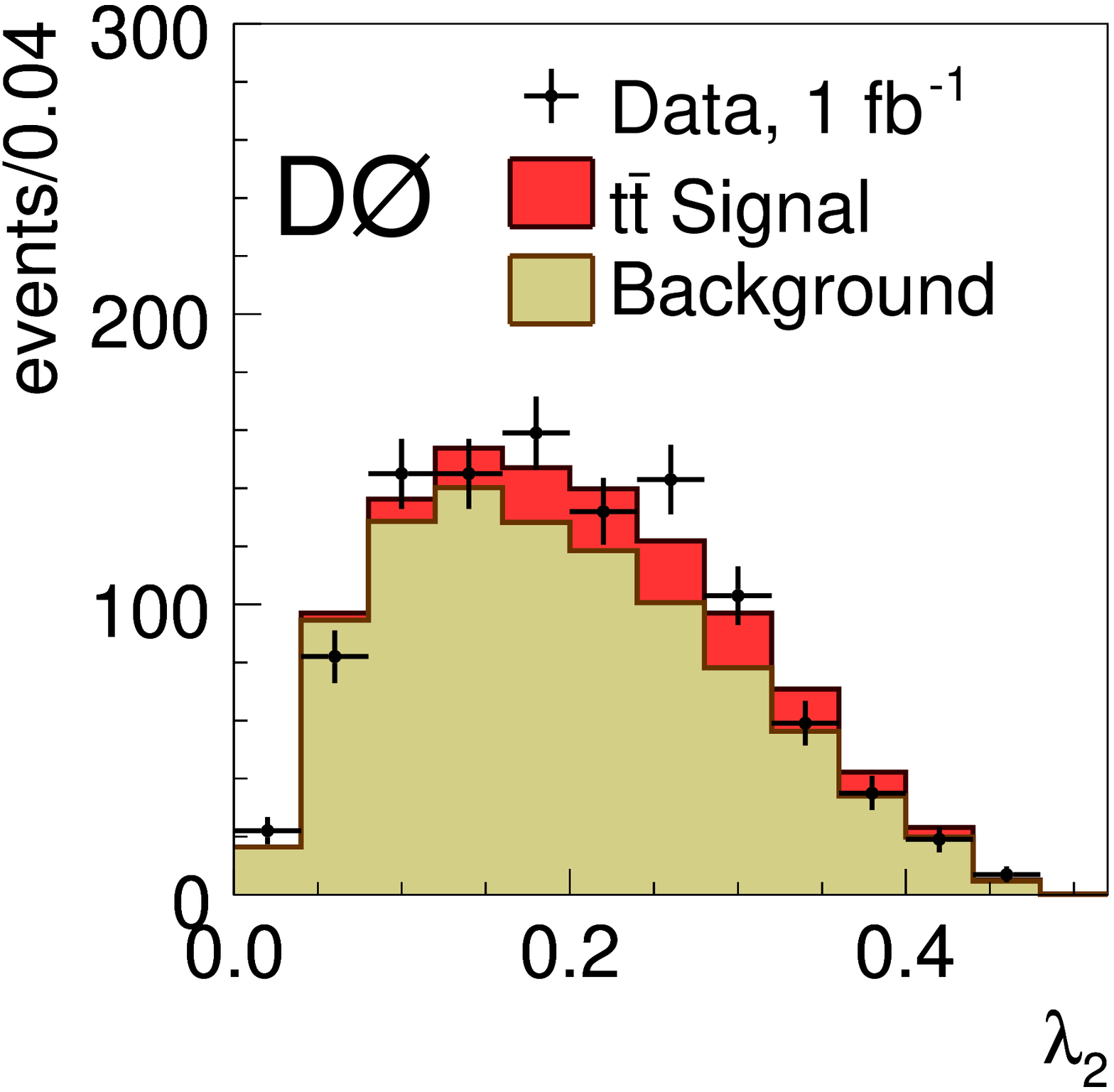}
\put(-130,138){\textsf{\textbf{(d)}}}%
\includegraphics*[width=2.25in,trim=20 5 15 45,clip=true]{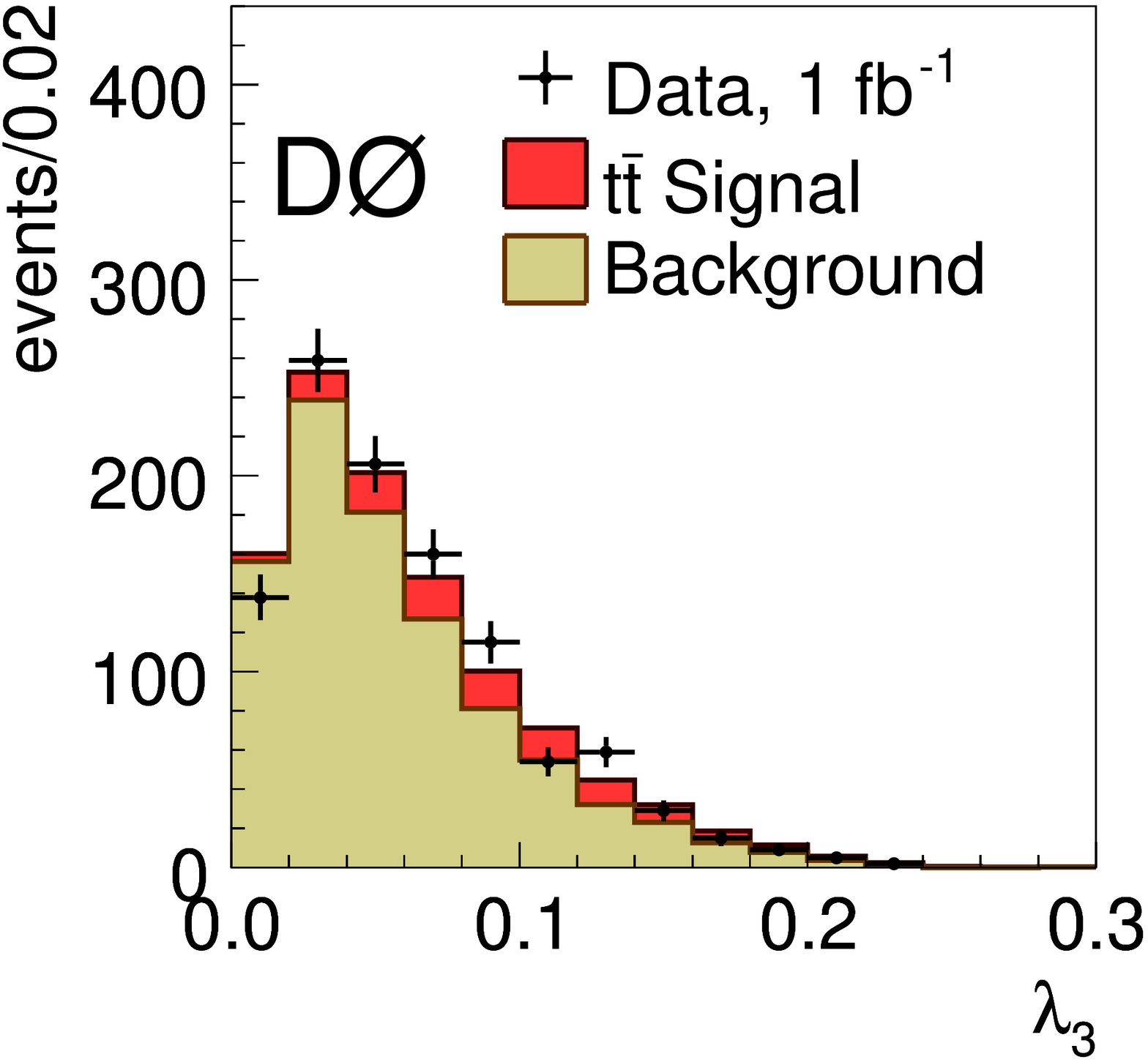}
\put(-130,138){\textsf{\textbf{(e)}}}%
\includegraphics*[width=2.25in,trim=20 5 15 45,clip=true]{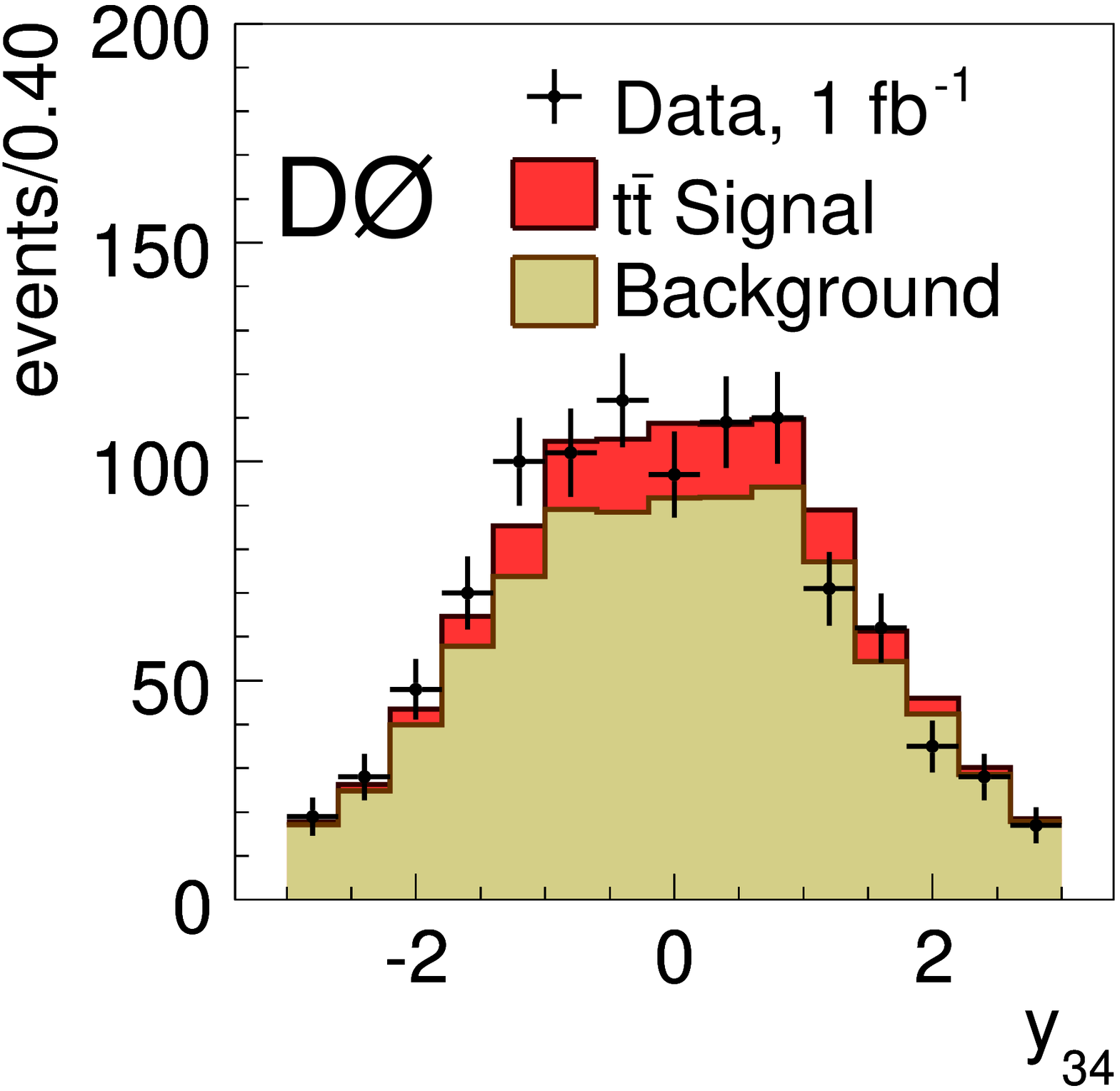}
\put(-130,138){\textsf{\textbf{(f)}}}%
\\
\includegraphics*[width=2.25in,trim=20 5 15 45,clip=true]{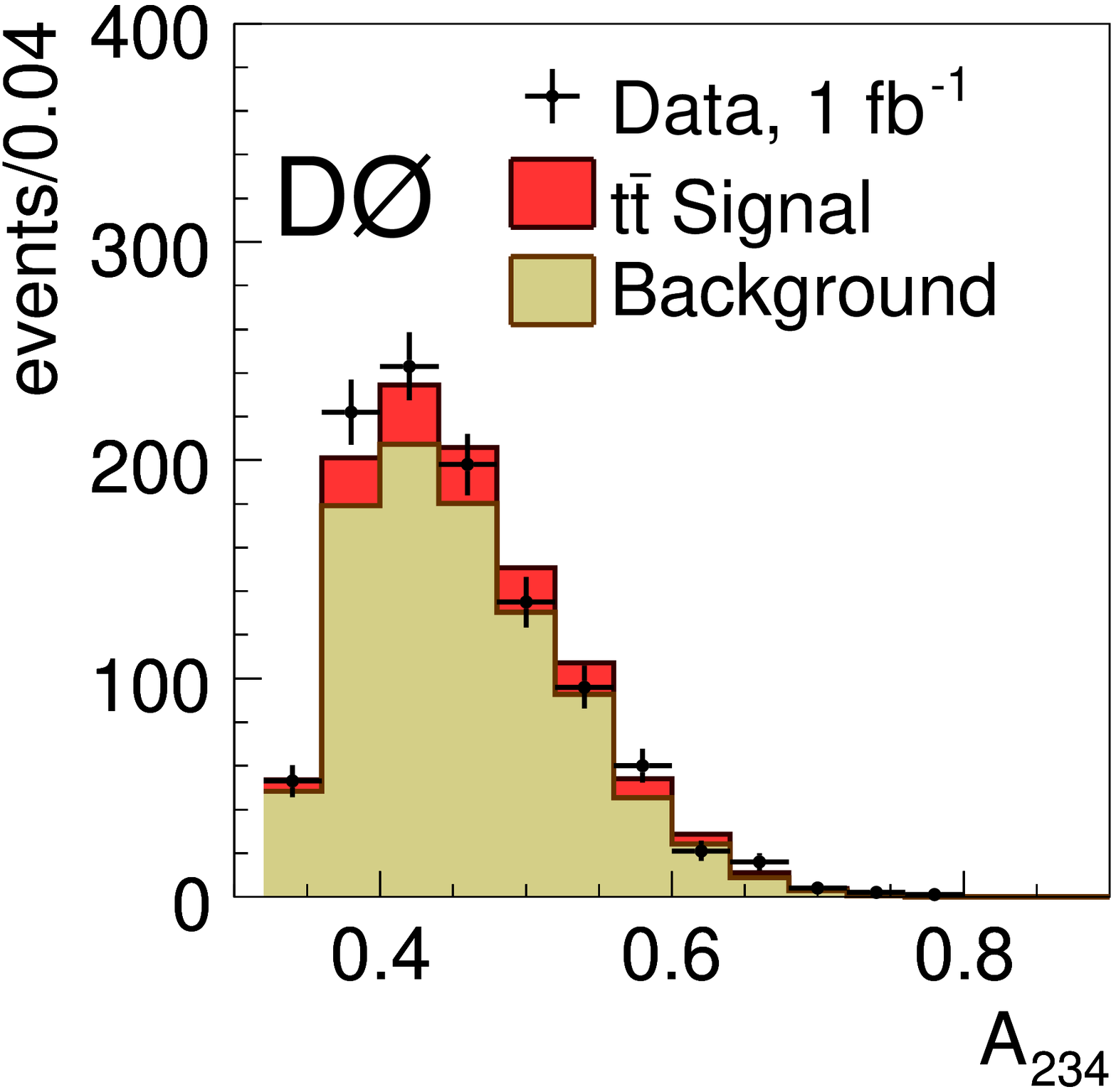}
\put(-130,138){\textsf{\textbf{(g)}}}%
\includegraphics*[width=2.25in,trim=20 5 15 45,clip=true]{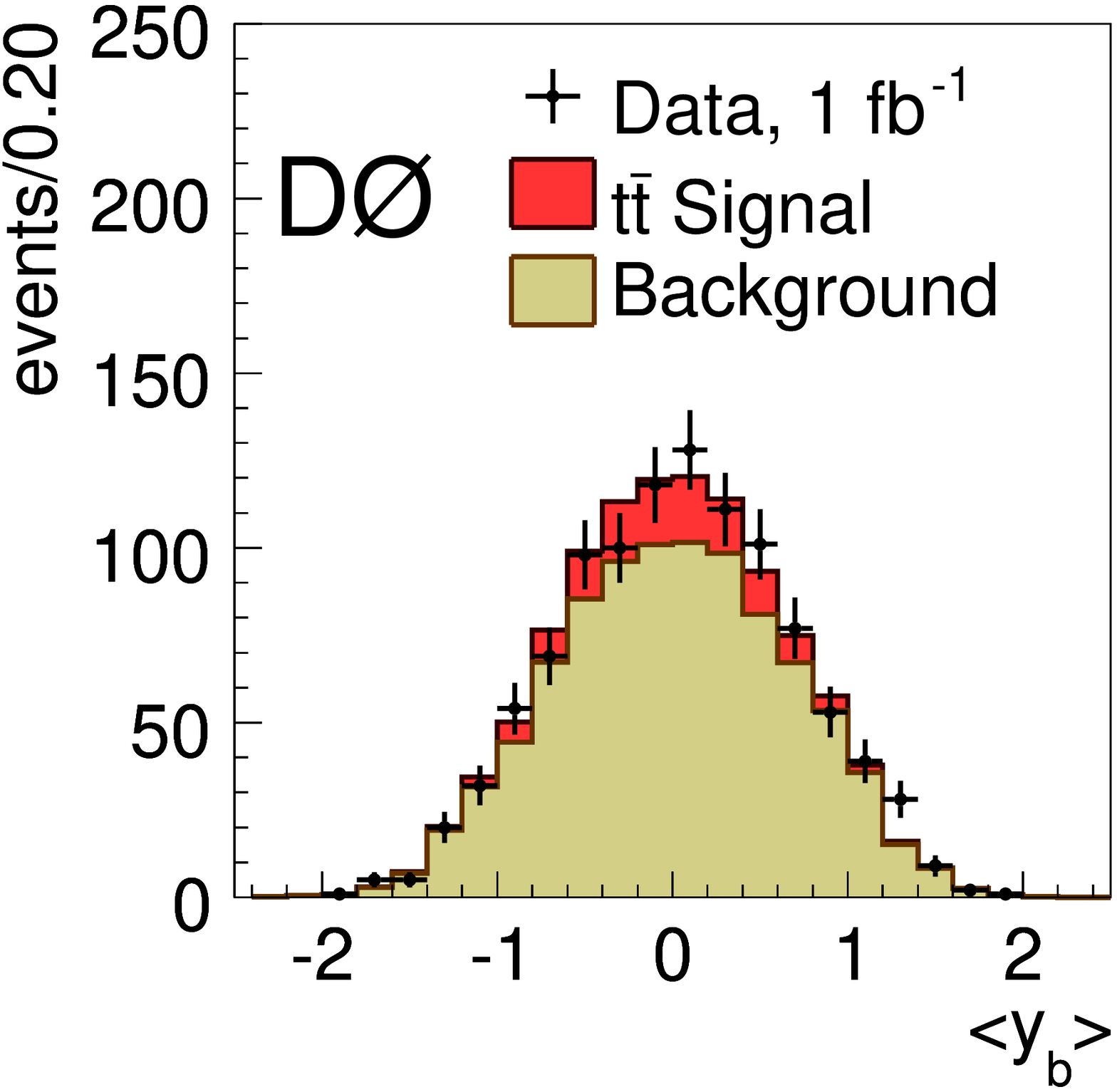}
\put(-130,138){\textsf{\textbf{(h)}}}%
\includegraphics*[width=2.25in,trim=20 5 15 45,clip=true]{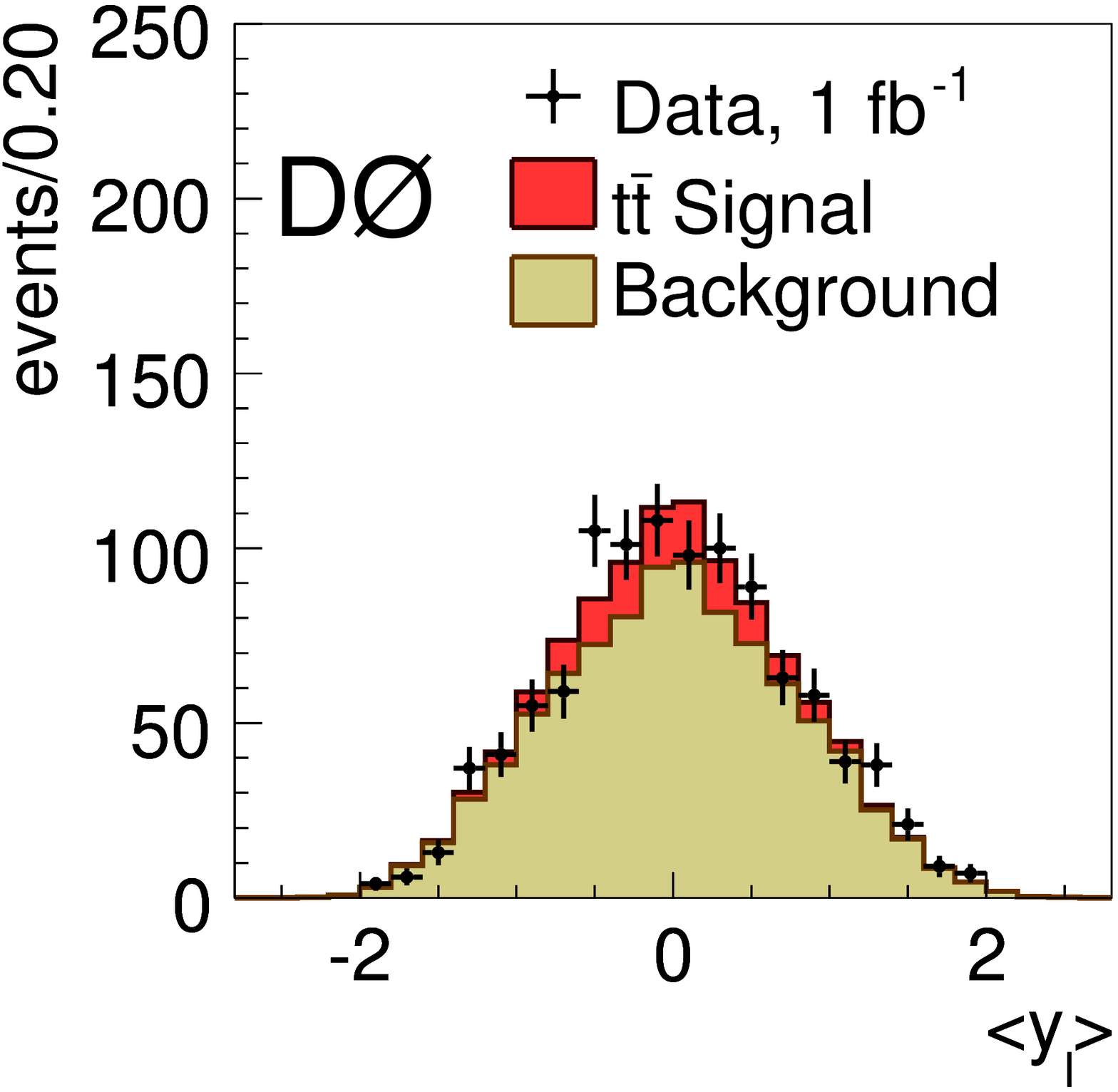}
\put(-130,138){\textsf{\textbf{(i)}}}%
\caption{Comparison between the data candidates and the sum of 
$t\bar{t}$ signal with $m_t=175$~GeV$/c^2$ and the data-based background
for the variables used in the likelihood discriminant.
Displayed error bars represent statistical uncertainties only.
\label{fig:llh_control_175}}
\end{figure*}

The signal and background likelihood templates were fit to the
likelihood output, shown in Fig.~\ref{fig:tmva_output}, for the
selected data events using {\sc tminuit}~\cite{minuit} from {\sc
  root}~\cite{root}.  Results from the fit are shown in
Fig.~\ref{fig:llh_result} and are in agreement with the
data.
The measured signal fractions are $(12.9\pm2.4)\%$ for
$m_t=170$~GeV$/c^2$ and $(12.5\pm2.3)\%$ for $m_t=175$~GeV$/c^2$.
Given 1051 data candidate events, this results in 136 and 131
$t\bar{t}$ events, respectively.
Distributions for the observables included in the likelihood, using
the signal and background fractions from the fit, are shown in
Fig.~\ref{fig:llh_control_175} for $m_t=175$~GeV$/c^2$.  There is
reasonable agreement between the data candidates and the sum of signal
and background, normalized to the fit results.

\begin{figure*}[t]
  \includegraphics[width=2.25in,trim=5 0 30 50,clip=true]{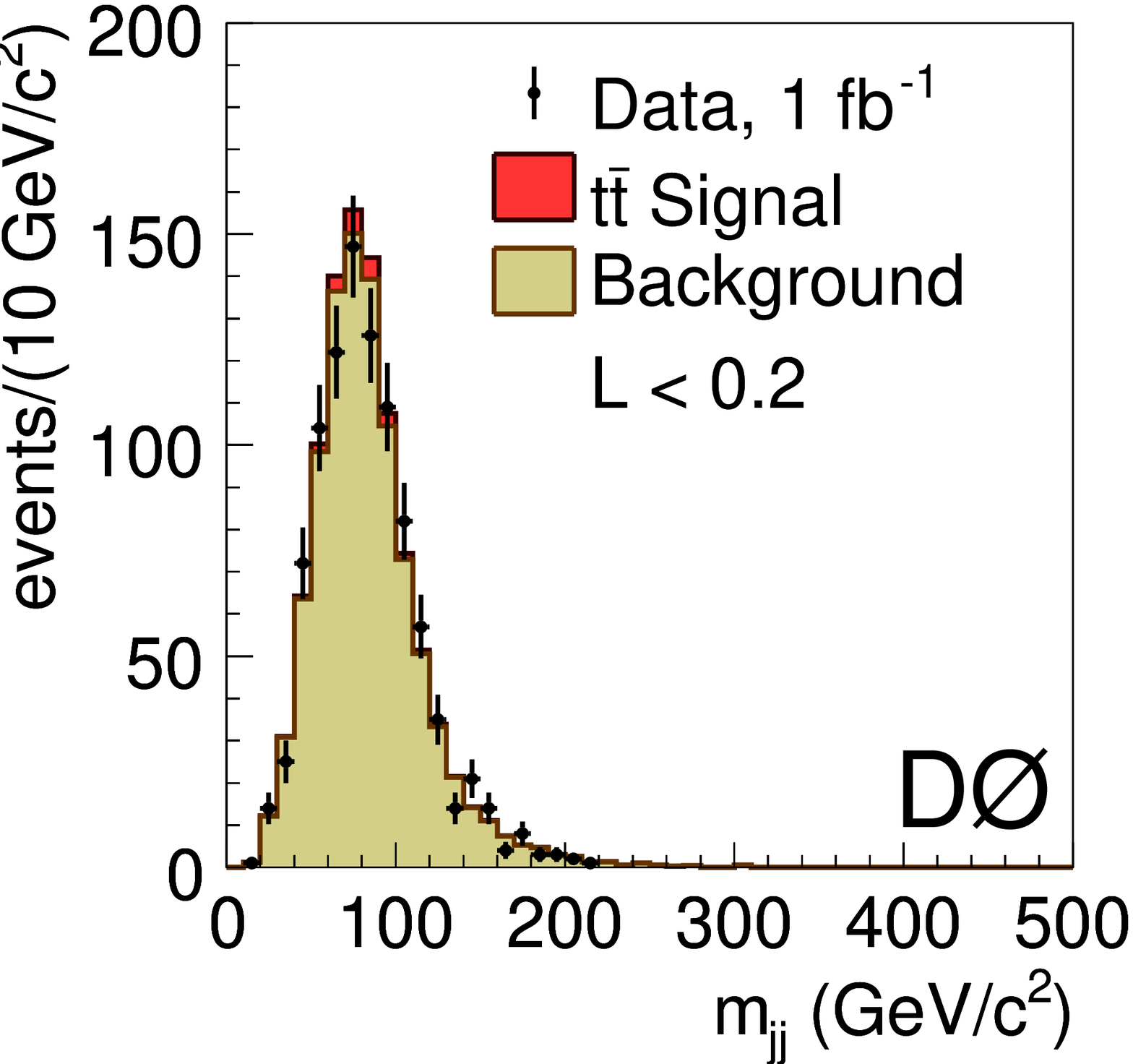}
\put(-125,140){\textsf{\textbf{(a)}}}%
  \includegraphics[width=2.25in,trim=5 0 30 50,clip=true]{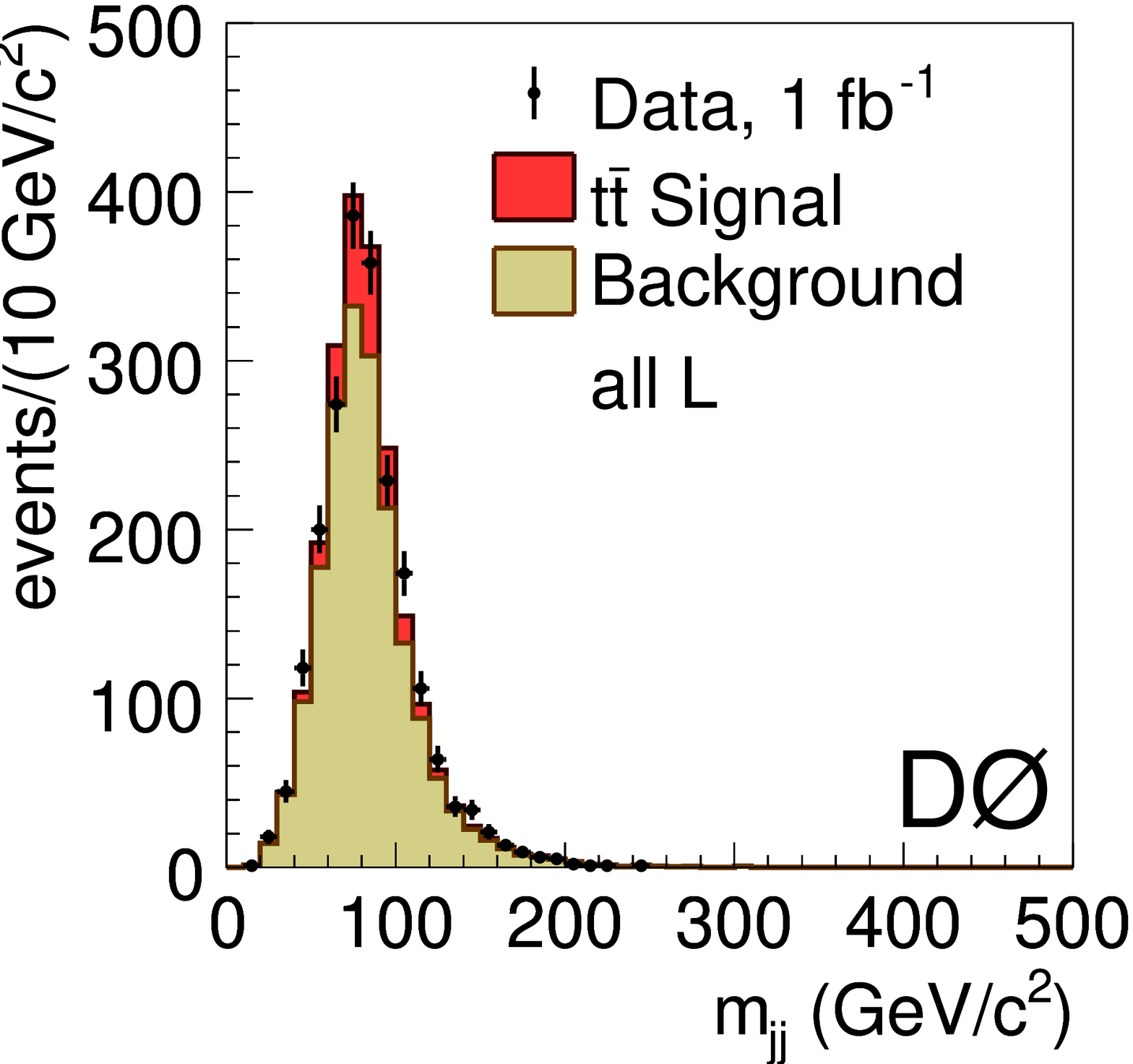}
\put(-125,140){\textsf{\textbf{(c)}}}%
  \includegraphics[width=2.25in,trim=5 0 30 50,clip=true]{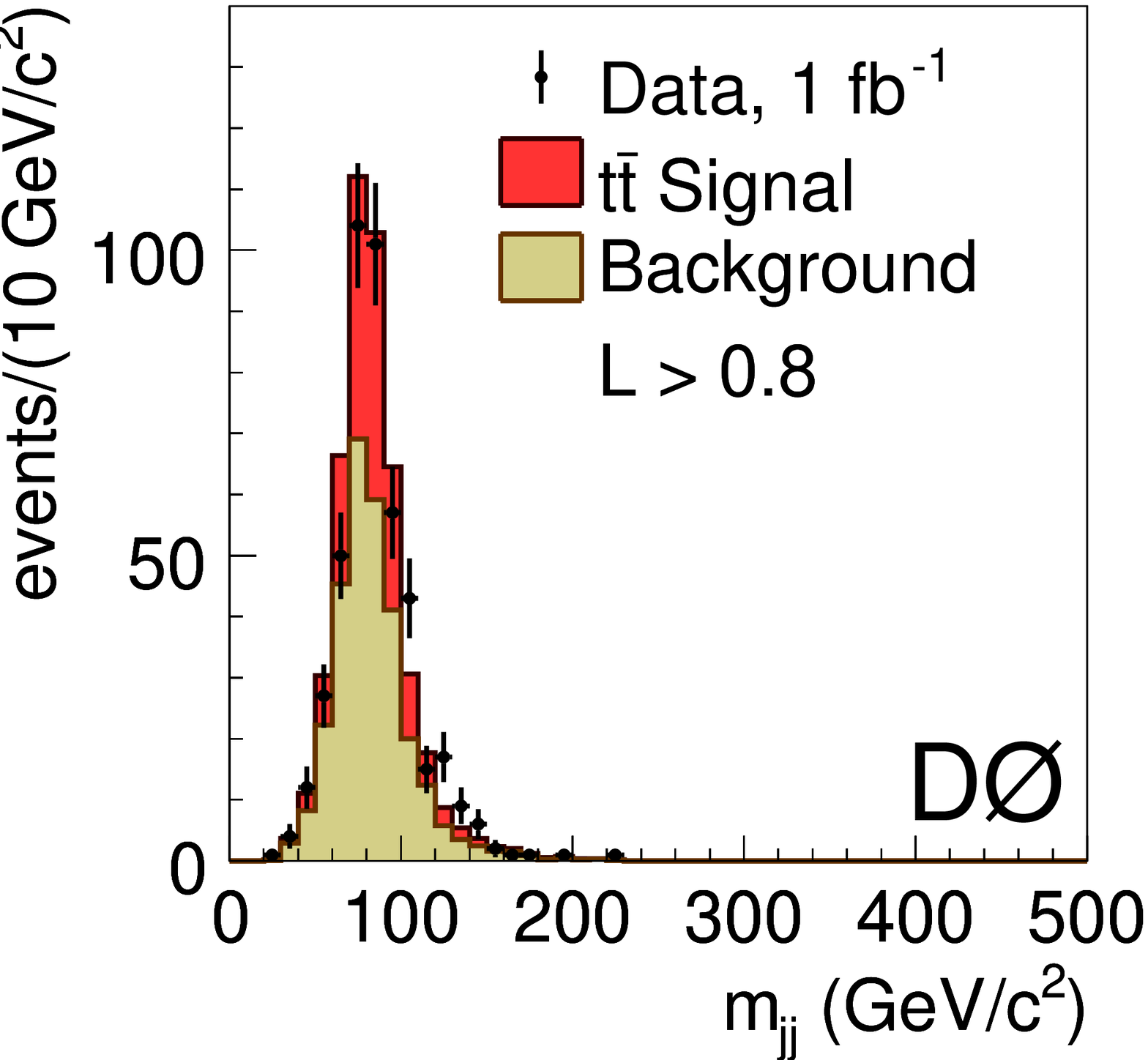}
\put(-125,140){\textsf{\textbf{(e)}}}%
\\
  \includegraphics[width=2.25in,trim=5 0 30 50,clip=true]{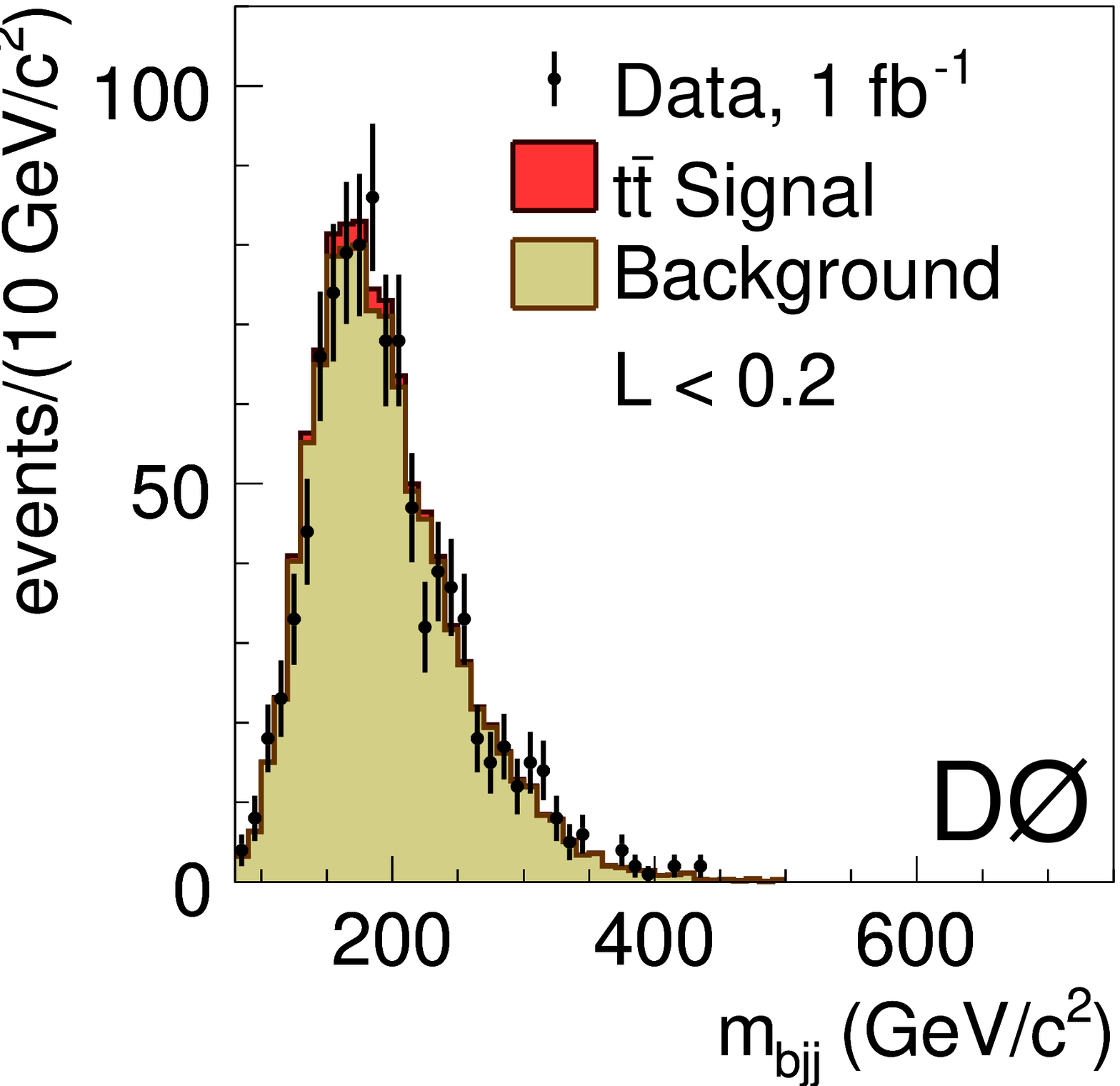}
\put(-125,140){\textsf{\textbf{(b)}}}%
  \includegraphics[width=2.25in,trim=5 0 30 50,clip=true]{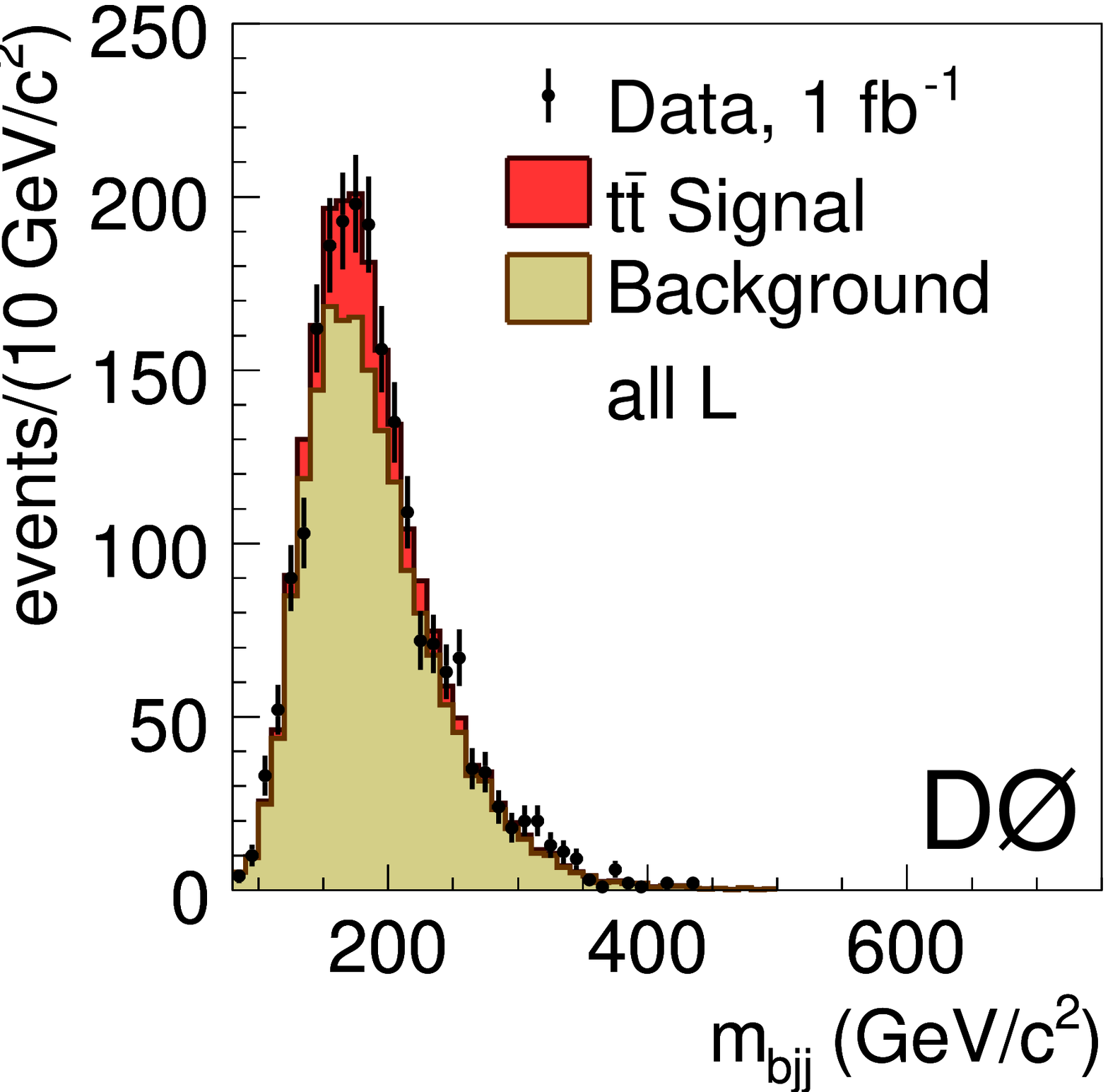}
\put(-125,140){\textsf{\textbf{(d)}}}%
  \includegraphics[width=2.25in,trim=5 0 30 50,clip=true]{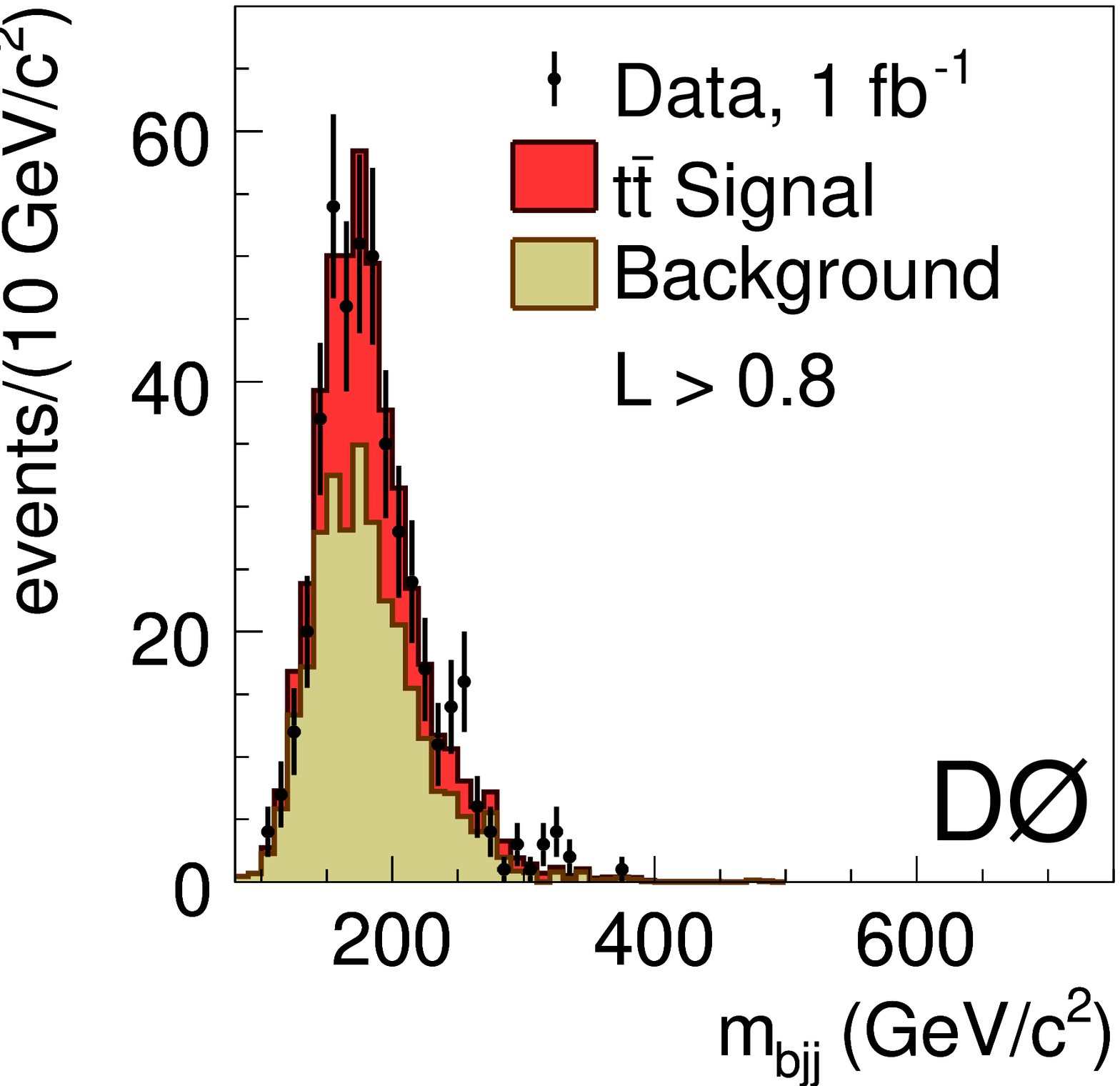}
\put(-125,140){\textsf{\textbf{(f)}}}%
\caption{Distributions for the reconstructed $W$ boson (top row) and
  top quark (bottom row) masses using the $m_t=175$~GeV$/c^2$ signal
  sample.  There are two entries per event.  
Displayed error bars represent statistical uncertainties only.
(c, d) Result from the
  purity fit. (a, b) additionally requiring $L<0.2$ to enhance
  background; (e, f) additionally requiring $L>0.8$ to enhance signal.
}
  \label{fig:llh_mass_175}
\end{figure*}

Jets in an event can be associated with the decays of individual top
quarks.  A $\chi^2$ was constructed comparing the dijet masses with
the $W$ boson mass and the two $bjj$ masses with each other.  The
combination with the lowest $\chi^2$ value was chosen.  The results
for the dijet mass and the $bjj$ mass are shown in
Figs.~\ref{fig:llh_mass_175}(c) and~(d).  There is good agreement
between data and the sum of signal and background.
The comparison is also made in a region of phase space dominated by
background ($L<0.2$) and one which has a significantly larger signal
fraction ($L>0.8$), also shown in Fig.~\ref{fig:llh_mass_175}.  The
distributions were not renormalized.  Both the background-dominated
and signal-enhanced distributions show reasonable agreement between
data and the sum of signal and background.


\subsection{Systematic Uncertainties}

The effects of systematic uncertainties and variations in input
variables were studied using ensemble tests.  Ten thousand
pseudo-experiments were run for each source of uncertainty.  
Each pseudo-experiment drew events from the systematically-shifted
signal and background distributions and was fit using the standard
signal and background likelihood templates.
With the exception of the two background-related systematics, all of
the systematic uncertainties are associated with the signal simulation
only.
All systematic uncertainties on the $t\bar{t}$ production cross
section measured with $m_t=175$~GeV$/c^2$ are summarised in
Table~\ref{tab:uncertainties}.  Many of these are described in more
detail in earlier sections of this paper.

This analysis relies on {\sc alpgen}+{\sc pythia} for the $t\bar{t}$
signal model used to determine the selection efficiency
(Table~\ref{tab:cutflow}) and the kinematic shapes included in the
likelihood determination (Fig.~\ref{fig:likelihood_input}).  It is
possible that the $t\bar{t}$ simulation does not properly reproduce
the properties of the $t\bar{t}$ system. Other analyses in the
lepton+jets and dilepton decay channels published by the D0
collaboration have found good agreement between the simulation and the
reconstructed data~\cite{lepjet1fb,dilepton1fb,singletopa,singletopb}.
Nevertheless, the simulation might mis-estimate the jet multiplicity
through differences in the QCD radiation or the underlying event.
The measured fraction of reconstructed $t\bar{t}$ events (using the
measured signal purities) with seven or more jets is $0.29\pm0.04$.
The signal events were weighted up and down by one standard deviation
in the statistical uncertainty on this ratio ($15\%$).
The entire analysis was repeated and the resulting difference in the
mean cross section applied as a systematic uncertainty.
The PDF in the simulation were also reweighted to correspond to
CTEQ6.5M.  The modified tolerance
approach~\cite{CTEQ6.5,bourilkov_2006_aa} was used to estimate the
effects of the PDF uncertainties on the measured cross section.
Both of these uncertainties, along with those related to the
reweighting of the heavy-flavor fragmentation function, luminosity
profile, and vertex distribution; are listed as the signal model
uncertainty in Table~\ref{tab:uncertainties}.
\begin{table}[t!]
  \caption{Uncertainties on the $t\bar{t}$ cross section categorized by source for the result 
corresponding to $m_t=175$~GeV$/c^2$.
    The uncertainties with $m_t=170$~GeV$/c^2$ are similar. }
  \label{tab:uncertainties}
\begin{ruledtabular}
\begin{tabular}{ld}
Source & \multicolumn{1}{c}{\text{Uncertainty (\%)}} \\ \hline 
Candidate statistics       & \pm18.5 \\
Background model           & \pm10.7 \\ 
Background model statistics & \pm 3.8 \\
Signal model               & \pm 3.2 \\
Signal model statistics    & \pm 0.5 \\
Trigger                    & \multicolumn{1}{c}{$-2.0$ \hglue1ex $+3.9$} \\
Jet identification efficiency  & \multicolumn{1}{c}{$-2.5$ \hglue1ex $+3.0$} \\
Jet taggability            & \pm 8.8 \\
Jet energy calibration     & \pm 10.8 \\
Jet energy resolution      & \multicolumn{1}{c}{$-3.1$ \hglue1ex $+2.2$} \\
$b$ tagging                & \multicolumn{1}{c}{$-8.6$ \hglue1ex $+9.2$} \\ \hline
Total statistical uncertainty    & \pm18.9 \\
Total systematic uncertainty     & \pm20.5 \\
Luminosity uncertainty     & \pm6.1 \\
\end{tabular}
\end{ruledtabular}
\end{table}
\begin{figure}[!t]
  %0.49
%  \includegraphics[width=0.49\textwidth,trim=20 20 20 35,clip=true]{\PATHsmallfonts/ttbar_xsec_theory_vs_measurement_4+40_lin.eps}
  \includegraphics[width=0.49\textwidth,trim=20 50 20 35,clip=true]{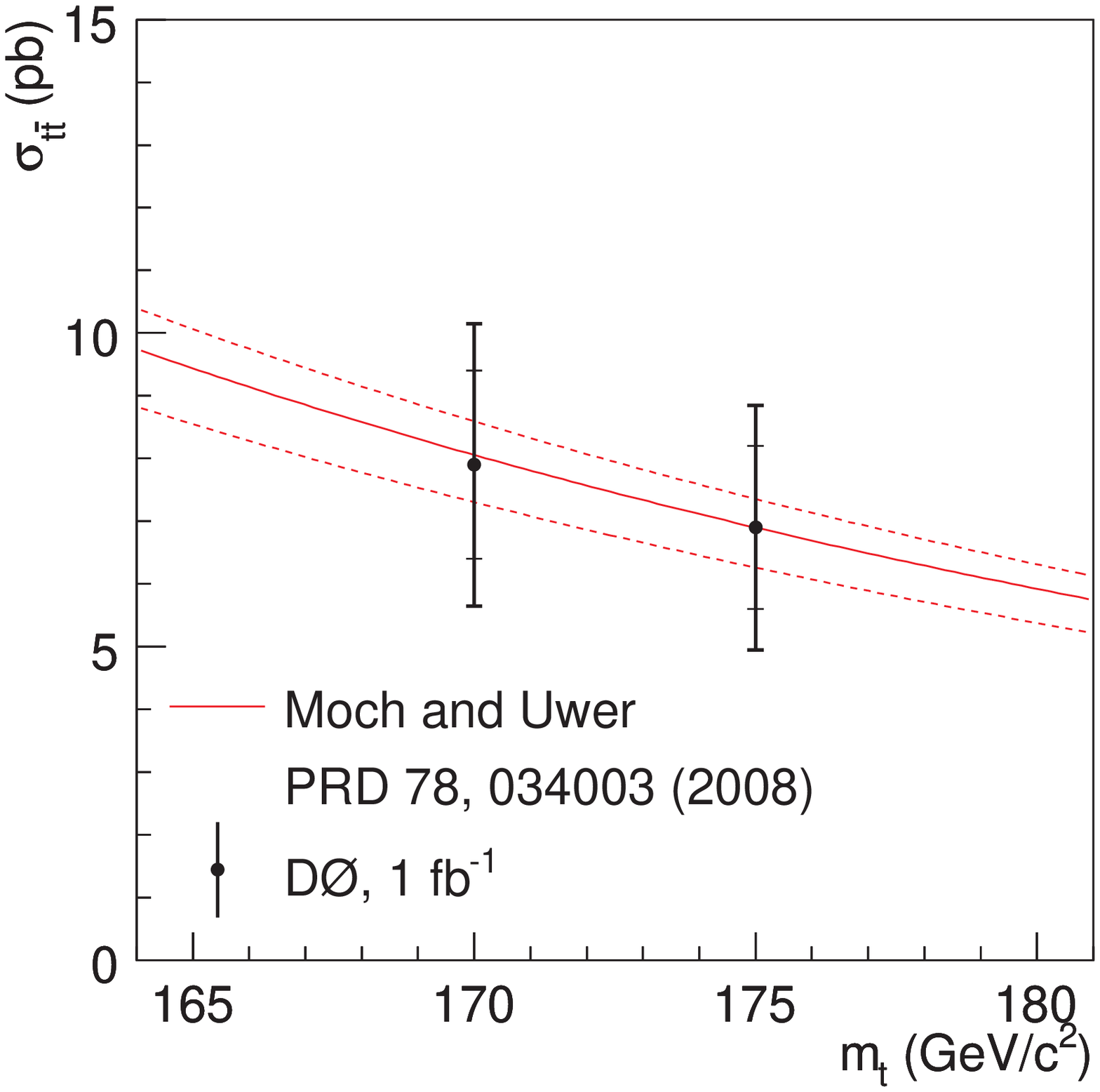}
  \caption{The measured $t\bar{t}$ production cross section ($\bullet$
    markers) together with the SM prediction (solid line)
    and uncertainty (dashed lines)~\cite{moch}.  Shown with the cross
    section measurements are the corresponding statistical uncertainty
    (inner error bars) and total uncertainty (outer error bars).}
  \label{fig:ttbar_xsec_theory_vs_measurement_lin}
\end{figure}

The dominant sources of systematic uncertainty in the $t\bar{t}$ cross
section measurement are the jet energy calibration ($10.8\%$),
construction of the data-based background ($10.7\%$), $b$ tagging
($9.2\%$), and jet taggability ($8.8\%$).  The total systematic
uncertainty is $20.5\%$.

\subsection{Cross Section Measurement}

The cross section is defined as
\begin{equation*}
\sigma_{t\bar{t}} = \frac{f N}{ {\cal L}{\varepsilon}},
\end{equation*}
where $f$ is the measured fraction of $t\bar{t}$ signal, $N$ is the
number of selected data events, $\cal L$ is the integrated luminosity,
and $\varepsilon$ is the inclusive $t\bar{t}$ efficiency given in
Table~\ref{tab:cutflow}.  This results in the following cross
sections:
\begin{align*}
\sigma_{t\bar{t}}^{170\,\text{GeV}/c^2}=7.9\pm1.5\,\text{(stat)}\pm1.6\,\text{(sys)}\pm0.5\,\text{(lum)  pb}\\
\sigma_{t\bar{t}}^{175\,\text{GeV}/c^2}=6.9\pm1.3\,\text{(stat)}\pm1.4\,\text{(sys)}\pm0.4\,\text{(lum)  pb}
\end{align*}
The statistical uncertainty includes the statistical uncertainties
associated with the signal and background templates.  The latter was
determined by re-fitting the data 100,000 times while allowing the
signal and background templates to vary according to their bin-to-bin
statistical uncertainties.  The resulting uncertainties are summarized
in Table~\ref{tab:uncertainties}.
Figure~\ref{fig:ttbar_xsec_theory_vs_measurement_lin} shows the
SM prediction together with the measured cross section
from this analysis. The SM expectation~\cite{moch} is in
agreement with the measured cross sections.

\section{Conclusions}

We presented the inclusive $t\bar{t}$ cross section measured in
$1$~fb${}^{-1}$ of $p\bar{p}$ interactions at $\sqrt{s}=1.96$~TeV.
The cross section was extracted using high-multiplicity jet events,
specifically events with at least six jets, two of them $b$~tagged.  A
model of the multijet background was created from lower
jet-multiplicity data.  A likelihood discriminant was used to separate
signal from background.  The cross section was obtained from a
likelihood fit to the discrimant distribution and was measured to be
$7.9\pm 2.2$~pb assuming $m_t=170$~GeV$/c^2$, and $6.9\pm 2.0$~pb
assuming $m_t=175$~GeV$/c^2$.  Both results agree with theoretical
expectations.

\begin{acknowledgments}
% acknowledgement_paragraph_r2.tex                         10/14/09
%
We thank the staffs at Fermilab and collaborating institutions, 
and acknowledge support from the 
DOE and NSF (USA);
CEA and CNRS/IN2P3 (France);
FASI, Rosatom and RFBR (Russia);
CNPq, FAPERJ, FAPESP and FUNDUNESP (Brazil);
DAE and DST (India);
Colciencias (Colombia);
CONACyT (Mexico);
KRF and KOSEF (Korea);
CONICET and UBACyT (Argentina);
FOM (The Netherlands);
STFC and the Royal Society (United Kingdom);
MSMT and GACR (Czech Republic);
CRC Program, CFI, NSERC and WestGrid Project (Canada);
BMBF and DFG (Germany);
SFI (Ireland);
The Swedish Research Council (Sweden);
and
CAS and CNSF (China).

\end{acknowledgments}


\begin{thebibliography}{99}
\bibitem{pseudorapidity} 
Pseudorapidity $\eta$ is defined as $\eta=-\ln[\tan(\theta/2)]$, where
$\theta$ is the polar angle with respect to the proton beam direction,
with origin at the center of the detector.
\bibitem{yphi}
Rapidity is defined as $y=-\ln[(E + p_z)/(E - p_z)]$, where $E$ is the
energy and $p_z$ is the momentum along the proton beam direction.
$\phi$ is defined as the azimuthal angle in the plane transverse to
the proton beam direction.
\bibitem{variations} $\not\!\!E_T/H_{T4}$ was varied between $0.12$
  and $0.08$, covering 68\% CL, to allow more or less phase space for
  additional jets.
\bibitem{topmassrange} Two top quark mass values were
  considered since the world-average mass varied between~$170.9\pm1.8$~GeV$/c^2$
  and~$174.2\pm3.4$~GeV$/c^2$ during the time period covered by this
  analysis.
\bibitem{lambda} 
Aplanarity is defined as $(3/2)\,\lambda_3$ and 
sphericity as $(3/2)\,(\lambda_2+\lambda_3)$.
\end{thebibliography}
\end{document}